\documentclass[11pt]{article}
\usepackage{graphicx,ifthen,color,algorithm,palatino}
\usepackage{amsmath,amsthm,amssymb,amsfonts,verbatim}
\usepackage{mathrsfs,accents,setspace}
\usepackage{eucal}
\def\smalldot{\mbox{\fontsize{0.1mm}{0.5em}\selectfont{$\bullet$}}}
\def\bBdot{\overset{\ \smalldot}{\bB}}

\def\bBdotdot{\overset{\smalldot\smalldot}{\bB}}

\def\bDdot{\overset{\ \smalldot}{\bD}}

\def\AtLev{\bA}
\def\ALoo{\bA_{11}}     \def\AUoo{\bA^{11}}
\def\ALotCo{\bA_{12,1}} \def\AUotCo{\bA^{12,1}} \def\AUotCoT{\bA^{12,1\,T}}
\def\ALotCt{\bA_{12,2}} \def\AUotCt{\bA^{12,2}} \def\AUotCtT{\bA^{12,2\,T}}
\def\ALotCi{\bA_{12,i}}	\def\AUotCi{\bA^{12,i}} 
\def\ALotCm{\bA_{12,m}}	\def\AUotCm{\bA^{12,m}} \def\AUotCmT{\bA^{12,m\,T}}
\def\AUotCij{\bA^{12,ij}}
\def\AUotCicommaj{\bA^{12,i,j}}
\def\ALttCo{\bA_{22,1}}	\def\AUttCo{\bA^{22,1}}
\def\ALttCt{\bA_{22,2}}	\def\AUttCt{\bA^{22,2}}
\def\ALttCi{\bA_{22,i}}	\def\AUttCi{\bA^{22,i}}
\def\ALttCm{\bA_{22,m}}	\def\AUttCm{\bA^{22,m}}
\def\AUttCij{\bA^{22,ij}}
\def\xveco{\bx_1}
\def\xvectCo{\bx_{2,1}}
\def\xvectCt{\bx_{2,2}}
\def\xvectCi{\bx_{2,i}}
\def\xvectCij{\bx_{2,ij}}
\def\xvectCm{\bx_{2,m}}
\def\aveco{\ba_1}
\def\avectCo{\ba_{2,1}}
\def\avectCt{\ba_{2,2}}
\def\avectCm{\ba_{2,m}}
\def\avectCi{\ba_{2,i}}
\def\bveco{\boldsymbol{b}_1}
\def\bvect{\boldsymbol{b}_2}
\def\bveci{\boldsymbol{b}_i}
\def\bvecm{\boldsymbol{b}_m}

\def\Bmato{\bB_1}
\def\Bmatt{\bB_2}
\def\Bmati{\bB_i}
\def\Bmatm{\bB_m}

\def\Bmatdoto{\bBdot_1}
\def\Bmatdott{\bBdot_2}
\def\Bmatdoti{\bBdot_i}
\def\Bmatdotm{\bBdot_m}

\def\cveczi{\bc_{0i}}
\def\cvecoi{\bc_{1i}}
\def\cvecti{\bc_{2i}}
\def\Cmatzi{\bC_{0i}}
\def\Cmatoi{\bC_{1i}}
\def\Cmatti{\bC_{2i}}
\def\sumim{\sum_{i=1}^m}

\def\bigX{{\LARGE\mbox{$\times$}}}
\def\nadj{{\tilde n}}
\def\oadj{{\tilde o}}
\newboolean{ShowFigures}
\newboolean{UnBlinded}
\newboolean{DoubleSpaced}
\setboolean{ShowFigures}{true}
\setboolean{UnBlinded}{true}
\setboolean{DoubleSpaced}{false}
\newtheorem{result}{\textbf{Result}}
\newcommand{\A}[1]{\bA_{#1}}

\newcommand{\Ainv}[1]{\bA^{#1}}
\newcommand{\ATinv}[1]{\bA^{#1\,T}}
\newcommand*{\dt}[1]{\accentset{\mbox{\smalldot}}{#1}}
\newcommand*{\ddt}[1]{\accentset{\mbox{\smalldot\smalldot}}{#1}}
\newcommand{\B}[1]{\bB_{#1}}

\newcommand{\dB}[1]{\dt{\boldsymbol{B}}_{#1}}
\newcommand{\ddB}[1]{\ddt{\boldsymbol{B}}_{#1}}

\def\const{\mbox{\rm const}}
\def\myand{\&\ }

\def\by{\boldsymbol{y}}
\def\bbeta{\boldsymbol{\beta}}
\def\bu{\boldsymbol{u}}
\def\bR{\boldsymbol{R}}
\def\bX{\boldsymbol{X}}
\def\bZ{\boldsymbol{Z}}
\def\bG{\boldsymbol{G}}
\def\bzero{\boldsymbol{0}}
\def\bmu{\boldsymbol{\mu}}
\def\bSigma{\boldsymbol{\Sigma}}
\def\bA{\boldsymbol{A}}
\def\bC{\boldsymbol{C}}
\def\bD{\boldsymbol{D}}
\def\bO{\boldsymbol{O}}
\def\bI{\boldsymbol{I}}
\def\bx{\boldsymbol{x}}
\def\smhalf{{\textstyle{\frac{1}{2}}}}
\def\bT{\boldsymbol{T}}
\def\bdeta{\boldsymbol{\eta}}
\def\vech{\mbox{\rm vech}}
\def\vecof{\mbox{\rm vec}}
\def\quarter{{\textstyle{1\over4}}}
\def\ba{\boldsymbol{a}}
\def\bb{\boldsymbol{b}}
\def\bc{\boldsymbol{c}}
\def\bd{\boldsymbol{d}}
\def\bM{\boldsymbol{M}}
\def\bLambda{\boldsymbol{\Lambda}}
\def\tr{\mbox{tr}}
\def\bomega{\boldsymbol{\omega}}
\def\bOmega{\boldsymbol{\Omega}}
\def\bB{\boldsymbol{B}}
\def\bQ{\boldsymbol{Q}}
\def\bR{\boldsymbol{R}}
\def\bh{\boldsymbol{h}}
\def\bH{\boldsymbol{H}}
\def\stackdum{\mathop{\mbox{\rm stack}}}
\def\stack#1{\stackdum_{#1}}
\def\blockdiagdum{\mathop{\mbox{\rm blockdiag}}}
\def\blockdiag#1{\blockdiagdum_{#1}}
\def\simind{\stackrel{{\tiny \mbox{ind.}}}{\sim}}

\def\Ssc{{\mathcal S}}
\def\diag{\mbox{diag}}
\def\bo{\boldsymbol{o}}
\def\jump{\vskip3mm\noindent}
\def\trans{^T}

\def\bib{\vskip12pt\par\noindent\hangindent=1 true cm\hangafter=1}
\def\pDens{\mathfrak{p}}
\def\qDens{\mathfrak{q}}
\def\pDensUnder{\underline{\pDens}}
\def\Gfull{G_{\mbox{\tiny full}}}
\def\Gdiag{G_{\mbox{\tiny diag}}}
\def\bAbG{\bA_{\mbox{\tiny$\bG$}}}
\def\bAbR{\bA_{\mbox{\tiny$\bR$}}}
\def\iCOMMAj{\,i,\,j}

\def\veryTinyBLUP{\mbox{\fontsize{1.5mm}{1em}\selectfont {\bf BLUP}}}
\def\veryTinyMFVB{\mbox{\fontsize{1.5mm}{1em}\selectfont {\bf MFVB}}}
\def\DBLUP{\bD_{\veryTinyBLUP}}
\def\DMFVB{\bD_{\veryTinyMFVB}}
\def\RBLUP{\bR_{\veryTinyBLUP}}
\def\RMFVB{\bR_{\veryTinyMFVB}}

\def\oMFVB{\bo_{\veryTinyMFVB}}
\def\Cov{\mbox{Cov}}
\def\qLone{q_1}
\def\qLtwo{q_2}
\def\hfill{\ \ \ \ }

\def\buLone{\bu^{\mbox{\tiny\rm L1}}}
\def\buLtwo{\bu^{\mbox{\tiny\rm L2}}}

\def\bSigmaLone{\bSigma^{\mbox{\tiny\rm L1}}}
\def\bSigmaLtwo{\bSigma^{\mbox{\tiny\rm L2}}}
\def\bSigmaLoneMinusOne{(\bSigma^{\mbox{\tiny\rm L1}})^{-1}}
\def\bSigmaLtwoMinusOne{(\bSigma^{\mbox{\tiny\rm L2}})^{-1}}
\def\bZLone{\bZ^{\mbox{\tiny\rm L1}}}

\def\bZLtwo{\bZ^{\mbox{\tiny\rm L2}}}

\def\bASigma{\bA_{\mbox{\tiny$\bSigma$}}}

\def\ThreeLevelNaturalToCommonParameters{\textsc{\footnotesize ThreeLevelNaturalToCommonParameters}}

\def\SscAlgOne{\Ssc_1}
\def\SscAlgTwo{\Ssc_2}
\def\SscAlgThree{\Ssc_3}
\def\SscAlgFour{\Ssc_4}
\def\SscAlgFive{\Ssc_5}
\def\SscAlgSix{\Ssc_6}
\def\SscAlgSeven{\Ssc_7}
\def\SscAlgEight{\Ssc_8}
\def\SscFirstInText{\Ssc_A}
\def\SscSecondInText{\Ssc_B}

\def\bomegaAlgTwoA{\bomega_{1}} 
\def\bomegaAlgTwoB{\bomega_{2}} 
\def\bOmegaAlgTwoC{\bOmega_{3}} 
\def\bomegaAlgTwoD{\bomega_{4i}} 
\def\bomegaAlgTwoE{\bomega_{5}}  
\def\bomegaAlgTwoF{\bomega_{6}}  
\def\bOmegaAlgTwoG{\bOmega_{7i}}
\def\bOmegaAlgTwoH{\bOmega_{8i}}

\def\bomegaAlgThreeA{\bomega_{9}}
\def\bomegaAlgThreeB{\bomega_{10}} 
\def\bomegaAlgThreeC{\bomega_{11}}

\def\omegaAlgFourA{\omega_{12}}   
\def\bomegaAlgFourB{\bomega_{13}}   
\def\bomegaAlgFourC{\bomega_{14}}  

\def\omegaAlgSixA{\bomega_{15}} 
\def\omegaAlgSixB{\bomega_{16}} 
\def\OmegaAlgSixC{\bOmega_{17}} 
\def\omegaAlgSixD{\bomega_{18i}} 
\def\omegaAlgSixE{\bomega_{19}} 
\def\omegaAlgSixF{\bomega_{20}} 
\def\OmegaAlgSixG{\bOmega_{21i}} 
\def\OmegaAlgSixH{\bOmega_{22i}} 
\def\omegaAlgSixI{\bomega_{23ij}} 
\def\omegaAlgSixJ{\bomega_{24}} 
\def\omegaAlgSixK{\bomega_{25}} 
\def\omegaAlgSixL{\bomega_{26}} 
\def\OmegaAlgSixM{\bOmega_{27ij}} 
\def\OmegaAlgSixN{\bOmega_{28ij}} 
\def\OmegaAlgSixO{\bOmega_{29ij}}

\def\omegaAlgSevenA{\bomega_{30}} 
\def\omegaAlgSevenB{\bomega_{31}} 
\def\omegaAlgSevenC{\omega_{32}} 
\def\omegaAlgSevenD{\bomega_{33i}}  
\def\omegaAlgSevenE{\bomega_{34i}} 
\def\omegaAlgSevenF{\bomega_{35i}} 

\def\omegaAlgEightA{\omega_{36}} 
\def\omegaAlgEightB{\bomega_{37}} 
\def\omegaAlgEightC{\omega_{38}} 
\def\omegaAlgEightD{\bomega_{39}} 
\def\omegaAlgEightE{\bomega_{40}} 
\def\omegaAlgEightF{\bomega_{41}} 

\def\bomegaAlgAoneA{\bomega_{42}}   
\def\bOmegaAlgAoneB{\bOmega_{43}}

\def\bomegaAlgAtwoA{\bomega_{44}}   
\def\bOmegaAlgAtwoB{\bOmega_{45}}   

\def\bomegaAlgAthreeA{\bomega_{46}}
\def\bOmegaAlgAthreeB{\bOmega_{47}}

\def\bomegaAlgAfourA{\bomega_{48}} 
\def\bOmegaAlgAfourB{\bOmega_{49}} 
\def\bomegaAlgAfourC{\bomega_{50}} 
\def\bOmegaAlgAfourD{\bOmega_{51}}  
\def\bOmegaAlgAfourE{\bOmega_{52}} 

\def\buLonei{\bu^{\mbox{\tiny\rm L1}}_{i}}
\def\buLtwoij{\bu^{\mbox{\tiny\rm L2}}_{ij}}

\def\iStt{i_{\mbox{\tiny stt}}}
\def\iEnd{i_{\mbox{\tiny end}}}

\def\biggerbdeta{\mbox{\large $\bdeta$}}

\def\biggerm{\mbox{\Large $m$}}

\def\MqSigmaLone{\bM_{\qDens(\bSigmaLone)}}
\def\MqSigmaLtwo{\bM_{\qDens(\bSigmaLtwo)}}

\def\LambdaqASigmaLone{\bLambda_{\qDens(\bASigmaLone)}}
\def\LambdaqASigmaLtwo{\bLambda_{\qDens(\bASigmaLtwo)}}

\def\mSUBpbetauGTObetau
{\biggerm_{\mbox{\footnotesize$\pDens(\bbeta,\bu|\bG)\rightarrow(\bbeta,\bu)$}}(\bbeta,\bu)}

\def\mSUBpybetauRTObetau
{\biggerm_{\mbox{\footnotesize$\pDens(\by|\bbeta,\bu,\bR)\rightarrow(\bbeta,\bu)$}}(\bbeta,\bu)}

\def\mSUBbetauTOpbetauSigma
{\biggerm_{\mbox{\footnotesize$(\bbeta,\bu)\rightarrow\pDens(\bbeta,\bu|\,\bSigma)$}}(\bbeta,\bu)}

\def\mSUBpbetauSigmaTObetau
{\biggerm_{\mbox{\footnotesize$\pDens(\bbeta,\bu|\,\bSigma)\rightarrow(\bbeta,\bu)$}}(\bbeta,\bu)}

\def\mSUBpbetauSigmaLoneLtwoTObetau
{\biggerm_{\mbox{\footnotesize$\pDens(\bbeta,\bu|\,\bSigmaLone,\bSigmaLtwo)\rightarrow(\bbeta,\bu)$}}(\bbeta,\bu)}

\def\etaSUBpbetauSigmaLoneLtwoTObetau
{\biggerbdeta_{\mbox{\footnotesize$\pDens(\bbeta,\bu|\,\bSigmaLone,\bSigmaLtwo)\rightarrow(\bbeta,\bu)$}}}

\def\mSUBpbetauSigmaLoneLtwoTOSigmaLone
{\biggerm_{\mbox{\footnotesize$\pDens(\bbeta,\bu|\,\bSigmaLone,\bSigmaLtwo)\rightarrow\bSigmaLone$}}(\bSigmaLone)}

\def\mSUBSigmaLoneTOpbetauSigmaLoneLtwo
{\biggerm_{\mbox{\footnotesize$\bSigmaLone\rightarrow\pDens(\bbeta,\bu|\,\bSigmaLone,\bSigmaLtwo)$}}(\bSigmaLone)}

\def\mSUBpbetauSigmaLoneLtwoTOSigmaLtwo
{\biggerm_{\mbox{\footnotesize$\pDens(\bbeta,\bu|\,\bSigmaLone,\bSigmaLtwo)\rightarrow\bSigmaLtwo$}}(\bSigmaLtwo)}

\def\etaSUBpbetauSigmaLoneLtwoTOSigmaLone
{\biggereta_{\mbox{\footnotesize$\pDens(\bbeta,\bu|\,\bSigmaLone,\bSigmaLtwo)\rightarrow\bSigmaLone$}}}

\def\etaSUBpbetauSigmaLoneLtwoCONNSigmaLone
{\biggereta_{\mbox{\footnotesize$\pDens(\bbeta,\bu|\,\bSigmaLone,\bSigmaLtwo)\leftrightarrow\bSigmaLone$}}}

\def\etaSUBpbetauSigmaLoneLtwoCONNSigmaLtwo
{\biggereta_{\mbox{\footnotesize$\pDens(\bbeta,\bu|\,\bSigmaLone,\bSigmaLtwo)\leftrightarrow\bSigmaLtwo$}}}

\def\etaSUBpbetauSigmaLoneLtwoCONNbetau
{\biggereta_{\mbox{\footnotesize$\pDens(\bbeta,\bu|\,\bSigmaLone,\bSigmaLtwo)\leftrightarrow(\bbeta,\bu)$}}}

\def\etaSUBSigmaLoneTOpbetauSigmaLoneLtwo
{\biggereta_{\mbox{\footnotesize$\bSigmaLone\rightarrow\pDens(\bbeta,\bu|\,\bSigmaLone,\bSigmaLtwo)$}}}

\def\etaSUBpbetauSigmaLoneLtwoTOSigmaLtwo
{\biggereta_{\mbox{\footnotesize$\pDens(\bbeta,\bu|\,\bSigmaLone,\bSigmaLtwo)\rightarrow\bSigmaLtwo$}}}

\def\etaSUBSigmaLtwoTOpbetauSigmaLoneLtwo
{\biggereta_{\mbox{\footnotesize$\bSigmaLtwo\rightarrow\pDens(\bbeta,\bu|\,\bSigmaLone,\bSigmaLtwo)$}}}

\def\etaSUBpbetauSigmaLoneLtwoTOSigmaLtwo
{\biggereta_{\mbox{\footnotesize$\pDens(\bbeta,\bu|\,\bSigmaLone,\bSigmaLtwo)\rightarrow\bSigmaLtwo$}}}

\def\etaSUBbetauTOpbetauSigmaLoneLtwo
{\biggereta_{\mbox{\footnotesize$(\bbeta,\bu)\rightarrow\pDens(\bbeta,\bu|\,\bSigmaLone,\bSigmaLtwo)$}}}

\def\mSUBbetauTOpbetauSigmaLoneLtwo
{\biggerm_{\mbox{\footnotesize$(\bbeta,\bu)\rightarrow\pDens(\bbeta,\bu|\,\bSigmaLone,\bSigmaLtwo)$}}(\bbeta,\bu)}

\def\mSUBpbetauSigmaTOSigma
{\biggerm_{\mbox{\footnotesize$\pDens(\bbeta,\bu|\,\bSigma)\rightarrow\bSigma$}}(\bSigma)}

\def\etaSUBpbetauSigmaTObetau
{\biggerbdeta_{\mbox{\footnotesize$\pDens(\bbeta,\bu|\,\bSigma)\rightarrow(\bbeta,\bu)$}}}

\def\etaSUBqbetau
{\biggerbdeta_{\mbox{\footnotesize$\qDens(\bbeta,\bu)$}}}

\def\mSUBbetauTOpybetausigsqEps
{\biggerm_{\mbox{\footnotesize$(\bbeta,\bu)\rightarrow\pDens(\by|\,\bbeta,\bu,\mysigeps^2)$}}(\bbeta,\bu)}

\def\mSUBpybetausigsqEpsTObetau
{\biggerm_{\mbox{\footnotesize$\pDens(\by|\,\bbeta,\bu,\mysigeps^2)\rightarrow(\bbeta,\bu)$}}(\bbeta,\bu)}

\def\mSUBpybetausigsqEpsTOsigsqEps
{\biggerm_{\mbox{\footnotesize$\pDens(\by|\,\bbeta,\bu,\mysigeps^2)\rightarrow\mysigeps^2$}}(\mysigeps^2)}

\def\mSUBsigsqEpsTOpybetausigsqEps
{\biggerm_{\mbox{\footnotesize$\mysigeps^2\rightarrow\pDens(\by|\,\bbeta,\bu,\mysigeps^2)$}}(\mysigeps^2)}

\def\etaSUBpybetausigsqEpsTObetau
{\biggerbdeta_{\mbox{\footnotesize$\pDens(\by|\,\bbeta,\bu,\mysigeps^2)\rightarrow(\bbeta,\bu)$}}}

\def\etaSUBbetauTOpybetausigsqEps
{\biggerbdeta_{\mbox{\footnotesize$(\bbeta,\bu)\rightarrow\pDens(\by|\,\bbeta,\bu,\mysigeps^2)$}}}

\def\etaSUBpbetauSigmaTObetau
{\biggerbdeta_{\mbox{\footnotesize$\pDens(\bbeta,\bu|\bSigma)\rightarrow(\bbeta,\bu)$}}}

\def\etaSUBpbetauSigmaCONNbetau
{\biggerbdeta_{\mbox{\footnotesize$\pDens(\bbeta,\bu|\bSigma)\leftrightarrow(\bbeta,\bu)$}}}

\def\etaSUBpbetauSigmaCONNSigma
{\biggerbdeta_{\mbox{\footnotesize$\pDens(\bbeta,\bu|\bSigma)\leftrightarrow\bSigma$}}}

\def\etaSUBbetauTOpbetauSigma
{\biggerbdeta_{\mbox{\footnotesize$(\bbeta,\bu)\rightarrow \pDens(\bbeta,\bu|\bSigma)$}}}

\def\etaSUBSigmaTOpbetauSigma
{\biggerbdeta_{\mbox{\footnotesize$\bSigma\rightarrow \pDens(\bbeta,\bu|\bSigma)$}}}

\def\mSUBSigmaTOpbetauSigma
{\biggerm_{\mbox{\footnotesize$\bSigma\rightarrow \pDens(\bbeta,\bu|\bSigma)$}}(\bSigma)}

\def\etaSUBpbetauSigmaTOSigma
{\biggerbdeta_{\mbox{\footnotesize$\pDens(\bbeta,\bu|\bSigma)\rightarrow\bSigma$}}}

\def\etaSUBpybetausigsqEpsCONNbetau
{\biggerbdeta_{\mbox{\footnotesize$\pDens(\by|\,\bbeta,\bu,\mysigeps^2)\leftrightarrow(\bbeta,\bu)$}}}

\def\etaSUBpybetausigsqEpsCONNsigsqEps
{\biggerbdeta_{\mbox{\footnotesize$\pDens(\by|\,\bbeta,\bu,\mysigeps^2)\leftrightarrow\mysigeps^2$}}}

\def\etaSUBpybetausigsqEpsTOsigsqEps
{\biggerbdeta_{\mbox{\footnotesize$\pDens(\by|\,\bbeta,\bu,\mysigeps^2)\rightarrow\mysigeps^2$}}}

\def\etaSUBbetauTOpybetausigsqEps
{\biggerbdeta_{\mbox{\footnotesize$(\bbeta,\bu)\rightarrow \pDens(\by|\,\bbeta,\bu,\mysigeps^2)$}}}

\def\etaSUBsigsqEpsTOpybetausigsqEps
{\biggerbdeta_{\mbox{\footnotesize$\mysigeps^2\rightarrow \pDens(\by|\,\bbeta,\bu,\mysigeps^2)$}}}

\def\biggereta{\mbox{\Large $\eta$}}
\def\biggerbdeta{\mbox{\Large $\bdeta$}}

\def\thickarrow{\longleftarrow}

\def\mysigeps{\sigma}

\def\bmat{\begin{bmatrix}}
\def\emat{\end{bmatrix}}

\def\bRes{\begin{Result}}
\def\eRes{\end{Result}}
\def\ASigma{\bA_{\mbox{\tiny$\bSigma$}}}
\def\bASigmaLone{\bA_{\mbox{\tiny$\bSigmaLone$}}}
\def\bASigmaLtwo{\bA_{\mbox{\tiny$\bSigmaLtwo$}}}

\def\sigsqeps{\sigma_{\varepsilon}^2}
\def\asigsq{a_{\sigma^2}}
\def\ssigsq{s_{\sigma^2}}
\def\nusigsq{\nu_{\sigma^2}}
\def\ASigma{\bA_{\mbox{\tiny{$\bSigma$}}}}
\def\nuSigma{\nu_{\mbox{\tiny{$\bSigma$}}}}
\def\nuSigmaLone{\nu_{\mbox{\tiny{$\bSigmaLone$}}}}
\def\nuSigmaLtwo{\nu_{\mbox{\tiny{$\bSigmaLtwo$}}}}
\def\sSigmaOne{s_{\mbox{\tiny{$\bSigma,1$}}}}

\def\sSigmaq{s_{\mbox{\tiny{$\bSigma,q$}}}}
\def\sSigmaOneLone{s_{\mbox{\tiny{$\bSigmaLone,1$}}}}
\def\sSigmaqLone{s_{\mbox{\tiny{$\bSigmaLone,\qLone$}}}}
\def\sSigmaOneLtwo{s_{\mbox{\tiny{$\bSigmaLtwo,1$}}}}
\def\sSigmaqLtwo{s_{\mbox{\tiny{$\bSigmaLtwo,\qLtwo$}}}}

\def\MqSigma{\bM_{\qDens(\bSigma^{-1})}}

\def\muq#1{\mu_{\qDens(#1)}}

\def\bmuq#1{\bmu_{\qDens(#1)}}

\def\sigsqeps{\mysigeps^2}

\def\SolveTwoLevelSparseMatrix{\textsc{\footnotesize SolveTwoLevelSparseMatrix}}
\def\SolveTwoLevelSparseLeastSquares{\textsc{\footnotesize SolveTwoLevelSparseLeastSquares}}
\def\SolveThreeLevelSparseMatrix{\textsc{\footnotesize SolveThreeLevelSparseMatrix}}
\def\SolveThreeLevelSparseLeastSquares{\textsc{\footnotesize SolveThreeLevelSparseLeastSquares}}
\def\TwoLevelNaturalToCommonParameters{\textsc{\footnotesize TwoLevelNaturalToCommonParameters}}

\begin{document}
\ifthenelse{\boolean{DoubleSpaced}}{\setstretch{1.5}}{}

\thispagestyle{empty}

\centerline{\Large\bf Streamlined Computing for Variational Inference}
\vskip2mm
\centerline{\Large\bf with Higher Level Random Effects}
\vskip7mm
\centerline{\normalsize\sc By Tui H. Nolan, Marianne Menictas and Matt P. Wand
\footnote{\rm Corresponding author: Professor Matt P. Wand, School of Mathematical
and Physical Sciences, University of Technology Sydney, P.O. Box 123, Broadway,
NSW 2007, Australia. e-mail: matt.wand@uts.edu.au}}
\vskip5mm
\centerline{\textit{University of Technology Sydney}}
\vskip6mm
\centerline{2nd July, 2020}

\vskip5mm
\centerline{\large {\sl Running title:} Streamlined Variational Inference} 
\vskip5mm

\centerline{\large\bf Abstract}
\vskip2mm

We derive and present explicit algorithms to facilitate streamlined
computing for variational inference for models containing higher level
random effects. Existing literature, such as Lee \myand Wand (2016), 
is such that streamlined variational inference is restricted to 
mean field variational Bayes algorithms for two-level random effects models.
Here we provide the following extensions:
(1) explicit Gaussian response mean field variational Bayes algorithms 
for three-level models, (2) explicit algorithms for the alternative 
variational message passing approach in the case of two-level 
and three-level models, and (3) an explanation of how arbitrarily 
high levels of nesting can be handled based on the recently 
published matrix algebraic results of the authors. A pay-off 
from (2) is simple extension to non-Gaussian response models. 
In summary, we remove barriers for streamlining variational 
inference algorithms based on either the mean field
variational Bayes approach or the variational message passing 
approach when higher level random effects are present.


\section{Introduction}\label{sec:intro}

Models involving higher level random effects commonly arise in 
a variety of contexts. The areas of study known as 
longitudinal data analysis (e.g.\ Fitzmaurice \textit{et al.}, 2008),
mixed models (e.g.\ Pinheiro \myand Bates, 2000), 
multilevel models (e.g.\ Goldstein, 2010), panel data analysis
(e.g.\ Baltagi, 2013) and small area estimation (e.g. Rao
\myand Molina, 2015) potentially each require the handling
of higher levels of nesting. Our main focus in this article 
is providing explicit algorithms that facilitate variational
inference for up to three-level random effects and a pathway
for handling even higher levels. Both direct and message
passing approaches to mean field variational Bayes are treated.

\begin{figure}[!ht]
\centering
{\includegraphics[width=\textwidth]{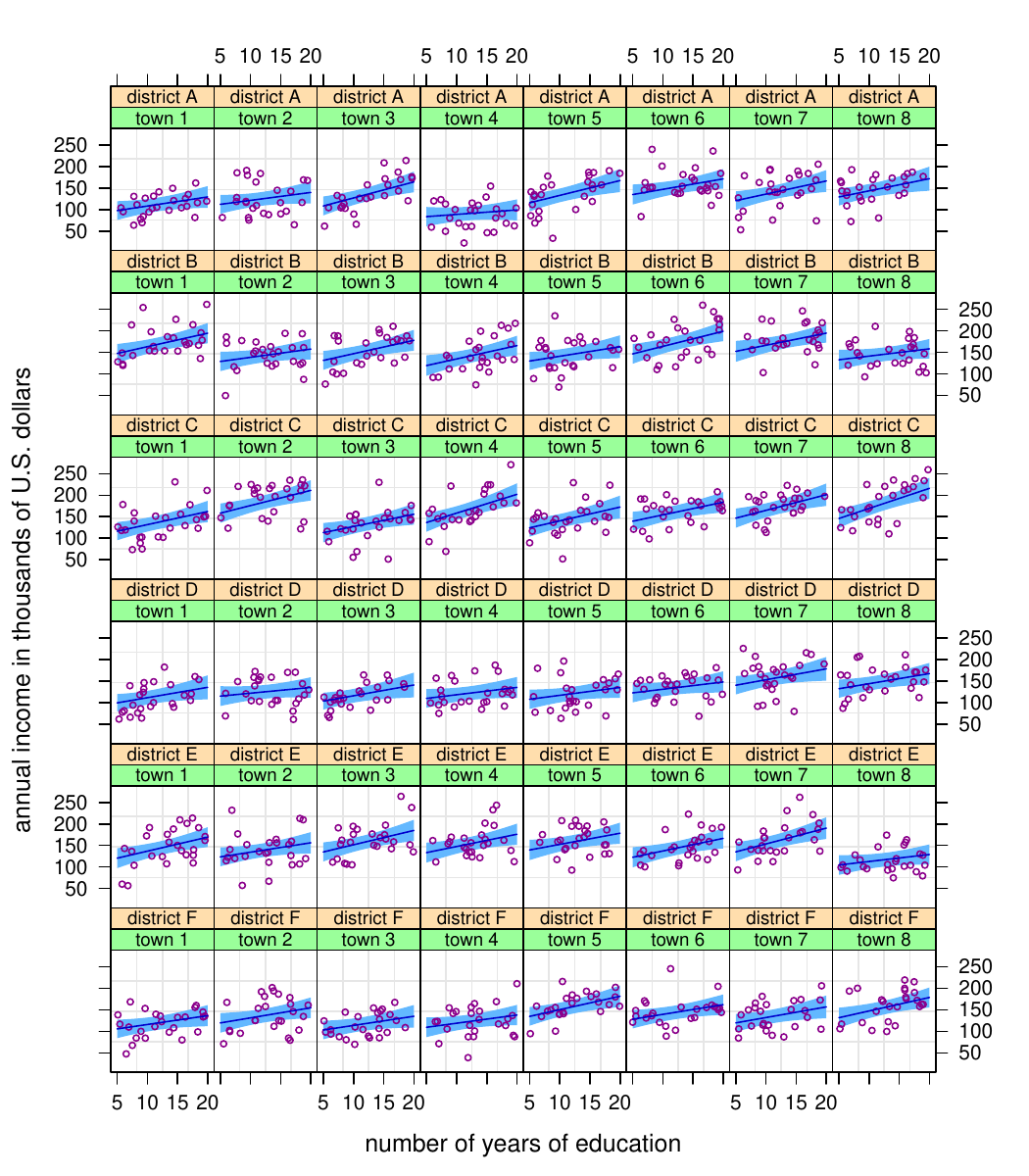}}
\caption{\it Simulated three-level data according to 
6 districts, each having 8 towns, each having 25 randomly chosen
residents. In each panel, the line corresponds to a mean field
variational Bayes fit, according to an appropriate multilevel model and 
the shaded region corresponds to pointwise 95\% credible intervals
for the mean response.}
\label{fig:residentProto} 
\end{figure}

A useful prototype setting for understanding the nature and computational
challenges is a fictitious sociology example in which residents (level 1 units)
are divided into different towns (level 2 units) and those towns are divided into
different districts (level 3 units). Following Goldstein (2010) we call these
three-level data, although note that Pinheiro \myand Bates (2000) 
use the term ``two-level'', corresponding to two levels of nesting, 
for the same setting. Figure \ref{fig:residentProto} displays 
simulated regression data generated according to this setting 
with a single predictor variable corresponding to years of education
and the response corresponding to annual income. In Figure \ref{fig:residentProto}, 
the number of districts is 6, the number of towns per district is 8
and the resident sample size within each town is 25.
%
%
In each panel of Figure \ref{fig:residentProto}, the line corresponds to 
the mean field variational Bayes fit of a three-level random intercepts and slopes
linear mixed model, as explained in Section \ref{sec:MFVB3lev}. 
Now suppose that the group and sample sizes are much larger with, 
say, 500 districts, 60 towns per district and 1,000 residents per town. Then 
na\"{\i}ve fitting is storage-greedy and computationally challenging
since the combined fixed and random effects design matrices have  
$1.83\times10^{12}$ entries of which at least 99.99\% equal zero.
A major contribution of this article is explaining how variational inference
can be achieved using only the 0.01\% non-zero design matrix components
with updates that are linear in the numbers of groups.

Our streamlined variational inference algorithms for higher level random effects
models rely on four theorems provided by Nolan \myand Wand (2020) concerning 
linear system solutions and sub-blocks of matrix inverses for two-level and
three-level sparse matrix problems which are the basis for the fundamental
Algorithms \ref{alg:SolveTwoLevelSparseMatrix}--\ref{alg:SolveThreeLevelSparseLeastSquares} 
in Appendix \ref{sec:matAlgs}. In that article, as well as here, we treat 
one higher level situation at a time. Even though four-level and even 
higher level situations may be of interest in future analysis, the required
theory is not yet in place. As we will see, covering both direct and message
passing approaches for just the two-level and three-level cases is quite a big task.
Nevertheless, our results and algorithms shed important light on streamlined
variational inference for general higher level random effects models.

After introducing the four fundamental algorithms 
in Section 3 and laying them out in Appendix \ref{sec:matAlgs}
we then derive an additional eight algorithms, labeled 
Algorithms \ref{alg:twoLevelMFVB}--\ref{alg:threeLevelVMPpen}, that
facilitate variational inference for two-level 
and three-level linear mixed models. The mean field variational Bayes approach 
is dealt with in Algorithms \ref{alg:twoLevelMFVB} and \ref{alg:threeLevelMFVB}. 
The remaining six algorithms are concerned with streamlined factor graph 
fragment updates according to the variational message passing infrastructure 
described in Wand (2017). As explained in Section 3.2 there, the message passing 
approach has  the advantage compartmentalization of variational inference algebra 
and code. Once a key fragment is identified, it only has to be derived and coded 
once and then can be used in models of arbitrarily large size.
The inherent complexity of streamlined variational inference for higher level
random effects models is such that the current article is restricted
to ordinary linear mixed models. Extensions such as generalized additive 
mixed models with higher level random effects and higher level 
group-specific curve models follow from \ref{alg:twoLevelMFVB}--\ref{alg:threeLevelVMPpen}, 
but are be treated elsewhere (e.g. Menictas, Nolan, Simpson \myand Wand, 2020).  
Section \ref{sec:closing} provides further details on this matter.

Our algorithms also build on previous work on streamlined variational
inference for similar classes of models described in Lee \myand Wand (2016). 
However, Lee \myand Wand (2016) only treated the two-level case, did not
employ QR decomposition enhancement and did not include any variational 
message passing algorithms. The current article is a systematic treatment
of higher level random effects models beyond the common two-level case.

Section \ref{sec:varInf} provides background material concerning 
variational inference. In Section \ref{sec:matrixOnly} we 
summarize issues involving matrix algebra and point to 
Appendix \ref{sec:matAlgs}. This appendix presents four algorithms for 
solving higher level sparse matrix problems which are fundamental 
for variational inference involving general models with hierarchical 
random effects structure. Streamlined variational inference for 
mixed models possessing two-level random effects structure is treated in 
Section \ref{sec:twoLevMods}, followed by treatment of the three-level situation 
in  Section \ref{sec:threeLevMods}. Derivations of all results and algorithms
given in Sections \ref{sec:twoLevMods} and \ref{sec:threeLevMods} are
deferred to Appendix \ref{sec:derivations}. Section \ref{sec:timing} demonstrates 
the speed advantages of streamlining for variational
inference in random effects models via some computational complexity
calculations and timing studies. 
Illustration for data from a large perinatal health study is given in Section \ref{sec:applic}.
In Section \ref{sec:closing} we close with some discussion about extensions to other settings.

\section{Variational Inference Background}\label{sec:varInf}

In keeping with the theme of this article, we will explain the essence 
of variational inference for a general class of Bayesian linear mixed models. 
Summaries of variational inference in wider statistical contexts
are given in Ormerod \myand Wand (2010) and Blei, Kucukelbir 
\myand McAuliffe (2017). 

\begin{figure}[!ht]
\centering
{\includegraphics[width=0.65\textwidth]{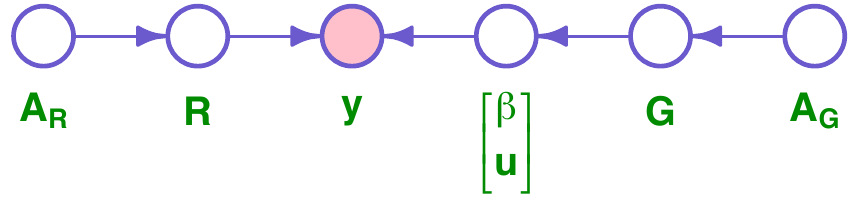}}
\caption{\it Directed acyclic graph representation of model (\ref{eq:lmmOne}).
The shading of the $\by$ node indicates that this vector of response values
is observed.}
\label{fig:lmmGandRDAG} 
\end{figure}

Suppose that the response data vector $\by$ is modeled according to 
a Bayesian version of the Gaussian linear mixed model (e.g. Robinson, 1991)
\begin{equation}
\by|\bbeta,\bu,\bR\sim N(\bX\bbeta+\bZ\bu,\bR),\quad 
\bu|\bG\sim N(\bzero,\bG),\quad\bbeta\sim N(\bmu_{\bbeta},\bSigma_{\bbeta})
\label{eq:lmmOne}
\end{equation}
for hyperparameters $\bmu_{\bbeta}$ and $\bSigma_{\bbeta}$ and
such that $\bbeta$ and $\bu|\bG$ are independent.
The $\bbeta$ and $\bu$ vectors are labeled \emph{fixed effects} and \emph{random effects},
respectively. Their corresponding \emph{design matrices} are $\bX$ and $\bZ$.
We will allow for the possibility that prior specification
for the covariance matrices $\bG$ and $\bR$ involves auxiliary 
covariance matrices $\bAbG$ and $\bAbR$ with conjugate Inverse G-Wishart
distributions (Wand, 2017). The prior specification of
$\bG$ and $\bR$ involves the specifications
\begin{equation}
\pDens(\bG|\bAbG),\quad \pDens(\bAbG),\quad \pDens(\bR|\bAbR)\quad\mbox{and}
\quad\pDens(\bAbR).
\label{eq:lmmTwo}
\end{equation}
Figure \ref{fig:lmmGandRDAG} is a directed acyclic graph representation
of (\ref{eq:lmmOne}) and (\ref{eq:lmmTwo}). The circles, usually called
\textit{nodes}, correspond to the model's random vectors and random matrices.
The arrows depict conditional independence relationships
(e.g. Bishop, 2006; Chapter 8).
%

Full Bayesian inference for the $\bbeta$, $\bG$ and $\bR$ and
the random effects $\bu$ involves the posterior density function
$\pDens(\bbeta,\bu,\bAbG,\bAbR,\bG,\bR|\by)$,  but typically is analytically 
intractable and Markov chain Monte Carlo approaches are required
for practical `exact' inference. Variational approximate inference 
involves mean field restrictions such as 
\begin{equation}
\pDens(\bbeta,\bu,\bAbG,\bAbR,\bG,\bR|\by)\approx \qDens(\bbeta,\bu,\bAbG,\bAbR)\,\qDens(\bG,\bR)
\label{eq:minimalProdRestric}
\end{equation}
for density functions $\qDens(\bbeta,\bu,\bAbG,\bAbR)$ and $\qDens(\bG,\bR)$, which we
call \emph{$\qDens$-densities}. The approximation at (\ref{eq:minimalProdRestric})
represents the minimal product restriction for which practical 
variational inference algorithms arise. However, as explained in Section 10.2.5 
of Bishop (2006), the graphical structure of Figure \ref{fig:lmmGandRDAG} induces further 
product density forms and the right-hand side of (\ref{eq:minimalProdRestric})
admits the further factorization 
\begin{equation}
\qDens(\bbeta,\bu)\qDens(\bAbG)\qDens(\bAbR)\qDens(\bG)\qDens(\bR).
\label{eq:qdensFullProd}
\end{equation}
With this product density form in place, the forms and optimal parameters for the 
$\qDens$-densities are obtained by minimising the Kullback-Leibler divergence
of the right-hand side of (\ref{eq:minimalProdRestric}) from its left-hand side.
The optimal $\qDens$-density parameters are interdependent and a coordinate
ascent algorithm (e.g. Algorithm 1 of Ormerod \myand Wand, 2010) is used
to obtain their solution. For example, the optimal $\qDens$-density for $(\bbeta,\bu)$,
denoted by $\qDens^*(\bbeta,\bu)$, is a Multivariate Normal density function with 
mean vector $\bmu_{\qDens(\bbeta,\bu)}$ and covariance matrix $\bSigma_{\qDens(\bbeta,\bu)}$.
The coordinate ascent algorithm is such that they are updated according to 
\begin{equation}
{\setlength\arraycolsep{3pt}
\begin{array}{rcl}
\bSigma_{\qDens(\bbeta,\bu)}&\thickarrow&\left\{\bC^T\,E_{\qDens}(\bR^{-1})\bC+
\left[  
\begin{array}{cc}
\bSigma_{\bbeta}^{-1} & \bO          \\
\bO                   & E_{\qDens}(\bG^{-1})
\end{array}
\right]\right\}^{-1}\\[4ex]
\quad\mbox{and}\quad
\bmu_{\qDens(\bbeta,\bu)}&\thickarrow&\bSigma_{\qDens(\bbeta,\bu)}\bC^T\,E_{\qDens}(\bR^{-1})
\left(\by +\left[\begin{array}{c}\bSigma^{-1}\bmu_{\bbeta}\\\bzero\end{array}\right]\right)
\end{array}
}
\label{eq:naiveLMMupdates}
\end{equation}
where $E_{\qDens}(\bG^{-1})$ and $E_{\qDens}(\bR^{-1})$ are the $\qDens$-density expectations of 
$\bG^{-1}$ and $\bR^{-1}$ and $\bC\equiv[\bX\ \bZ]$. If, for example,
(\ref{eq:lmmOne}) corresponds to a mixed model with three-level random effects
such that $\bR=\mysigeps^2\,\bI$ then, as pointed out in Section \ref{sec:intro}, 
with 60 groups at level 2 and 500 groups at level 3  the matrix $\bC$ has
almost 2 trillion entries of which 99.99\% are zero. Moreover, 
$\bSigma_{\qDens(\bbeta,\bu)}$ is a $61,002\times61,002$ matrix of which only about
0.016\% of its approximately 3.7 billion entries are required for
variational inference under mean field restriction (\ref{eq:minimalProdRestric}).
Avoiding the wastage of the na\"{\i}ve updates given by (\ref{eq:naiveLMMupdates}) 
is the crux of this article and dealt with in the upcoming sections.
The updates for $E_{\qDens}(\bG^{-1})$ and $E_{\qDens}(\bR^{-1})$ depend on parameterizations
of $\bG$ and $\bR$. For example, $\bR=\mysigeps^2\bI$ for some $\mysigeps^2>0$ throughout
Sections \ref{sec:twoLevMods} and \ref{sec:threeLevMods}. However, these
covariance parameter updates are relatively simple and free of storage and
computational efficiency issues. Similar comments apply to the updates for
the $\qDens$-density parameters of $\bAbG$ and $\bAbR$.

\begin{figure}[!ht]
\centering
{\includegraphics[width=0.75\textwidth]{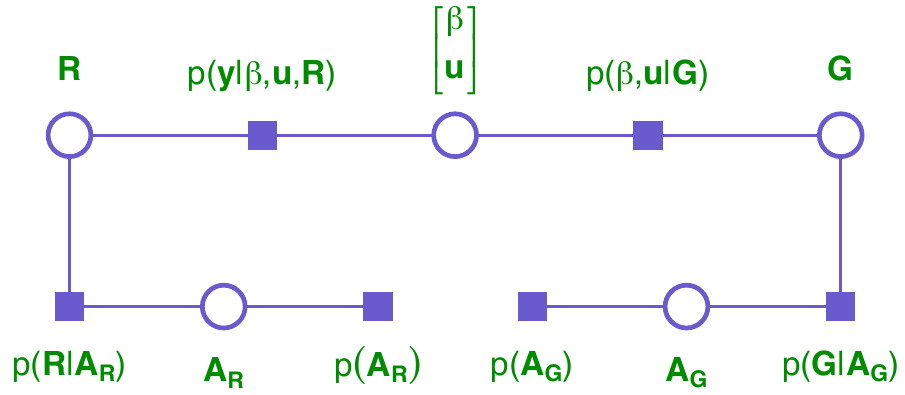}}
\caption{\it Factor graph representation of the product structure
of (\ref{eq:fullLMMdens}) with the solid rectangles corresponding to
the factors and open circles corresponding to the unobserved random
vectors and random matrices of the Bayesian linear mixed model given by 
(\ref{eq:lmmOne}) and (\ref{eq:lmmTwo}), known as stochastic nodes. 
Edges join each factor to the stochastic nodes that are present in
the factor.}
\label{fig:lmmGandRfacGraph} 
\end{figure}

An alternative approach to obtaining $\bmu_{\qDens(\bbeta,\bu)}$,
the relevant sub-blocks of $\bSigma_{\qDens(\bbeta,\bu)}$ and the covariance
and auxiliary variable $q$-parameter updates is to use the notion of 
\emph{message passing} on a \emph{factor graph}. The relevant
factor graph for model (\ref{eq:lmmOne}), according to the 
product density form (\ref{eq:qdensFullProd}), is shown in Figure 
\ref{fig:lmmGandRfacGraph}.
%

\noindent
The circles in Figure \ref{fig:lmmGandRfacGraph} correspond to the parameters
in each factor of (\ref{eq:qdensFullProd}) and are referred to 
as \emph{stochastic nodes}. The squares correspond to the factors of 
\begin{equation}
\pDens(\by,\bbeta,\bu,\bAbG,\bAbR,\bG,\bR)=\pDens(\by|\bbeta,\bu,\bR)\,\pDens(\bbeta,\bu|\bG)
\,\pDens(\bG|\bAbG)\,\pDens(\bR|\bAbR)\,\pDens(\bAbG)\,\pDens(\bAbR),
\label{eq:fullLMMdens}
\end{equation}
with factorization according to the conditional independence structure
apparent from Figure \ref{fig:lmmGandRDAG}. Then, as explained in e.g.\ Minka (2005),
the $\qDens$-density of $(\bbeta,\bu)$ can be expressed as 
$$\qDens(\bbeta,\bu)\propto\mSUBpybetauRTObetau\,\,\mSUBpbetauGTObetau$$
where 
$$\mSUBpybetauRTObetau\quad\mbox{and}\quad\mSUBpbetauGTObetau$$
are known as \emph{messages}, with the subscripts indicating
that they are passed from $\pDens(\by|\bbeta,\bu,\bR)$ to $(\bbeta,\bu)$ and
$\pDens(\bbeta,\bu|\bG)$ to $(\bbeta,\bu)$, respectively. Messages are simply 
functions of the stochastic node to which the message is passed and, 
for mean field variational inference, are formed according to rules 
listed in Minka (2005) and Section 2.5 of Wand (2017). To compartmentalize
algebra and coding for variational message passing, Wand (2017)
advocates the use of \emph{fragments}, which are sub-graphs
of a factor graph containing a single factor and each of its
neighboring stochastic nodes. In Sections 4 and 5 of Wand (2017), 
eight important fragments are identified and treated including
those needed for a wide range of linear mixed models.
However, in the interests of brevity, Wand (2017) ignored
issues surrounding potentially very large and sparse matrices
in the message parameter vectors. In Sections \ref{sec:twoLevMods} 
and \ref{sec:threeLevMods} of this article, we explain how the messages 
passed to the $(\bbeta,\bu)$ node can be streamlined to avoid massive sparse
matrices.

A core component of the message passing approach to 
variational inference is exponential family forms,
sufficient statistics and natural parameters.
For a $d\times 1$ Multivariate Normal random vector 
$$\bx\sim N(\bmu,\bSigma)$$
this involves re-expression of its density function according to 
{\setlength\arraycolsep{1pt}
\begin{eqnarray*}
\pDens(\bx)&=&(2\pi)^{-d/2}|\bSigma|^{-1/2}
\exp\{-\smhalf(\bx-\bmu)^T\bSigma^{-1}(\bx-\bmu)\}\\[1ex]
&=&\exp\{\bT(\bx)^T\bdeta - A(\bdeta)-\textstyle{\frac{d}{2}}\log(2\pi)\}
\end{eqnarray*}
}
where
$$
\bT(\bx)\equiv\left[
\begin{array}{c}
\bx \\
\vech(\bx\,\bx^T)
\end{array}
\right]
\quad\mbox{and}\quad
\bdeta\equiv
\left[
\begin{array}{c}
\bdeta_1 \\
\bdeta_2 
\end{array}
\right]
\equiv
\left[
\begin{array}{c}
\bSigma^{-1}\bmu \\
-\smhalf\,\bD_d^T\vecof(\bSigma^{-1})
\end{array}
\right]
\label{eq:natural}
$$
are, respectively, the \emph{sufficient statistic} and \emph{natural parameter} vectors.
The matrix $\bD_d$, known as the \emph{duplication matrix of order $d$}, 
is the $d^2\times\{\smhalf d(d+1)\}$ matrix containing only zeroes and
ones such that $\bD_d\,\vech(\bA)=\vecof(\bA)$ for any symmetric $d\times d$ matrix $\bA$.
The function  
$$A(\bdeta)=\,-\quarter\,\bdeta_1^T
\left\{\vecof^{-1}(\bD_d^{+T}\bdeta_2)\right\}^{-1}\bdeta_1
-\smhalf
\log\big|-2\,\vecof^{-1}(\bD_d^{+T}\bdeta_2)\big|$$
is the \emph{log-partition} function, where $\bD_d^+\equiv(\bD_d^T\bD_d)^{-1}\bD_d^T$ is
the Moore-Penrose inverse of $\bD_d$ and is such that $\bD_d^+\vecof(\bA)=\vech(\bA)$ whenever
$\bA$ is symmetric. The inverse of the natural parameter transformation is given by
$$
\bmu=-\smhalf\left\{\vecof^{-1}(\bD_d^{+T}\bdeta_2)\right\}^{-1}\bdeta_1\quad\mbox{and}\quad
\bSigma=-\smhalf\left\{\vecof^{-1}(\bD_d^{+T}\bdeta_2)\right\}^{-1}.
$$

The $\vecof$ and $\vech$ matrix operators are reasonably well-established
(e.g. Gentle, 2007). If $\ba$ is a $d^2\times1$ vector then $\vecof^{-1}(\ba)$ is the
$d\times d$ matrix such that $\vecof\big(\vecof^{-1}(\ba)\big)=\ba$. We also require $\vecof$ inversion
of non-square matrices. If $\ba$ is a $(d_1d_2)\times 1$
vector then $\vecof^{-1}_{d_1\times d_2}(\ba)$ is the $d_1\times d_2$ matrix such
that $\vecof\big(\vecof^{-1}_{d_1\times d_2}(\ba)\big)=\ba$.

The other major distributional family used throughout this article is a
generalization of the Inverse Wishart distribution known as the \emph{Inverse G-Wishart}
distribution. It corresponds to the matrix inverses of random matrices that have
a \emph{G-Wishart} distribution (e.g. Atay-Kayis \myand Massam, 2005; 
Maestrini \myand Wand, 2020). For any positive integer $d$, let $G$ be 
an undirected graph with $d$ nodes labeled $1,\ldots,d$ and 
set $E$ consisting of sets of pairs of nodes that are connected 
by an edge. We say that the symmetric $d\times d$ matrix $\bM$ \emph{respects} 
$G$ if 
$$\bM_{ij}=0\quad\mbox{for all}\quad \{i,j\}\notin E.$$
A $d\times d$ random matrix $\bX$ has an Inverse G-Wishart distribution
with graph $G$ and parameters $\xi>0$ and symmetric $d\times d$ 
matrix $\bLambda$, written
$$\bX\sim\mbox{Inverse-G-Wishart}(G,\xi,\bLambda)$$
if and only if the density function of $\bX$ satisfies
$$\pDens(\bX)\propto |\bX|^{-(\xi+2)/2}\exp\{-\smhalf\tr(\bLambda\,\bX^{-1})\}$$
over arguments $\bX$ such that $\bX$  is symmetric and positive definite 
and $\bX^{-1}$ respects $G$. Two important special cases are 
$$G=\Gfull\equiv\mbox{totally connected $d$-node graph},$$
for which the Inverse G-Wishart distribution coincides with the ordinary
Inverse Wishart distribution, and
$$G=\Gdiag\equiv\mbox{totally disconnected $d$-node graph},$$
for which the Inverse G-Wishart distribution coincides with
a product of independent Inverse Chi-Squared random variables.
The subscripts of $\Gfull$ and $\Gdiag$ reflect the 
fact that $\bX^{-1}$ is a full matrix and $\bX^{-1}$ is
a diagonal matrix in each special case.

The $G=\Gfull$ case corresponds to the ordinary Inverse Wishart
distribution. However, with message passing in mind, we 
work with the more general Inverse G-Wishart family.

In the $d=1$ special case the graph $G=\Gfull=\Gdiag$ 
and the Inverse G-Wishart distribution reduces to the
Inverse Chi-Squared distribution. We write
$$x\sim\mbox{Inverse-$\chi^2$}(\xi,\lambda)$$
for this  $\mbox{Inverse-G-Wishart}(G,\xi,\lambda)$
special case with $d=1$ and $\lambda>0$ scalar.

Finally, we remark on the $\pDens$ and $\qDens$ notation 
used for density functions in this article. In the 
variational inference literature these letters have
become very commonplace to denote the density functions
corresponding to the model and the density
functions of parameters according to the mean field
approximation, with $\pDens$ for the former and $\qDens$
for the latter. However, the same letters are commonly
used as dimension variables in the mixed models literature
(e.g. Pinheiro \myand Bates, 2000). Therefore we use
ordinary $p$ and $q$ as dimension variables and scripted
versions of these letters ($\pDens$ and $\qDens$) for 
density functions.

\section{Matrix Algebraic Background}\label{sec:matrixOnly}

For matrices $\bM_1,\ldots,\bM_d$ we define:
$$\stack{1 \le i \le d}(\bM_i)\equiv
\left[
\begin{array}{c}
\bM_1\\
\vdots\\
\bM_d
\end{array}
\right]
\quad\mbox{and}\quad
\blockdiag{1 \le i \le d}(\bM_i)\equiv
\left[
\begin{array}{cccc}
\bM_1 & \bO   & \cdots & \bO\\
\bO   & \bM_2 & \cdots & \bO\\
\vdots& \vdots& \ddots & \vdots\\
\bO   & \bO& \cdots &\bM_d 
\end{array}
\right]
$$
with the first of these definitions requiring that $\bM_i$, $1\le i\le d$, each 
having the same number of columns. Such notation is very useful for 
defining matrices that appear in higher level random effects 
models. 

A key observation in this work is the fact that streamlining
of variational inference algorithms for higher level random
effects models can be achieved by recognition and isolation
of a few fundamental algorithms, which we call \emph{multilevel
sparse matrix problem} algorithms. These algorithms, based
on the results of Nolan \myand Wand (2020), are similar
to those used traditionally for fitting
frequentist random effects (Pinheiro \myand Bates, 2000).
For each level there are two types of sparse
matrix solution algorithms: one that applies to general forms
and one that uses a QR-decomposition enhancement for a 
particular form that arises commonly for models containing 
random effects. Both types are needed for variational inference.

Appendix \ref{sec:matAlgs} provides the details of the multilevel
sparse matrix problem algorithms used in the upcoming 
variational inference algorithms. There are four such matrix algebraic
algorithms:
\begin{center}
\begin{tabular}{lll}
\SolveTwoLevelSparseMatrix         &\ \ \ \ \ &Algorithm \ref{alg:SolveTwoLevelSparseMatrix} \\ 
\SolveTwoLevelSparseLeastSquares   &\ \ \ \ \ &Algorithm \ref{alg:SolveTwoLevelSparseLeastSquares} \\ 
\SolveThreeLevelSparseMatrix       &\ \ \ \ \ &Algorithm \ref{alg:SolveThreeLevelSparseMatrix} \\  
\SolveThreeLevelSparseLeastSquares &\ \ \ \ \ &Algorithm \ref{alg:SolveThreeLevelSparseLeastSquares} 
\end{tabular}
\end{center}
We use these four descriptive names in the variational inference algorithms
that begin in the next section.

\section{Two-Level Models}\label{sec:twoLevMods}

We now present streamlined algorithms for two-level linear 
mixed models.

\subsection{Mean Field Variational Bayes}

Consider the following Bayesian model:
\begin{equation}
\begin{array}{c}
\by_i|\bbeta,\bu_i,\mysigeps^2\simind N(\bX_i\bbeta+\bZ_i\,\bu_i,\mysigeps^2\,\bI),
\quad \bu_i|\bSigma\simind N(\bzero,\bSigma),\quad 1\le i\le m,\\[2ex]
\bbeta\sim N(\bmu_{\bbeta},\bSigma_{\bbeta}),\quad
\mysigeps^2|\asigsq\sim\mbox{Inverse-$\chi^2$}(\nusigsq,1/\asigsq),\\[2ex]
\asigsq\sim\mbox{Inverse-$\chi^2$}(1,1/(\nusigsq\ssigsq^2)),\\[2ex]
\bSigma|\ASigma\sim\mbox{Inverse-G-Wishart}\big(\Gfull,\nuSigma+2q-2,\ASigma^{-1}\big),\\[2ex]
\ASigma\sim\mbox{Inverse-G-Wishart}(\Gdiag,1,\bLambda_{\ASigma}),\quad
\bLambda_{\ASigma}\equiv\{\nuSigma\diag(\sSigmaOne^2,\ldots,\sSigmaq^2)\}^{-1},
\end{array}
\label{eq:twoLevelGaussRespBaye}
\end{equation}
where matrix dimensions, for $1\le i\le m$, are as follows:
$$
\by_i\ \mbox{is $n_i\times 1$},\ \ \bX_i\ \mbox{is $n_i\times p$},\ \ \bbeta\ \mbox{is $p\times1$},
\ \ \bZ_i\ \mbox{is $n_i\times q$},\ \ \bu_i\ \mbox{is $q\times 1$}\ \mbox{and}\ 
\bSigma\ \ \mbox{is $q\times q$}.
$$
Also, for example, $\bu_i|\bSigma\simind N(\bzero,\bSigma)$ is shorthand for the $\bu_i$ being independently
distributed $N(\bzero,\bSigma)$ random vectors conditional on $\bSigma$.
Next define the matrices
$$
\by\equiv\left[  
\begin{array}{c}
\by_1\\
\vdots\\
\by_m
\end{array}
\right],
\ 
\bX\equiv\left[  
\begin{array}{c}
\bX_1\\
\vdots\\
\bX_m
\end{array}
\right],
\  
\bZ\equiv\blockdiag{1\le i\le m}(\bZ_i),\
\bu\equiv\left[  
\begin{array}{c}
\bu_1\\
\vdots\\
\bu_m
\end{array}
\right]
\ \mbox{and}\ 
\bC\equiv[\bX\ \bZ].
$$
The hyperparameters $\bmu_{\bbeta}(p\times1)$ and $\bSigma_{\bbeta}(p\times p)$
are such that $\bSigma_{\bbeta}$ is symmetric and positive definite and 
$\nusigsq,\nuSigma,\ssigsq,\sSigmaOne,\ldots,\sSigmaq>0$. 
Note that (\ref{eq:twoLevelGaussRespBaye}) implies
that the prior on $\mysigeps$ is Half-Cauchy with scale parameter $\ssigsq$ 
and the prior on $\bSigma$ is within the class described in Huang \myand Wand (2013).
As explained in Huang \myand Wand (2013), such priors allow standard deviation and 
correlation parameters to have arbitrary non-informativeness. 

Now consider the following mean field restriction on the joint posterior
density function of all parameters in (\ref{eq:twoLevelGaussRespBaye}):
\begin{equation}
\pDens(\bbeta,\bu,\asigsq,\ASigma,\mysigeps^2,\bSigma|\by)
\approx \qDens(\bbeta,\bu,\asigsq,\ASigma)\,\qDens(\mysigeps^2,\bSigma)
\label{eq:producRestrict}
\end{equation}
where, generically, each $\qDens$ represents a density function of the 
random vector indicated by its argument. Then application of the
minimum Kullback-Leibler divergence equations (e.g. equation (10.9) of Bishop, 2006) 
leads to the optimal $\qDens$-density functions for the parameters of interest being as follows:
\begin{equation}
{\setlength\arraycolsep{1pt}
\begin{array}{ll}
&\qDens^*(\bbeta,\bu)\ \mbox{has a $N\big(\bmu_{\qDens(\bbeta,\bu)},
\bSigma_{\qDens(\bbeta,\bu)}\big)$ density function},\\[1ex]
&\qDens^*(\mysigeps^2)\ \mbox{has an $\mbox{Inverse-$\chi^2$}
\big(\xi_{\qDens(\mysigeps^2)},\lambda_{\qDens(\mysigeps^2)}\big)$ density function}\\[1ex]
\mbox{and\ }&\qDens^*(\bSigma)\ \mbox{has an 
$\mbox{Inverse-G-Wishart}(\Gfull,\xi_{\qDens(\bSigma)},\bLambda_{\qDens(\bSigma)})$ density function.}\\
\end{array}
}
\label{eq:twoLevMFVBforms}
\end{equation}
The optimal $\qDens$-density parameters are determined via an 
iterative coordinate ascent algorithm, with details deferred 
to Appendix \ref{sec:DerivTwoLevMFVB}.
Algorithm 2 of Lee \myand Wand (2016) is a na\"{\i}ve 
mean field variational Bayes algorithm for a class of
two-level Gaussian response linear mixed models that includes
model (\ref{eq:twoLevelGaussRespBaye}) as a special case.
Subsequent algorithms in Lee \myand Wand (2016) achieve
streamlining. In the current article, we offer an alternative
approach, based on Algorithms \ref{alg:twoLevelMFVB} and 
\ref{alg:threeLevelMFVB}, that handle higher level random effects 
in a natural way.

Note that updates for
$\bmu_{\qDens(\bbeta,\bu)}$ and $\bSigma_{\qDens(\bbeta,\bu)}$ may be written
\begin{equation}
\begin{array}{ll}
&\bmu_{\qDens(\bbeta,\bu)}\leftarrow(\bC^T\RMFVB^{-1}\bC+\DMFVB)^{-1}(\bC^T\RMFVB^{-1}\by + \oMFVB)\\[2ex]
\mbox{and}\ \ &\bSigma_{\qDens(\bbeta,\bu)}\leftarrow(\bC^T\RMFVB^{-1}\bC+\DMFVB)^{-1}
\end{array}
\label{eq:muSigmaMFVBupd}
\end{equation}
where 
\begin{equation}
\RMFVB\equiv\mu_{\qDens(1/\mysigeps^2)}^{-1}\bI,
\quad
\DMFVB\equiv\left[
\begin{array}{cc}
\bSigma_{\bbeta}^{-1} & \bO               \\[1ex]
\bO & \bI_m\otimes\bM_{\qDens(\bSigma^{-1})}
\end{array}
\right]
\quad\mbox{and}\quad
\oMFVB\equiv\left[
\begin{array}{c}
\bSigma_{\bbeta}^{-1}\bmu_{\bbeta}\\[1ex]
\bzero
\end{array}
\right].
\label{eq:MFVBmatDefns}
\end{equation}
For increasingly large sample sizes the matrix $\bSigma_{\qDens(\bbeta,\bu)}$ 
becomes untenably massive. Fortunately, only the following relatively
small sub-blocks of $\bSigma_{\qDens(\bbeta,\bu)}$ are required for
variational inference concerning $\mysigeps^2$ and $\bSigma$: 
\begin{equation}
{\setlength\arraycolsep{1pt}
\begin{array}{rcl}
&&\bSigma_{\qDens(\bbeta)}=\mbox{top left-hand $p\times p$ sub-block 
of $(\bC^T\RMFVB^{-1}\bC+\DMFVB)^{-1}$},\\[1ex]
&&\bSigma_{\qDens(\bu_i)}=\mbox{subsequent $q\times q$ diagonal sub-blocks of 
$(\bC^T\RMFVB^{-1}\bC+\DMFVB)^{-1}$}\\
&&\qquad\qquad\mbox{below $\bSigma_{\qDens(\bbeta)}$,\ $1\le i\le m$, and}\\[1ex]
&&E_{\qDens}\{(\bbeta-\bmu_{\qDens(\beta)})(\bu_i-\bmu_{\qDens(\bu_i)})^T\}=
\mbox{subsequent $p\times q$ sub-blocks of} \\[1ex]
&&\qquad\qquad\qquad\qquad\mbox{$(\bC^T\RMFVB^{-1}\bC+\DMFVB)^{-1}$
to the right of $\bSigma_{\qDens(\bbeta)}$,\ $1\le i\le m$.}
\end{array}
}
\label{eq:CovMFVB}
\end{equation}

\noindent
For a streamlined mean field variational Bayes algorithm, we appeal to:

\jump\noindent
\begin{result} 
The mean field variational Bayes updates of $\bmu_{\qDens(\bbeta,\bu)}$ and each
of the sub-blocks of $\bSigma_{\qDens(\bbeta,\bu)}$ listed in (\ref{eq:CovMFVB}) are 
expressible as a two-level sparse matrix least squares problem 
(see Appendix \ref{sec:twoLevSMA}) of the form:
$$\left\Vert\bb-\bB\bmu_{\qDens(\bbeta,\bu)}
\right\Vert^2
$$
where $\bb$ and the non-zero sub-blocks of $\bB$, according to the notation
in (\ref{eq:BandbForms}), are, for $1\le i\le m$,
$$\bveci\equiv\left[\begin{array}{c}   
\mu_{\qDens(1/\mysigeps^2)}^{1/2}\by_i\\[1ex]
m^{-1/2}\bSigma_{\bbeta}^{-1/2}\bmu_{\bbeta}\\[1ex]
\bzero
\end{array}
\right],
\quad\Bmati\equiv\left[\begin{array}{c}   
\mu_{\qDens(1/\mysigeps^2)}^{1/2}\bX_i\\[1ex]
m^{-1/2}\bSigma_{\bbeta}^{-1/2}\\[1ex]
\bO
\end{array}
\right]
\quad\mbox{and}\quad
\Bmatdoti\equiv
\left[\begin{array}{c}   
\mu_{\qDens(1/\mysigeps^2)}^{1/2}\bZ_i\\[1ex]
\bO\\[1ex]
\bM_{\qDens(\bSigma^{-1})}^{1/2}
\end{array}
\right],
$$
with each of these matrices having $\nadj_i=n_i+p+q$ rows.
The solutions are, according to the notation in (\ref{eq:AtLevxa})
and (\ref{eq:AtLevInv}),
$$\bmu_{\qDens(\bbeta)}=\xveco,\quad\bSigma_{\qDens(\bbeta)}=\AUoo$$ 
\textit{and}
$$\bmu_{\qDens(\bu_i)}=\xvectCi,\ \ \bSigma_{\qDens(\bu_i)}=\AUttCi,
\ \ E_{\qDens}\{(\bbeta-\bmu_{\qDens(\bbeta)})(\bu_i-\bmu_{\qDens(\bu_i)})^T\}=\AUotCi,
\quad 1\le i\le m.
$$
\label{res:twoLevelMFVB}
\end{result}

Result \ref{res:twoLevelMFVB} implies that the \SolveTwoLevelSparseLeastSquares\ algorithm 
listed in Algorithm \ref{alg:SolveTwoLevelSparseLeastSquares} applies for handling 
the $\bmu_{\qDens(\bbeta,\bu)}$ and $\bSigma_{\qDens(\bbeta,\bu)}$ sub-block updates.
A derivation is in Appendix \ref{sec:resTwoDeriv}.
This results in Algorithm \ref{alg:twoLevelMFVB} for streamlined mean field variational
Bayes for the two-level Gaussian response linear mixed model. A derivation is given in
Appendix \ref{sec:DerivTwoLevMFVB}. 

An important aspect of Result \ref{res:twoLevelMFVB} and 
Algorithm \ref{alg:twoLevelMFVB}
is that the vector $(\bbeta,\bu)$ is treated as an entity
in the updates. This contrasts with block Markov chain Monte
Carlo sampling schemes where sub-vectors of $(\bbeta,\bu)$ are
updated separately. In the case of variational inference, block updating 
of the sub-vectors of $(\bbeta,\bu)$ corresponds to the imposition of
more stringent product restrictions on the $\qDens$-density of 
$(\bbeta,\bu)$ and degradation of accuracy.

Algorithm \ref{alg:twoLevelMFVB} uses the mean field variational Bayes
approximate marginal log-likelihood $\log\{\pDensUnder(\by;\qDens)\}$ in its stopping criterion. 
For model (\ref{eq:twoLevelGaussRespBaye}) this is given by
\begin{equation}
\log\{\underline{\pDens}(\by;\qDens)\}=E_{\qDens}
\{\log\,\pDens(\by,\bbeta,\bu,\sigma^2,\asigsq,\bSigma,\ASigma)
-\qDens(\bbeta,\bu,\sigma^2,\asigsq,\bSigma,\ASigma)\}.
\label{eq:logMLrawTwoLev}
\end{equation}
An explicit streamlined expression for $\log\{\pDensUnder(\by;\qDens)\}$ and 
corresponding derivation is given in Nolan (2020).

\begin{algorithm}[!th]
\begin{center}
\begin{minipage}[t]{154mm}
\begin{small}
\begin{itemize}
\setlength\itemsep{4pt}
\item[] Data Inputs: $\by_i(n_i\times1),\ \bX_i(n_i\times p),\ \bZ_i(n_i\times q),\ 
1\le i\le m$.
\item[] Hyperparameter Inputs: $\bmu_{\bbeta}(p\times1)$, 
$\bSigma_{\bbeta}(p\times p)\ \mbox{symmetric and positive definite}$,\\
\null$\qquad\qquad\qquad\qquad\qquad\quad\ssigsq,\nusigsq,\sSigmaOne,\ldots,\sSigmaq,\nuSigma>0$.
\item[] Initialize: $\muq{1/\sigsqeps} > 0$, $\muq{1/\asigsq} > 0$,
                  $\bM_{\qDens(\bSigma^{-1})}(q\times q),\ \bM_{\qDens(\ASigma^{-1})}(q\times q)$ 
                  both symmetric and\\
\null$\qquad\qquad$ positive definite. 
\item[] $\xi_{\qDens(\mysigeps^2)}\thickarrow\nu_{\sigma^2}+\sumim n_i$\ \ \ ;\ \ \ 
$\xi_{\qDens(\bSigma)}\thickarrow \nuSigma+2q-2+m$\ \ \ ;\ \ \ 
$\xi_{\qDens(a_{\sigma^2})}\thickarrow\nusigsq+1$\ \ \ ;\ \ \ $\xi_{\qDens(\ASigma)}\thickarrow\nuSigma+q$
\\[0.5ex]
\item[] Cycle: 
\begin{itemize}
\item[] For $i = 1,\ldots, m$:	
\begin{itemize}	       
 \item[] $\bveci\thickarrow\left[\begin{array}{c}   
\mu_{\qDens(1/\mysigeps^2)}^{1/2}\by_i\\[1ex]
m^{-1/2}\bSigma_{\bbeta}^{-1/2}\bmu_{\bbeta}\\[1ex]
\bzero
\end{array}
\right],\ 
\Bmati\thickarrow\left[\begin{array}{c}   
\mu_{\qDens(1/\mysigeps^2)}^{1/2}\bX_i\\[1ex]
m^{-1/2}\bSigma_{\bbeta}^{-1/2}\\[1ex]
\bzero
\end{array}
\right],\
\Bmatdoti\thickarrow\left[\begin{array}{c}   
\mu_{\qDens(1/\mysigeps^2)}^{1/2}\bZ_i\\[1ex]
\bzero\\[1ex]
\bM_{\qDens(\bSigma^{-1})}^{1/2}\\[1ex]
\end{array}
\right].
$
\end{itemize}
\item[] $\SscAlgOne\thickarrow\SolveTwoLevelSparseLeastSquares\Big(\big\{(\bveci,\Bmati,\Bmatdoti):1\le i\le m\big\}\Big)$
\item[] $\bmu_{\qDens(\bbeta)}\thickarrow\mbox{$\xveco$ component of $\SscAlgOne$}$
\ \ \ ;\ \ \ $\bSigma_{\qDens(\bbeta)}\thickarrow\mbox{$\AUoo$ component of $\SscAlgOne$}$
\item[] $\lambda_{\qDens(\sigsqeps)}\thickarrow\muq{1/\asigsq}$\ \ ;\ \ 
$\Lambda_{\qDens(\bSigma)}\thickarrow\bM_{\qDens(\ASigma^{-1})}$
\item[] For $i = 1,\ldots, m$:
\begin{itemize}	
\item[] $\bmu_{\qDens(\bu_i)}\thickarrow\mbox{$\xvectCi$ component of $\SscAlgOne$}$\ \ \ ;\ \ \ 
$\bSigma_{\qDens(\bu_i)}\thickarrow\mbox{$\AUttCi$ component of $\SscAlgOne$}$
\item[] $E_{\qDens}\{(\bbeta-\bmu_{\qDens(\bbeta)})(\bu_i-\bmuq{\bu_i})^T\}\thickarrow
                 \mbox{$\AUotCi$ component of $\SscAlgOne$}$
\item[] $\lambda_{\qDens(\sigsqeps)}\thickarrow \lambda_{\qDens(\sigsqeps)}
+\big\Vert\by_i-\bX_i\bmu_{\qDens(\bbeta)}-\bZ_i\bmu_{\qDens(\bu_i)}\big\Vert^2$
\item[] $\lambda_{\qDens(\sigsqeps)}\thickarrow \lambda_{\qDens(\sigsqeps)}
+\mbox{tr}(\bX_i^T\bX_i\bSigma_{\qDens(\bbeta)})
+\mbox{tr}(\bZ_i^T\bZ_i\bSigma_{\qDens(\bu_i)})$
\item[] $\lambda_{\qDens(\sigsqeps)}\thickarrow \lambda_{\qDens(\sigsqeps)}
+2\,\mbox{tr}\big[\bZ_i^T\bX_iE_{\qDens}\{(\bbeta-\bmu_{\qDens(\bbeta)})
                 (\bu_i-\bmuq{\bu_i})^T\}\big]$
\item[] $\bLambda_{\qDens(\bSigma)}\thickarrow\bLambda_{\qDens(\bSigma)}+\bmu_{\qDens(\bu_i)}\bmu_{\qDens(\bu_i)}\trans 
              + \bSigma_{\qDens(\bu_i)}$
\end{itemize}   
\item[] $\muq{1/\sigsqeps} \thickarrow \xi_{\qDens(\mysigeps^2)}/\lambda_{\qDens(\sigsqeps)}$\ \ \ ;\ \ \  
$\MqSigma \thickarrow(\xi_{\qDens(\bSigma)}-q+1)\,\bLambda^{-1}_{\qDens(\bSigma)}$ 	 
\item[] $\lambda_{\qDens(\asigsq)}\thickarrow\muq{1/\sigsqeps}+1/(\nusigsq\ssigsq^2)$\ \ \ ;\ \ \ 
$\muq{1/\asigsq} \thickarrow \xi_{\qDens(\asigsq)}/\lambda_{\qDens(\asigsq)}$ 
\item[] $\bLambda_{\qDens(\ASigma)}\thickarrow 
\diag\big\{\mbox{diagonal}\big(\bM_{\qDens(\bSigma^{-1})}\big)\big\}+\{\nuSigma\diag(\sSigmaOne^2,\ldots,\sSigmaq^2)\}^{-1}$
\item[] $\bM_{\qDens(\ASigma^{-1})}\thickarrow \xi_{\qDens(\ASigma)}\bLambda_{\qDens(\ASigma)}^{-1}$.
\end{itemize} 
\item[] until the increase in $\log\{\underline{\pDens}(\by;\qDens)\}$ is negligible.
\item[] Outputs: $\bmu_{\qDens(\bbeta)}$,\ $\bSigma_{\qDens(\bbeta)}$,\  
$\big\{\big(\bmu_{\qDens(\bu_i)},\bSigma_{\qDens(\bu_i)},E_{\qDens}\{(\bbeta-\bmu_{\qDens(\bbeta)})(\bu_i-\bmuq{\bu_i})^T\}\big)
:1\le i\le m\big\}$
\item[]$\qquad\qquad$
$\xi_{\qDens(\mysigeps^2)},\lambda_{\qDens(\mysigeps^2)},\xi_{\qDens(\bSigma)},\bLambda_{\qDens(\bSigma)}$ 
\end{itemize}
\end{small}
\end{minipage}
\end{center}
\caption{\it QR-decomposition-based streamlined algorithm for obtaining mean field variational
Bayes approximate posterior density functions for the parameters in the two-level linear
mixed model (\ref{eq:twoLevelGaussRespBaye}) with product density restriction
(\ref{eq:producRestrict}).}
\label{alg:twoLevelMFVB} 
\end{algorithm}

\subsection{Variational Message Passing}

\begin{figure}[!ht]
\centering
{\includegraphics[width=0.65\textwidth]{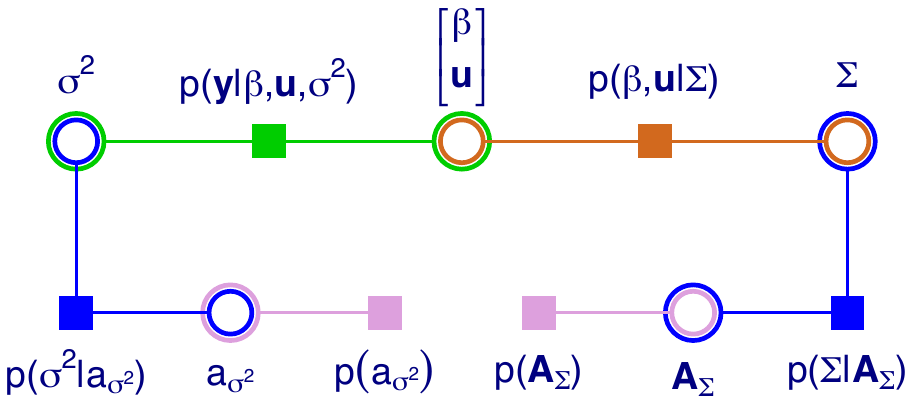}}
\caption{\it Factor graph representation of the Bayesian two-level 
Gaussian response linear mixed model (\ref{eq:twoLevelGaussRespBaye}).}
\label{fig:twoLevFacGraph} 
\end{figure}

We now turn attention to the variational message passing alternative.
Note that the joint density function of all of the 
random variables and random vectors in the Bayesian two-level Gaussian 
response linear mixed model (\ref{eq:twoLevelGaussRespBaye})
admits the following factorization:
\begin{equation}
\pDens(\by,\bbeta,\bu,\mysigeps^2,\bSigma,\asigsq,\bASigma)
=\pDens(\by|\bbeta,\bu,\mysigeps^2)\pDens(\mysigeps^2|\asigsq)\pDens(\asigsq)\pDens(\bbeta,\bu|\bSigma)\pDens(\bSigma|\bASigma)\pDens(\bASigma).
\label{eq:factorAlg}
\end{equation}
Figure \ref{fig:twoLevFacGraph} shows a factor graph representation
of (\ref{eq:factorAlg}) with color-coding of 
\emph{fragment} types, according to the nomenclature in Wand (2017).

Each of these fragments is treated in Section 4.1 of Wand (2017).
However, the updates for the Gaussian likelihood fragment, shown in green
in Figure \ref{fig:twoLevFacGraph}, and the Gaussian penalization
fragment, shown in brown in Figure \ref{fig:twoLevFacGraph},
are given in simple na\"{\i}ve forms in Wand (2017) without matrix
algebraic streamlining. The next two subsections overcome this deficiency.

\subsection{Streamlined Gaussian Likelihood Fragment Updates}\label{sec:streamGaussLikFrag}

We now focus on the Gaussian likelihood fragment, shown in green in 
Figure \ref{fig:twoLevFacGraph}. As presented in Section 4.1.5 of Wand (2017), the
messages passed between $\pDens(\by|\bbeta,\bu,\mysigeps^2)$ and  $(\bbeta,\bu)$ 
involve  Multivariate Normal distributions with natural 
parameter vectors containing 
\begin{equation}
p+mq+\smhalf(p+mq)(p+mq+1)
\label{eq:fullDimnVal}
\end{equation}
unique entries. Since the sizes of these vectors grow quadratically with the number of groups,
message passing suffers from burdensome storage and computational
demands. We overcome this problem by noticing that messages passed to and from
$\pDens(\by|\bbeta,\bu,\mysigeps^2)$ are within \emph{reduced} Multivariate Normal
families.

\begin{figure}[!ht]
\centering
{\includegraphics[width=0.5\textwidth]{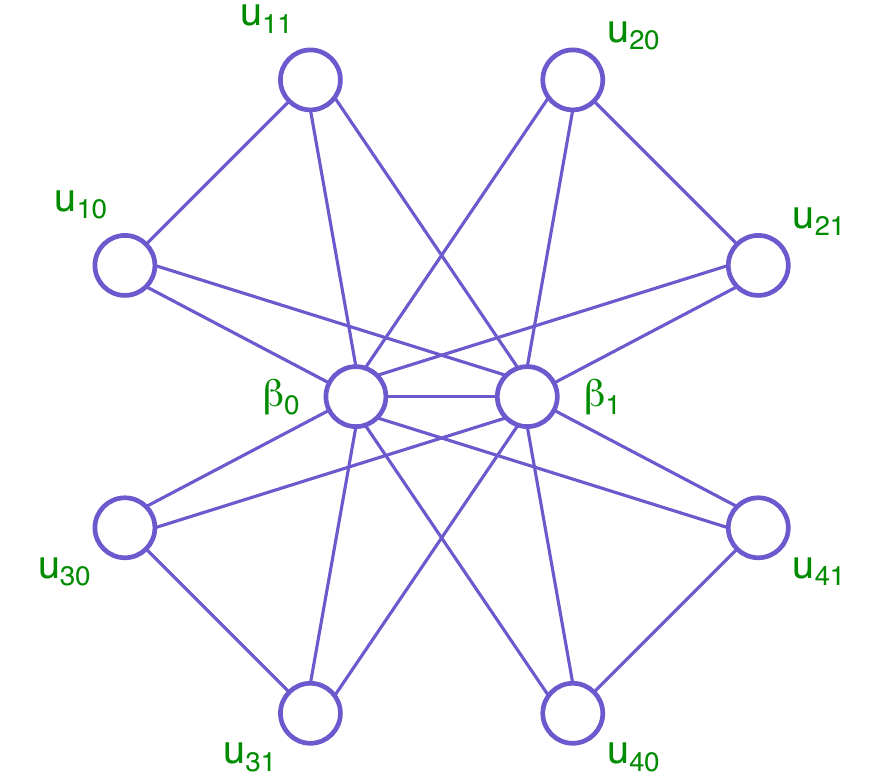}}
\caption{\it
Undirected probabilistic graph with edges coding the 
conditional dependencies of the entries of $(\bbeta,\bu)$ given the
rest for the case $p=q=2$ and $m=4$.}
\label{fig:butterflyGraph} 
\end{figure}

Note that the full conditional density function of $(\bbeta,\bu)$
is Multivariate Normal with inverse covariance matrix
$$\Cov(\bbeta,\bu|\mbox{rest})^{-1}=\mysigeps^{-2}\bC^T\bC+\mbox{blockdiag}(
\bSigma_{\bbeta}^{-1},\bI_m\otimes\bSigma^{-1}),
$$
where `rest' denotes all other random variables in the model,
is a two-level sparse matrix. The same is true for $\bSigma_{\qDens(\bbeta,\bu)}^{-1}$,
the inverse covariance matrix of the mean field approximate posterior density
function of $(\bbeta,\bu)$. In the variational message passing approach this sparseness 
transfers to reduced exponential family forms being sufficient. For example,
in the case of $p=q=2$ the messages passed between $\pDens(\by|\bbeta,\bu,\mysigeps^2)$ and
$(\bbeta,\bu)=(\beta_0,\beta_1,u_{10},u_{11},\ldots,u_{m0},u_{m1})$
have the generic exponential family forms:
\begin{equation}
\setlength\arraycolsep{1pt}{
\begin{array}{rcl}
&&\exp\Big\{\eta_{\mbox{\tiny$\beta_0$}}+\eta_{\mbox{\tiny$\beta_1$}}\beta_1
+{\displaystyle\sumim}\,(\eta_{\mbox{\tiny$u_{i0}$}}u_{i0}+\eta_{\mbox{\tiny$u_{i1}$}}u_{i1})
+\eta_{\mbox{\tiny$\beta_0^2$}}\beta_0^2+\eta_{\mbox{\tiny$\beta_1^2$}}\beta_1^2+
\displaystyle{\sumim}\,(\eta_{\mbox{\tiny$u_{i0}^2$}}\ u_{i0}^2+\eta_{\mbox{\tiny$u_{i1}^2$}}u_{i1}^2)\\[2ex]
&&\qquad\qquad+\displaystyle{\sum_{i=1}^m}(\eta_{\mbox{\tiny$\beta_0u_{i0}$}}\beta_0\,u_{i0}
                                                +\eta_{\mbox{\tiny$\beta_0u_{i1}$}}\beta_0\,u_{i1}
                                                +\eta_{\mbox{\tiny$\beta_1u_{i0}$}}\beta_1\,u_{i0}
                                                +\eta_{\mbox{\tiny$\beta_1u_{i1}$}}\beta_1\,u_{i1})\Big\}.
\end{array}
}
\label{eq:fullFourm}
\end{equation}
Therefore, it is natural to insist that all messages passed
to $(\bbeta,\bu)$ from factors outside of the two-level Gaussian likelihood fragment
are within the same reduced exponential family. Under such a conjugacy
constraint, the natural parameter vectors of messages passed to and from 
$(\bbeta,\bu)$ have length
$$p+\smhalf\,p(p+1)+m\{q+\smhalf\,q(q+1)+pq\}$$
which is linear in $m$ and considerably lower than (\ref{eq:fullDimnVal})
when the number of groups is large. The reduced exponential family has 
an attractive graph theoretic representation. The full Multivariate 
Normal distribution, in which sparseness is ignored, has dimension $p+mq$. 
The probabilistic undirected graph that respects independence of any pair of random variables conditional
on the rest for the $N(\bmu,\bSigma)$ distribution is an undirected graph with an edge 
between the $\ell$th and $\ell'$th nodes if and only if $(\bSigma^{-1})_{\ell\ell'}\ne 0$
(e.g. Rue \myand Held, 2005). The restricted exponential family corresponds to 
removal of edges in a fully connected $(p+mq)$-node graph. Figure
\ref{fig:butterflyGraph} depicts the reduced graph in the case of
$p=q=2$ and $m=4$. The fully connected graph has 45
edges, whereas the reduced graph corresponding to the restricted
exponential family has only 21 edges. For
general $p$, $q$ and $m$ the numbers of edges are, respectively,
$\smhalf(p+mq)(p+mq-1)$ and $\smhalf\,p(p-1)+m\{\smhalf\,q(q-1)+pq\}$. 
So, for example, if $p=q=2$ and $m=10,000$ then the number of edges 
in the reduced graph is about 50,000 compared with about 200 million 
in the full graph.

The message from $\pDens(\by|\bbeta,\bu,\mysigeps^2)$ to $(\bbeta,\bu)$ is
\begin{equation}
\begin{array}{l}
\mSUBpybetausigsqEpsTObetau=\\[1ex]
\null\qquad\qquad
\exp\left\{
\left[
{\setlength\arraycolsep{0pt}
\begin{array}{c}
\ \ \ \ \ \ \ \ \ \  \bbeta\\
\ \ \ \ \ \vech(\bbeta\bbeta^T)\\[1ex]
\displaystyle{\stack{1\le i\le m}}\left[
{\setlength\arraycolsep{0pt}
\begin{array}{c}
\bu_i\\
\vech(\bu_i\bu_i^T)\\
\vecof(\bbeta\bu_i^T)
\end{array}
}
\right]
\end{array}
}
\right]^T\null\hskip-3mm\etaSUBpybetausigsqEpsTObetau
\right\}
\end{array}
\label{eq:restrictedMsgFirst}
\end{equation}
with natural parameter vector $\etaSUBpybetausigsqEpsTObetau$
of length
\begin{equation}
p+\smhalf p(p+1)+m\{q+\smhalf q(q+1)+ pq\}.
\label{eq:lenYip}
\end{equation}
Under conjugacy, the reverse message $\mSUBbetauTOpybetausigsqEps$ has the same algebraic
form as (\ref{eq:restrictedMsgFirst}) with natural parameter vector 
$\etaSUBbetauTOpybetausigsqEps$ also of length (\ref{eq:lenYip}).

\jump
\noindent
\begin{result}
The variational message passing updates of 
the quantities $\bmu_{\qDens(\bbeta)}$, $\bmu_{\qDens(\bu_i)}$, $1\le i\le m$,
and the sub-blocks of $\bSigma_{\qDens(\bbeta,\bu)}$ 
listed in (\ref{eq:CovMFVB}) with $\qDens$-density 
expectations with respect to the normalization of
$$\mSUBpybetausigsqEpsTObetau\,\mSUBbetauTOpybetausigsqEps$$
are expressible as a two-level sparse matrix problem 
(see Appendix \ref{sec:twoLevSMA}) with
$$
\AtLev=
-2\left[
\begin{array}{cc}
\vecof^{-1}(\bD_p^{+T}\bdeta_{1,2}) & 
\Big[\smhalf\,\displaystyle{\stack{1\le i\le m}\{\vecof^{-1}(\bdeta_{2,3,i})^T\}}\Big]^T 
\\[3ex]
\smhalf\,\displaystyle{\stack{1\le i\le m}\{\vecof^{-1}(\bdeta_{2,3,i})^T\}} &
\displaystyle{\blockdiag{1\le i\le m}\{\vecof^{-1}(\bD_q^{+T}\bdeta_{2,2,i})\}} 
\end{array}
\right]
$$
and
$$
\ba\equiv
\left[
\arraycolsep=2.2pt\def\arraystretch{1.6}
\begin{array}{c}
\setstretch{4.5}
\bdeta_{1,1} \\[1ex]
\displaystyle{\stack{1\le i\le m}}(\bdeta_{2,1,i})
\end{array}
\right]
\quad\mbox{where}\quad
\left[
\begin{array}{c}
\bdeta_{1,1}\ (p\times 1)\\[1ex]
\bdeta_{1,2}\ (\smhalf p(p+1)\times 1)\\[1ex]
\displaystyle{\stack{1\le i\le m}}\left[
\begin{array}{ll}
\bdeta_{2,1,i}&(q\times 1)\\
\bdeta_{2,2,i}&(\smhalf q(q+1)\times 1)\\
\bdeta_{2,3,i}&(pq\times 1)
\end{array}
\right]
\end{array}
\right]
$$
is the partitioning of $\etaSUBpybetausigsqEpsCONNbetau$ that defines 
$\bdeta_{1,1}$, $\bdeta_{1,2}$ and $\{(\bdeta_{2,1,i},\bdeta_{2,2,i},\bdeta_{2,3,i}):1\le i\le m\}$.
The solutions, according to the notation in (\ref{eq:AtLevxa}) and (\ref{eq:AtLevInv}), are
$\bmu_{\qDens(\bbeta)}=\xveco$, $\bSigma_{\qDens(\bbeta)}=\AUoo$ \textit{and}
$$\bmu_{\qDens(\bu_i)}=\xvectCi,\ \ \bSigma_{\qDens(\bu_i)}=\AUttCi,
\ \ E_{\qDens}\{(\bbeta-\bmu_{\qDens(\beta)})(\bu_i-\bmu_{\qDens(\bu_i)})^T\}=\AUotCi,
\quad 1\le i\le m.
$$
\label{res:twoLevelVMPlik}
\end{result}

\noindent
{\bf Remark.} Variational message passing differs from mean field variational
Bayes in that its two-level sparse matrix problem is not expressible in 
a least squares form.
\jump

The process of converting a generic reduced natural parameter vector $\bdeta_{\qDens(\bbeta,\bu)}$ 
to the corresponding $\bmu_{\qDens(\bbeta,\bu)}$ vector and important sub-blocks
of $\bSigma_{\qDens(\bbeta,\bu)}$, as illustrated by Result \ref{res:twoLevelVMPlik}, 
is fundamental to streamlining of variational message passing for two-level linear mixed models. 
We call this procedure the \TwoLevelNaturalToCommonParameters\ algorithm and
list required steps as Algorithm \ref{alg:TwoLevelNaturalToCommonParameters}.

\begin{algorithm}[!th]
\begin{center}
\begin{minipage}[t]{154mm}
\begin{small}
\begin{itemize}
\setlength\itemsep{4pt}
\item[] Inputs: $p,q,m,\etaSUBqbetau$
\item[] $\bomegaAlgTwoA\thickarrow\mbox{first $p$ entries of $\etaSUBqbetau$}$
\item[] $\bomegaAlgTwoB\thickarrow\mbox{next $\smhalf\,p(p+1)$ entries of $\etaSUBqbetau$}$
\ ;\ $\bOmegaAlgTwoC\thickarrow -2\vecof^{-1}(\bD_p^{+T}\bomegaAlgTwoB)$
\item[] $\iStt\thickarrow p + \smhalf\,p(p+1)+1$\ \ \ ;\ \ \ $\iEnd\thickarrow\iStt + q-1$
\item[] For $i=1,\ldots,m$:
\begin{itemize}
\setlength\itemsep{0pt}
\item[] $\bomegaAlgTwoD\thickarrow\mbox{sub-vector of}\ \etaSUBqbetau$\ \mbox{with entries $\iStt$ to $\iEnd$ inclusive}
\item[] $\iStt\thickarrow\iEnd+1$\ \ \ ;\ \ \ $\iEnd\thickarrow\iStt + \smhalf\,q(q+1)-1$
\item[] $\bomegaAlgTwoE\thickarrow\mbox{sub-vector of}\ \etaSUBqbetau$\ \mbox{with entries $\iStt$ to $\iEnd$ inclusive}
\item[] $\iStt\thickarrow\iEnd+1$\ \ \ ;\ \ \ $\iEnd\thickarrow\iStt + pq-1$
\item[] $\bomegaAlgTwoF\thickarrow\mbox{sub-vector of}\ \etaSUBqbetau$\ \mbox{with entries $\iStt$ to $\iEnd$ inclusive}
\item[] $\iStt\thickarrow\iEnd+1$\ \ \ ;\ \ \ $\iEnd\thickarrow\iStt + q-1$
\item[] $\bOmegaAlgTwoG\thickarrow\,-2\,\vecof^{-1}(\bD_q^{+T}\bomegaAlgTwoE)$\ \ \ ;\ \ \ 
$\bOmegaAlgTwoH\thickarrow\,-\vecof^{-1}_{p\times q}(\bomegaAlgTwoF)$
\end{itemize}
\item[] $\SscAlgTwo\thickarrow
\SolveTwoLevelSparseMatrix\Big(\bomegaAlgTwoA,\bOmegaAlgTwoC,
\big\{(\bomegaAlgTwoD,\bOmegaAlgTwoG,\bOmegaAlgTwoH):1\le i\le m\big\}\Big)$ 
\item[] $\bmu_{\qDens(\bbeta)}\thickarrow\mbox{$\xveco$ component of $\SscAlgTwo$}$\ \ ;\ \
$\bSigma_{\qDens(\bbeta)}\thickarrow\mbox{$\AUoo$ component of $\SscAlgTwo$}$
\item[] For $i=1,\ldots,m$:
\begin{itemize}
\item[]$\bmu_{\qDens(\bu_i)}\thickarrow\mbox{$\xvectCi$ component of $\SscAlgTwo$}$
\ \ ;\ \ $\bSigma_{\qDens(\bu_i)}\thickarrow\mbox{$\AUttCi$ component of $\SscAlgTwo$}$
\item[] $E_{\qDens}\{(\bbeta-\bmu_{\qDens(\bbeta)}\}(\bu_i-\bmu_{\qDens(\bu_i)})^T\}
\thickarrow\mbox{$\AUotCi$ component of $\SscAlgTwo$}$
\end{itemize}
\item[] Outputs: $\bmu_{\qDens(\bbeta)},\bSigma_{\qDens(\bbeta)},
\big\{\big(\bmu_{\qDens(\bu_i)},\bSigma_{\qDens(\bu_i)},
E_{\qDens}\{(\bbeta-\bmu_{\qDens(\bbeta)}\}(\bu_i-\bmu_{\qDens(\bu_i)})^T\}):\ 1\le i\le m\big\}$
\end{itemize}
\end{small}
\end{minipage}
\end{center}
\caption{\textit{The} \TwoLevelNaturalToCommonParameters\ \textit{algorithm for 
conversion of a two-level reduced natural parameter vector to its corresponding
common parameters.}}
\label{alg:TwoLevelNaturalToCommonParameters} 
\end{algorithm}

\vskip3mm
It is easily shown (Appendix \ref{sec:drvTwoLevLikFrag}) that messages between 
$\pDens(\by|\bbeta,\bu,\mysigeps^2)$ and $\mysigeps^2$ have Inverse Chi-Squared forms.
For example,
\begin{equation}
\mSUBpybetausigsqEpsTOsigsqEps=\exp
\left\{
\left[
\begin{array}{c}
1/\mysigeps^2\\[1ex]
\log(\mysigeps^2)
\end{array}
\right]^T\etaSUBpybetausigsqEpsTOsigsqEps
\right\}.
\label{eq:GaussLikToSigsqMsg}
\end{equation}

Algorithm \ref{alg:twoLevelVMPlik} lists parameter
updates for the two-level Gaussian likelihood fragment with streamlining
according to the restricted exponential family form
(\ref{eq:restrictedMsgFirst}). Note that it makes
use of \SolveTwoLevelSparseMatrix\ (Algorithm \ref{alg:SolveTwoLevelSparseMatrix})
since the natural parameter updates correspond to a two-level sparse matrix problem 
\emph{without} least squares representation. Appendix \ref{sec:drvTwoLevLikFrag} 
provides details on the derivation of Algorithm \ref{alg:twoLevelVMPlik}.

As in Wand (2017), Algorithm \ref{alg:twoLevelVMPlik} uses the notation
\begin{equation}
\bdeta_{f\leftrightarrow\theta}\equiv\bdeta_{f\rightarrow\theta}+\bdeta_{\theta\rightarrow f}.
\label{eq:doubArrowNotat}
\end{equation}

\begin{algorithm}[!th]
\begin{center}
\begin{minipage}[t]{165mm}
\textbf{Data Inputs:} $\by_i(n_i\times1),\ \bX_i(n_i\times p),\ \bZ_i(n_i\times q),\ 
1\le i\le m$ 
\jump\noindent
\textbf{Parameter Inputs:} $\etaSUBpybetausigsqEpsTObetau$,\ $\etaSUBbetauTOpybetausigsqEps$,\ 
 $\etaSUBpybetausigsqEpsTOsigsqEps$,\\
\null\ \ \ \ \ \ \ \ \ \ \ \ \ \ \ \ \ \ \ \ \ \ \ \ \  \ \ $\etaSUBsigsqEpsTOpybetausigsqEps$\\
\textbf{Updates:}
\begin{itemize}
\setlength\itemsep{0pt}
\item[]$\mu_{\qDens(1/\mysigeps^2)}\thickarrow
\Big(\big(\etaSUBpybetausigsqEpsCONNsigsqEps\big)_1+1\Big)\Big/\big(\etaSUBpybetausigsqEpsCONNsigsqEps\big)_2$
\item[] $\SscAlgThree\thickarrow
\TwoLevelNaturalToCommonParameters\Big(p,q,m,\etaSUBpybetausigsqEpsCONNbetau\Big)$ 
\item[] $\bmu_{\qDens(\bbeta)}\thickarrow\mbox{$\bmu_{\qDens(\bbeta)}$ component of $\SscAlgThree$}$\ \ ;\ \
$\bSigma_{\qDens(\bbeta)}\thickarrow\mbox{$\bSigma_{\qDens(\bbeta)}$ component of $\SscAlgThree$}$
\item[]$\bomegaAlgThreeA\thickarrow\bzero_p$\ \ ;\ \ 
$\bomegaAlgThreeB\thickarrow\bzero_{\frac{1}{2}\,p(p+1)}$\ \ ;\ \ 
$\bomegaAlgThreeC\thickarrow 0$
\item[] For $i=1,\ldots,m$:
\begin{itemize}
\item[] $\bomegaAlgThreeA\thickarrow\bomegaAlgThreeA+\bX_i^T\by_i$\ \ \ ;\ \ \ 
        $\bomegaAlgThreeB\thickarrow\bomegaAlgThreeB-\smhalf\bD_p^T\vecof(\bX_i^T\bX_i)$
\item[]$\bmu_{\qDens(\bu_i)}\thickarrow\mbox{$\bmu_{\qDens(\bu_i)}$ component of $\SscAlgThree$}$
\ \ ;\ \ $\bSigma_{\qDens(\bu_i)}\thickarrow\mbox{$\bSigma_{\qDens(\bu_i)}$ component of $\SscAlgThree$}$
\item[] $E_{\qDens}\{(\bbeta-\bmu_{\qDens(\bbeta)})(\bu_i-\bmu_{\qDens(\bu_i)})^T\}\thickarrow
\mbox{$E_{\qDens}\{(\bbeta-\bmu_{\qDens(\bbeta)})(\bu_i-\bmu_{\qDens(\bu_i)})^T\}$ component}$
\item[] $\qquad\qquad\qquad\qquad\qquad\qquad\qquad\quad\mbox{ of $\SscAlgThree$}$
\item[]$\bomegaAlgThreeC\thickarrow\bomegaAlgThreeC 
-\smhalf\Vert\by_i-\bX_i\bmu_{\qDens(\bbeta)} - \bZ_i\bmu_{\qDens(\bu_i)}\Vert^2$
\item[] $\bomegaAlgThreeC\thickarrow\bomegaAlgThreeC-\smhalf\tr(\bSigma_{\qDens(\bbeta)}\bX_i^T\bX_i)-\smhalf\tr(\bSigma_{\qDens(\bu_i)}\bZ_i^T\bZ_i)$
\item[] $\qquad\qquad-\tr[\{\bZ_i^T\bX_iE_{\qDens}\{(\bbeta-\bmu_{\qDens(\bbeta)})(\bu_i-\bmu_{\qDens(\bu_i)})^T\}]$
\end{itemize}
\item[] $\etaSUBpybetausigsqEpsTObetau\thickarrow\mu_{\qDens(1/\mysigeps^2)}
\left[
\begin{array}{c}
\bomegaAlgThreeA\\
\bomegaAlgThreeB\\
{\displaystyle{\stack{1\le i\le m}}}
\left[\begin{array}{c}
\bZ_i^T\by_i\\
-\smhalf\bD_q^T\vecof(\bZ_i^T\bZ_i)\\
-\vecof(\bX_i^T\bZ_i)\\
\end{array}\right]
\end{array}
\right]$
\item[] $\etaSUBpybetausigsqEpsTOsigsqEps\thickarrow
\left[
\begin{array}{c}
-\smhalf{\displaystyle\sum_{i=1}^m} n_i\\[1ex]
\bomegaAlgThreeC
\end{array}
\right]$
\end{itemize}
\textbf{Parameter Outputs:} $\etaSUBpybetausigsqEpsTObetau$, $\etaSUBpybetausigsqEpsTOsigsqEps$.
\jump
\end{minipage}
\end{center}
\caption{\it The inputs, updates and outputs of the matrix algebraic streamlined Gaussian likelihood fragment
for two-level models.}
\label{alg:twoLevelVMPlik} 
\end{algorithm}

\subsection{Streamlined Gaussian Penalization Fragment Updates}

Next we turn our attention to the Gaussian penalization fragment
when the random effects vector has two-level structure.
The relevant fragment is shown in brown in Figure \ref{fig:twoLevFacGraph}.

As shown in Appendix \ref{sec:drvTwoLevPenFrag}, the message from $\pDens(\bbeta,\bu|\bSigma)$
to $(\bbeta,\bu)$ has the generic form (\ref{eq:fullFourm}) but
with even more vanishing terms than the message passed from 
$\pDens(\by|\bbeta,\bu,\mysigeps^2)$. However, with conjugacy in mind, we
work with messages having the same form as (\ref{eq:restrictedMsgFirst}).
This implies that 
$$\mSUBpbetauSigmaTObetau=
\exp\left\{
\left[
{\setlength\arraycolsep{0pt}
\begin{array}{c}
\ \ \ \ \ \ \ \ \ \  \bbeta\\
\ \ \ \ \ \vech(\bbeta\bbeta^T)\\[1ex]
{\displaystyle\stack{1\le i\le m}}\left[
{\setlength\arraycolsep{0pt}
\begin{array}{c}
\bu_i\\
\vech(\bu_i\bu_i^T)\\
\vecof(\bbeta\bu_i^T)
\end{array}
}
\right]
\end{array}
}
\right]^T\etaSUBpbetauSigmaTObetau
\right\}
$$
with natural parameter vector $\etaSUBpybetausigsqEpsTObetau$ also of length
(\ref{eq:lenYip}). The reverse message has an analogous form.

\jump
\noindent
\begin{result}
The variational message passing updates of 
the quantities  $\bmu_{\qDens(\bu_i)}$ and $\bSigma_{\qDens(\bu_i)}$, $1\le i\le m$,
with $\qDens$-density expectations with respect to the normalization of
$$\mSUBpbetauSigmaTObetau\,\mSUBbetauTOpbetauSigma$$
are expressible as a two-level sparse matrix problem (see Appendix \ref{sec:twoLevSMA}) 
with
$$
\AtLev=
-2\left[
\begin{array}{cc}
\vecof^{-1}(\bD_p^{+T}\bdeta_{1,2}) & 
\Big[\smhalf\,\displaystyle{\stack{1\le i\le m}\{\vecof^{-1}(\bdeta_{2,3,i})^T\}}\Big]^T 
\\[3ex]
\smhalf\,\displaystyle{\stack{1\le i\le m}\{\vecof^{-1}(\bdeta_{2,3,i})^T\}} &
\displaystyle{\blockdiag{1\le i\le m}\{\vecof^{-1}(\bD_q^{+T}\bdeta_{2,2,i})\}} 
\end{array}
\right]
$$
and
$$
\ba\equiv
\left[
\arraycolsep=2.2pt\def\arraystretch{1.6}
\begin{array}{c}
\setstretch{4.5}
\bdeta_{1,1}     \\[1ex]
\displaystyle{\stack{1\le i\le m}}(\bdeta_{2,1,i})\\
\end{array}
\right]
\quad\mbox{where}\quad
\left[
\begin{array}{c}
\bdeta_{1,1}\ (p\times 1)\\[1ex]
\bdeta_{1,2}\ (\smhalf p(p+1)\times 1)\\[1ex]
\displaystyle{\stack{1\le i\le m}}
\left[
\begin{array}{ll}
\bdeta_{2,1,i}&(q\times 1)\\
\bdeta_{2,2,i}&(\smhalf q(q+1)\times 1)\\
\bdeta_{2,3,i}&(pq\times 1)
\end{array}
\right]
\end{array}
\right]
$$
is the partitioning of $\etaSUBpbetauSigmaCONNbetau$ that defines 
$\bdeta_{1,1}$, $\bdeta_{1,2}$ and $\{(\bdeta_{2,1,i},\bdeta_{2,2,i},\bdeta_{2,3,i}):1\le i\le m\}$.
The solutions are, according to the notation in (\ref{eq:AtLevxa}) and (\ref{eq:AtLevInv}),
$$\bmu_{\qDens(\bu_i)}=\xvectCi\quad\textit{and}\quad\bSigma_{\qDens(\bu_i)}=\AUttCi,\quad 1\le i\le m.$$
\label{res:twoLevelVMPpen}
\end{result}

As shown in Appendix \ref{sec:drvTwoLevLikFrag}, the message from $\pDens(\bbeta,\bu|\bSigma)$ to $\bSigma$
has the Inverse-G-Wishart form
$$\mSUBpbetauSigmaTOSigma=
\exp\left\{
\left[
\begin{array}{c}
\log|\bSigma|\\[1ex]
\vech(\bSigma^{-1})
\end{array}
\right]^T\etaSUBpbetauSigmaTOSigma
\right\}.
$$
Conjugacy considerations dictate that the message from $\bSigma$ to \
$\pDens(\bbeta,\bu|\bSigma)$ is within the same exponential family.

Algorithm \ref{alg:twoLevelVMPpen} lists the natural parameter
updates for the Gaussian penalization fragment for two-level random effects.
Notation such as $\etaSUBpbetauSigmaCONNSigma$ is as defined by (\ref{eq:doubArrowNotat}).
See Appendix \ref{sec:drvTwoLevLikFrag} for its derivation.

\begin{algorithm}[!th]
\begin{center}
\begin{minipage}[t]{165mm}
\textbf{Hyperparameter Inputs:} $\bmu_{\bbeta}(p\times 1)$, $\bSigma_{\bbeta}(p\times p)$, $m$, $q$\\
\textbf{Parameter Inputs:} $\etaSUBpbetauSigmaTObetau$,\ $\etaSUBbetauTOpbetauSigma$,\ $\etaSUBpbetauSigmaTOSigma$,
$\etaSUBSigmaTOpbetauSigma$\\
\textbf{Updates:}
\begin{itemize}
\setlength\itemsep{1pt}
\item[] $\omegaAlgFourA\thickarrow\mbox{first entry of $\etaSUBpbetauSigmaCONNSigma$}$
\ ; 
$\bomegaAlgFourB\thickarrow\mbox{remaining entries of $\etaSUBpbetauSigmaCONNSigma$}$
\item[] $\bM_{\qDens(\bSigma^{-1})}\thickarrow
\big\{\omegaAlgFourA + \smhalf(q+1)\big\}\{\vecof^{-1}\big(\bD_q^{+T}\bomegaAlgFourB\big)\}^{-1}$
\item[] $\SscAlgFour\thickarrow
\TwoLevelNaturalToCommonParameters\Big(p,q,m,\etaSUBpbetauSigmaCONNbetau\Big)$
\item[]$\bomegaAlgFourC\thickarrow\bzero_{\frac{1}{2}\,q(q+1)}$
\item[] For $i=1,\ldots,m$:
\begin{itemize}
\setlength\itemsep{0pt}
\item[] $\bmu_{\qDens(\bu_i)}\thickarrow\mbox{$\bmu_{\qDens(\bu_i)}$ component of $\SscAlgFour$}$
\ \ ;\ \ $\bSigma_{\qDens(\bu_i)}\thickarrow\mbox{$\bSigma_{\qDens(\bu_i)}$ component of $\SscAlgFour$}$
\item[] $\bomegaAlgFourC\thickarrow\bomegaAlgFourC -\smhalf\,
\bD_q^T\vecof\Big(\bmu_{\qDens(\bu_i)}\bmu_{\qDens(\bu_i)}^T +\bSigma_{\qDens(\bu_i)}\Big)$
\end{itemize}
\item[] $\etaSUBpbetauSigmaTObetau\thickarrow
\left[
\begin{array}{c}
\bSigma_{\bbeta}^{-1}\bmu_{\bbeta}\\[2ex]
-\smhalf\bD_p^T\vecof(\bSigma_{\bbeta}^{-1})\\[2ex]
{\displaystyle{\stack{1\le i\le m}}}\left[\begin{array}{c}
\bzero_q\\[1ex]
-\smhalf\bD_q^T\vecof\big(\bM_{\qDens(\bSigma^{-1})}\big)\\
\bzero_{pq}\\
\end{array}\right]
\end{array}
\right]$
\item[] $\etaSUBpbetauSigmaTOSigma\thickarrow
\left[
\begin{array}{c}
-\smhalf m\\[1ex]
\bomegaAlgFourC
\end{array}
\right]$
\end{itemize}
\textbf{Parameter Outputs:} $\etaSUBpbetauSigmaTObetau$\ ,\ $\etaSUBpbetauSigmaTOSigma$.
\end{minipage}
\end{center}
\caption{\it The inputs, updates and outputs of the matrix algebraic streamlined Gaussian 
penalization fragment for two-level models.}
\label{alg:twoLevelVMPpen} 
\end{algorithm}

\subsection{$\qDens$-Density Determination After Variational Message Passing Convergence}\label{sec:qDensTwoLev}

After convergence of the variational message passing iterations, determination of 
$\qDens$-density parameters of interest requires some additional non-trivial steps,
essentially involving mapping particular natural parameter vectors to common parameters
of interest. We will explain this in the context of inference for the parameters
in (\ref{eq:twoLevelGaussRespBaye}) and its Figure \ref{fig:twoLevFacGraph} factor graph representation.

For the fixed and random effects parameters we need to first carry out:
{\setlength\arraycolsep{1pt}
\begin{eqnarray*}
\etaSUBqbetau&\thickarrow& \etaSUBpybetausigsqEpsTObetau+\etaSUBpbetauSigmaTObetau\\[1ex]
\SscFirstInText&\thickarrow& \TwoLevelNaturalToCommonParameters\Big(p,q,m,\etaSUBqbetau\Big)
\end{eqnarray*}
}
and then unpack $\SscFirstInText$ to obtain the mean and important covariance matrix sub-blocks:

\centerline{
$\bmu_{\qDens(\bbeta)}$,\ $\bSigma_{\qDens(\bbeta)}$,\  
$\Big\{\bmu_{\qDens(\bu_i)},\bSigma_{\qDens(\bu_i)},E_{\qDens}\{(\bbeta-\bmu_{\qDens(\bbeta)})(\bu_i-\bmuq{\bu_i})^T\}
:1\le i\le m\Big\}.$
}

\noindent
of the $N(\bmu_{\qDens(\bbeta,\bu)},\bSigma_{\qDens(\bbeta,\bu)})$ optimal $\qDens$-density function.

The error variance $\sigma^2$ has its optimal $\qDens$-density function being
that of an $\mbox{Inverse-}\chi^2\big(\xi_{\qDens(\sigma^2)},\lambda_{\qDens(\sigma^2)}\big)$ 
distribution, and its parameters are determined from the steps:
{\setlength\arraycolsep{1pt}
\begin{eqnarray*}
\biggerbdeta_{\qDens(\sigma^2)}&\thickarrow&\etaSUBpybetausigsqEpsTOsigsqEps+\biggerbdeta_{\pDens(\sigma^2|\asigsq)\to\sigma^2}\\[1ex]
\xi_{\qDens(\sigma^2)}&\thickarrow& -2\big(\biggerbdeta_{\qDens(\sigma^2)}\big)_1-2,\ \ ;\ \ 
\lambda_{\qDens(\sigma^2)}\thickarrow -2\big(\biggerbdeta_{\qDens(\sigma^2)}\big)_2
\end{eqnarray*}
}
where $\big(\biggerbdeta_{\qDens(\sigma^2)}\big)_j$ denotes the $j$th entry of the vector 
$\biggerbdeta_{\qDens(\sigma^2)}$ for $j=1,2$.

Finally, the random effects covariance matrix $\bSigma$ has its optimal $\qDens$-density function being
that of an $\mbox{Inverse-G-Wishart}\big(\Gfull,\xi_{\qDens(\bSigma)},\bLambda_{\qDens(\bSigma)}\big)$ 
distribution. The steps for determining its parameters after variational message passing
convergence are:
{\setlength\arraycolsep{1pt}
\begin{eqnarray*}
\biggerbdeta_{\qDens(\bSigma)}&\thickarrow&\etaSUBpbetauSigmaTOSigma+\biggerbdeta_{\pDens(\bSigma|\ASigma)\to\bSigma}\\[1ex]
\xi_{\qDens(\bSigma)}&\thickarrow& -2\big(\biggerbdeta_{\qDens(\bSigma)}\big)_1-2,\ \ ;\ \ 
\bLambda_{\qDens(\bSigma)}\thickarrow -2\,\vecof^{-1}\Big(\bD_q^{+T}\big(\biggerbdeta_{\qDens(\bSigma)}\big)_2\Big)
\end{eqnarray*}
}
where $\big(\biggerbdeta_{\qDens(\bSigma)}\big)_1$ denotes the first entry of $\biggerbdeta_{\qDens(\bSigma)}$
and $\big(\biggerbdeta_{\qDens(\bSigma)}\big)_2$ denotes its remaining entries.

\subsection{Generalized Linear Mixed Model Extensions}

In this article  we focus on Gaussian response linear mixed models. 
The general principles also apply to non-Gaussian response models within
the generalized linear mixed models framework. For the variational message passing
approach Algorithm \ref{alg:twoLevelVMPpen} is applicable for generalized linear mixed models
as well since it involves nodes of the factor graph that are isolated from the likelihood
factor. However, Algorithm \ref{alg:twoLevelVMPlik} is specific to the Gaussian likelihood
factor and extension to non-Gaussian likelihood cases is the subject of
ongoing research. 

\section{Three-Level Models}\label{sec:threeLevMods}

We now return to the three-level situation illustrated by Figure \ref{fig:residentProto}
and derive algorithms for streamlined variational inference based on Algorithms 
\ref{alg:SolveThreeLevelSparseMatrix} and \ref{alg:SolveThreeLevelSparseLeastSquares}.

\subsection{Mean Field Variational Bayes}\label{sec:MFVB3lev}

A Bayesian version of the three-level linear mixed model treated
in the previous subsection is
\begin{equation}
\begin{array}{c}
\by_{ij}|\bbeta,\buLone_i,\buLtwo_{ij},\mysigeps^2\simind N(\bX_{ij}\bbeta+\bZLone_{ij}\,\buLone_i
+\bZLtwo_{ij}\,\buLtwo_{ij},\mysigeps^2\,\bI),\\[2ex]
\left[
\begin{array}{c}
\buLone_i\\
\buLtwo_{ij}
\end{array}
\right]\Big|\bSigmaLone,\bSigmaLtwo
\simind N\left(\left[\begin{array}{c}
\bzero\\[1ex]
\bzero
\end{array}\right],
\left[
\begin{array}{cc}
\bSigmaLone & \bO \\
\bO         &\bSigmaLtwo 
\end{array}
\right]
\right),
\quad 1\le i\le m,\ 1\le j\le n_i,\\[3ex]
\bbeta\sim N(\bmu_{\bbeta},\bSigma_{\bbeta}),\quad\mysigeps^2|\asigsq\sim\mbox{Inverse-$\chi^2$}(\nusigsq,1/\asigsq),\\[2ex]
\asigsq\sim\mbox{Inverse-$\chi^2$}(1,1/(\nusigsq\ssigsq^2)),\\[2ex]
\bSigmaLone|\bASigmaLone\sim\mbox{Inverse-G-Wishart}\big(\Gfull,\nuSigmaLone+2\qLone-2,(\bASigmaLone)^{-1}\big),\\[2ex]
\bASigmaLone\sim\mbox{Inverse-G-Wishart}(\Gdiag,1,\{\nuSigmaLone\diag(\sSigmaOneLone^2,\ldots,\sSigmaqLone^2)\}^{-1}),\\[2ex]
\bSigmaLtwo|\bASigmaLtwo\sim\mbox{Inverse-G-Wishart}\big(\Gfull,\nuSigmaLtwo+2\qLtwo-2,(\bASigmaLtwo)^{-1}\big),\\[2ex]
\bASigmaLtwo\sim\mbox{Inverse-G-Wishart}(\Gdiag,1,\{\nuSigmaLtwo\diag(\sSigmaOneLtwo^2,\ldots,\sSigmaqLtwo^2)\}^{-1}).
\end{array}
\label{eq:threeLevelGaussRespBaye}
\end{equation}
where hyperparameters such as $\nuSigmaLone>0$ and $\sSigmaOneLone,\ldots,\sSigmaqLone>0$ are defined analogously to the
two-level case.

The minimal mean field restriction needed for a tractable variational inference algorithm is
\begin{equation}
\pDens(\bbeta,\bu,\asigsq,\bASigmaLone,\bASigmaLtwo,\mysigeps^2,\bSigmaLone,\bSigmaLtwo|\by)
\approx\qDens(\bbeta,\bu,\asigsq,\bASigmaLone,\bASigmaLtwo)\,\qDens(\mysigeps^2,\bSigmaLone,\bSigmaLtwo).
\label{eq:threeLevProducRestrict}
\end{equation}
The optimal $\qDens$-densities have forms analogous to those given in (\ref{eq:twoLevMFVBforms}) but
with 
$$\qDens^*(\bSigmaLone)\ \mbox{an}\ \mbox{Inverse-G-Wishart}\Big(\Gfull,\xi_{\qDens(\bSigmaLone)},
\bLambda_{\qDens(\bSigmaLone)}\Big)$$
density function. A similar result holds for $\qDens^*(\bSigmaLtwo)$.

As in the two-level case, only the following relatively small
sub-blocks of $\bSigma_{\qDens(\bbeta,\bu)}$ are required for 
variational inference concerning $\sigma^2$, $\bSigmaLone$ and $\bSigmaLtwo$:
\begin{equation}
\begin{array}{c}
\bSigma_{\qDens(\bbeta)},\quad\bSigma_{\qDens(\buLone_i)},
\quad\bSigma_{\qDens(\buLtwo_{ij})},\quad
E_{\qDens}\{(\bbeta-\bmu_{\qDens(\bbeta)})(\buLonei-\bmuq{\buLonei})^T\}
\\[2ex]
E_{\qDens}\{(\bbeta-\bmu_{\qDens(\bbeta)})(\buLtwoij-\bmuq{\buLtwoij})^T\}
\quad\mbox{and}\quad 
E_{\qDens}\{(\buLonei-\bmuq{\buLonei})(\buLtwoij-\bmuq{\buLtwoij})^T\}
\end{array}
\label{eq:subBlocksThreeLev}
\end{equation}
for $1\le i\le m$ and $1\le j\le n_i$. 
Result \ref{res:threeLevelMFVB} is the three-level analog of 
Result \ref{res:twoLevelMFVB} in that it provides a link between
the three-level sparse matrix least squares problems and updates for $\bmu_{\qDens(\bbeta,\bu)}$
and the important sub-blocks of $\bSigma_{\qDens(\bbeta,\bu)}$.

\jump\noindent
\begin{result}
The mean field variational Bayes updates of $\bmu_{\qDens(\bbeta,\bu)}$ and each
of the sub-blocks of $\bSigma_{\qDens(\bbeta,\bu)}$ corresponding to (\ref{eq:subBlocksThreeLev}) 
are expressible as a three-level sparse matrix least squares problem (see Appendix \ref{sec:threeLevSMA}) 
of the form:
$$\left\Vert\bb-\bB\bmu_{\qDens(\bbeta,\bu)}
\right\Vert^2
$$
where $\bb$ and the non-zero sub-blocks of $\bB$, according to the notation
given by (\ref{eq:BandbGenThreeLev}), are for $1\le j\le n_i,\ 1\le i\le m$:
$$\bb_{ij}\equiv
\left[\begin{array}{c}   
\mu_{\qDens(1/\mysigeps^2)}^{1/2}\by_{ij}\\[2ex]
\Big(\displaystyle{\sum_{i=1}^{m}}\,n_i\Big)^{-1/2}\bSigma_{\bbeta}^{-1/2}\bmu_{\bbeta}\\[2ex]
\bzero\\[1ex]
\bzero
\end{array}
\right],
\quad\bB_{ij}\equiv
\left[\begin{array}{c}   
\mu_{\qDens(1/\mysigeps^2)}^{1/2}\bX_{ij}\\[2ex]
\Big(\displaystyle{\sum_{i=1}^{m}}\,n_i\Big)^{-1/2}\bSigma_{\bbeta}^{-1/2}\\[2ex]
\bO\\[2ex]
\bO
\end{array}
\right],
$$
$$
\bBdot_{ij}\equiv
\left[\begin{array}{c}   
\mu_{\qDens(1/\mysigeps^2)}^{1/2}\bZLone_{ij}\\[2ex]
\bO\\[2ex]
n_i^{-1/2}\Big(\bM_{\qDens(\bSigmaLoneMinusOne)}\Big)^{1/2}\\[2ex]
\bO
\end{array}
\right]
\quad\mbox{and}\quad
\bBdotdot_{ij}\equiv
\left[\begin{array}{c}   
\mu_{\qDens(1/\mysigeps^2)}^{1/2}\bZLtwo_{ij}\\[2ex]
\bO\\[2ex]
\bO\\[2ex]
\Big(\bM_{\qDens(\bSigmaLtwoMinusOne)}\Big)^{1/2}
\end{array}
\right]
$$
with each of these matrices having $\oadj_{ij}=o_{ij}+p+q_1+q_2$ rows.
The solutions are, according to notation illustrated by
(\ref{eq:3levSubBlockNotat1})--(\ref{eq:3levSubBlockNotat3}),
$$\bmu_{\qDens(\bbeta)}=\xveco,\ \bSigma_{\qDens(\bbeta)}=\AUoo,$$
$$\bmu_{\qDens(\buLonei)}=\xvectCi,\ \bSigma_{\qDens(\buLonei)}=\AUttCi,\ 
E_{\qDens}\{(\bbeta-\bmu_{\qDens(\bbeta)})(\buLonei-\bmuq{\buLonei})^T\}=
\AUotCi\ \textit{for}\ 1\le i\le m$$
and
$$\bmu_{\qDens(\buLtwoij)}=\bx_{2, ij},\ 
\bSigma_{\qDens(\buLtwoij)}=\bA^{22, ij},\ 
E_{\qDens}\{(\bbeta-\bmu_{\qDens(\bbeta)})(\buLtwoij-\bmuq{\buLtwoij})^T\}
=\bA^{12, ij},$$
$$E_{\qDens}\{(\buLonei-\bmuq{\buLonei})(\buLtwoij-\bmuq{\buLtwoij})^T\}
=\bA^{12,\iCOMMAj}\ \textit{for}\ 1\le i\le m,\ 1\le j\le n_i.
$$
\label{res:threeLevelMFVB}
\end{result}
\null\vskip2mm

Algorithm \ref{alg:threeLevelMFVB} provides a streamlined mean field
variational Bayes algorithm for approximate fitting and inference
for (\ref{eq:threeLevelGaussRespBaye}). An explicit streamlined expression for the
stopping criterion, $\log\{\pDensUnder(\by;\qDens)\}$, is given in 
Nolan (2020). We are not aware of any previously published variational 
inference algorithms that achieve streamlined inference for mixed models 
with three-level random effects.

\begin{algorithm}[!th]
  \begin{center}
    \begin{minipage}[t]{154mm}
      \begin{small}
        \begin{itemize}
          \setlength\itemsep{4pt}
          \item[] Data Inputs: $\by_{ij}(o_{ij}\times1),\ \bX_{ij}(o_{ij}\times p),\
                  \bZLone_{ij} (o_{ij}\times \qLone), \bZLtwo_{ij} (o_{ij}\times \qLtwo),
                  \ 1\le i\le m, \ 1 \le j \le n_{i}$.
          \item[] Hyperparameter Inputs: $\bmu_{\bbeta}(p\times1)$, 
                  $\bSigma_{\bbeta}(p\times p)\ \mbox{symmetric and positive definite}$,\\
      \null$\qquad\qquad\qquad\qquad\qquad\quad\ssigsq,\nusigsq,\sSigmaOneLone,\ldots,\sSigmaqLone,\nuSigmaLone,
           \sSigmaOneLtwo,\ldots,\sSigmaqLtwo,\nuSigmaLtwo>0$
          \item[] Initialize: $\muq{1/\sigsqeps} > 0$, $\muq{1/\asigsq} > 0$,
                              $\bM_{\qDens((\bSigmaLone)^{-1})}(q_{1}\times q_{1}), \
                               \bM_{\qDens((\bSigmaLtwo)^{-1})}(q_{2}\times q_{2}),$
          \item[] $\bM_{\qDens(\bASigmaLone^{-1})}(q_{1}\times q_{1}), \
                   \bM_{\qDens(\bASigmaLtwo^{-1})}(q_{2}\times q_{2})$ symmetric and positive definite,
          \item[] $\xi_{\qDens(\mysigeps^2)}\thickarrow\nu_{\sigma^2}+\displaystyle{\sumim\sum_{j=1}^{n_i}}o_{ij}$\ \ \ ;\ \ \
                  $\xi_{\qDens(\bSigmaLone)}\thickarrow \nuSigmaLone + 2 q_{1} -2 + m$\ \ \ ;\ \ \
                  $\xi_{\qDens(\bSigmaLtwo)}\thickarrow \nuSigmaLtwo + 2 q_{2} -2 + \displaystyle{\sumim} n_i$
          \item[] $\xi_{\qDens(a_{\sigma^2})}\thickarrow\nusigsq+1$\ \ \ ;\ \ \
                  $\xi_{\qDens(\bASigmaLone)}\thickarrow\nuSigmaLone+q_{1}$\ \ \ ;\ \ \
                  $\xi_{\qDens(\bASigmaLtwo)}\thickarrow\nuSigmaLtwo+q_{2}$
          \item[] Cycle:
          \begin{itemize}
            \item[] For $i = 1,\ldots, m$:
            \begin{itemize}
              \item[] For $j = 1,\ldots, n_{i}$:
              \begin{itemize}
                \item[] $
                          \bb_{ij}\thickarrow
                          \left[
                            \begin{array}{c}
                              \mu_{\qDens(1/\sigsqeps)}^{1/2}\by_{ij}\\[1ex]
                             \Big(\displaystyle{\sum_{i=1}^{m}}\,n_i\Big)^{-1/2}\bSigma_{\bbeta}^{-1/2}\bmu_{\bbeta}\\[1ex]
                              \bzero \\[1ex]
                              \bzero \\[1ex]
                            \end{array}
                          \right]\ \ ; \ \ 
                          \bB_{ij}\thickarrow
                          \left[
                            \begin{array}{c}
                              \mu_{\qDens(1/\sigsqeps)}^{1/2}\bX_{ij} \\[1ex]
                            \Big(\displaystyle{\sum_{i=1}^{m}}\,n_i\Big)^{-1/2}\bSigma_{\bbeta}^{-1/2}\\[1ex]
                              \bO \\[1ex]
                              \bO \\[1ex]
                            \end{array}
                          \right],
                          $
                  \item[]$
                          \bBdot_{ij}\thickarrow
                          \left[
                            \begin{array}{c}
                              \mu_{\qDens(1/\sigsqeps)}^{1/2}\bZLone_{ij}\\[1ex]
                              \bO \\[1ex]
                              n_i^{-1/2}\bM_{\qDens((\bSigmaLone)^{-1})}^{1/2} \\[1ex]
                              \bO \\
                            \end{array}
                          \right]
                          \ \ ; \ \ 
                          \bBdotdot_{ij}\thickarrow
                          \left[
                            \begin{array}{cc}
                              \mu_{\qDens(1/\sigsqeps)}^{1/2}\bZLtwo_{ij} \\[1ex]
                              \bO \\[1ex]
                              \bO \\[1ex]
                              \bM_{\qDens((\bSigmaLtwo)^{-1})}^{1/2} \\[1ex]
                            \end{array}
                          \right]
                          $
              \end{itemize}
            \end{itemize}
            \item[] $\SscAlgFive\thickarrow\SolveThreeLevelSparseLeastSquares\\
                    \null\qquad\quad\Big(\big\{(\bb_{ij},\bB_{ij},\bBdot_{ij},\bBdotdot_{ij}):
                    1\le i\le m, 1\le j\le n_i\big\}\Big)$
            \item[] $\bmu_{\qDens(\bbeta)}\thickarrow\mbox{$\xveco$ component of $\SscAlgFive$}$
            \ \ \ ;\ \ \ $\bSigma_{\qDens(\bbeta)}\thickarrow\mbox{$\AUoo$ component of $\SscAlgFive$}$
            \item[] $\lambda_{\qDens(\sigsqeps)}\thickarrow\muq{1/\asigsq}$\ \ ;\ \
            $\bLambda_{\qDens(\bSigmaLone)}\thickarrow\bM_{\qDens(\bASigmaLone^{-1})}$\ \ ;\ \
            $\bLambda_{\qDens(\bSigmaLtwo)}\thickarrow\bM_{\qDens(\bASigmaLtwo^{-1})}$
            \item[] For $i = 1,\ldots, m$:
            \begin{itemize}
              \item[] $\bmu_{\qDens(\buLonei)}\thickarrow\mbox{$\xvectCi$ component of $\SscAlgFive$}$\ \ \ ;\ \ \
              $\bSigma_{\qDens(\buLonei)}\thickarrow\mbox{$\AUttCi$ component of $\SscAlgFive$}$
              \item[] $E_{\qDens}\{(\bbeta-\bmu_{\qDens(\bbeta)})(\buLonei-\bmuq{\buLonei})^T\}\thickarrow
                               \mbox{$\AUotCi$ component of $\SscAlgFive$}$
              \item[] $\bLambda_{\qDens(\bSigmaLone)}\thickarrow\bLambda_{\qDens(\bSigmaLone)} +
                       \bmu_{\qDens(\buLonei)}\bmu_{\qDens(\buLonei)}\trans + \bSigma_{\qDens(\buLonei)}$
              \item[] For $j = 1,\ldots, n_{i}$:
                \begin{itemize}
                    \item[] $\bmu_{\qDens(\buLtwoij)}\thickarrow\mbox{$\bx_{2, ij}$ component of $\SscAlgFive$}$\ \ \ ;\ \ \
                            $\bSigma_{\qDens(\buLtwoij)}\thickarrow\mbox{$\bA^{22, ij}$ component of $\SscAlgFive$}$
                    \item[] $E_{\qDens}\{(\bbeta-\bmu_{\qDens(\bbeta)})(\buLtwoij-\bmuq{\buLtwoij})^T\}\thickarrow
                             \mbox{$\bA^{12, ij}$ component of $\SscAlgFive$}$
                    \item[] $E_{\qDens}\{(\buLonei-\bmuq{\buLonei})(\buLtwoij-\bmuq{\buLtwoij})^T\}\thickarrow
                             \mbox{$\bA^{12, i, \, j}$ component of $\SscAlgFive$}$
                    \item[] $\lambda_{\qDens(\sigsqeps)} \thickarrow \lambda_{\qDens(\sigsqeps)}
                             + \big\Vert\by_{ij}-\bX_{ij}\bmu_{\qDens(\bbeta)}-\bZLone_{ij}
                             \bmu_{\qDens(\buLonei)} - \bZLtwo_{ij} \bmu_{\qDens(\buLtwoij)} \big\Vert^2$
                    \item[] $\lambda_{\qDens(\sigsqeps)}\thickarrow \lambda_{\qDens(\sigsqeps)}
                            +\mbox{tr}(\bX_{ij}^T\bX_{ij}\bSigma_{\qDens(\bbeta)})
                            +\mbox{tr}\{(\bZLone_{ij})^T\bZLone_{ij}\bSigma_{\qDens(\buLonei)}\}$
                    \item[] \textsl{continued on a subsequent page}\ $\ldots$
                \end{itemize}
              \end{itemize}
            \end{itemize}
          \end{itemize}
        \end{small}
      \end{minipage}
    \end{center}
  \caption{\it QR-decomposition-based streamlined algorithm for obtaining mean field variational
  Bayes approximate posterior density functions for the parameters in the three-level linear
  mixed model (\ref{eq:threeLevelGaussRespBaye}) with product density restriction
  (\ref{eq:threeLevProducRestrict}). The algorithm description requires more than one page
and is continued on a subsequent page.}
  \label{alg:threeLevelMFVB}
\end{algorithm}

\setcounter{algorithm}{4}
\begin{algorithm}[!th]
  \begin{center}
    \begin{minipage}[t]{154mm}
      \begin{small}
        \begin{itemize}
          \setlength\itemsep{4pt}
          \item[]
          \begin{itemize}
            \item[]
            \begin{itemize}
              \item[]
                \begin{itemize}
                    \item[] $ \lambda_{\qDens(\sigsqeps)}\thickarrow \lambda_{\qDens(\sigsqeps)}
                              +\mbox{tr}((\bZLtwo_{ij})^T\bZLtwo_{ij}\bSigma_{\qDens(\buLtwoij)})$
                    \item[] $\lambda_{\qDens(\sigsqeps)}\thickarrow \lambda_{\qDens(\sigsqeps)}
                             +2\,\mbox{tr}\big[(\bZLone_{ij})^T\bX_{ij} E_{\qDens}\{(\bbeta-\bmu_{\qDens(\bbeta)})
                              (\buLonei-\bmuq{\buLonei})^T\}\big]$
                    \item[] $\lambda_{\qDens(\sigsqeps)}\thickarrow \lambda_{\qDens(\sigsqeps)}
                             +2\,\mbox{tr}\big[(\bZLtwo_{ij})^T\bX_{ij} E_{\qDens}\{(\bbeta-\bmu_{\qDens(\bbeta)})
                              (\buLtwoij-\bmuq{\buLtwoij})^T\}\big]$
                    \item[] $\lambda_{\qDens(\sigsqeps)}\thickarrow \lambda_{\qDens(\sigsqeps)}
                             +2\,\mbox{tr}\big[(\bZLone_{ij})^T\bZLtwo_{ij} E_{\qDens}\{(\buLonei-\bmuq{\buLonei})
                              (\buLtwoij-\bmuq{\buLtwoij})^T\}\big]$
                  \item[] $\bLambda_{\qDens(\bSigmaLtwo)}\thickarrow\bLambda_{\qDens(\bSigmaLtwo)} +
                       \bmu_{\qDens(\buLtwoij)}\bmu_{\qDens(\buLtwoij)}\trans + \bSigma_{\qDens(\buLtwoij)}$

                \end{itemize}
            \end{itemize}
            \item[] $\muq{1/\sigsqeps} \thickarrow \xi_{\qDens(\mysigeps^2)}/\lambda_{\qDens(\sigsqeps)}$
            \item[] $\MqSigmaLone \thickarrow(\xi_{\qDens(\bSigmaLone)}-q_{1}+1)\, \bLambda^{-1}_{\qDens(\bSigmaLone)}$
            \ \ \ ;\ \ \
            $\MqSigmaLtwo \thickarrow(\xi_{\qDens(\bSigmaLtwo)}-q_{2}+1)\, \bLambda^{-1}_{\qDens(\bSigmaLtwo)}$
            \item[] $\lambda_{\qDens(\asigsq)}\thickarrow\muq{1/\sigsqeps}+1/(\nusigsq\ssigsq^2)$\ \ \ ;\ \ \
            $\muq{1/\asigsq} \thickarrow \xi_{\qDens(\asigsq)}/\lambda_{\qDens(\asigsq)}$
            \item[] $\LambdaqASigmaLone \thickarrow \diag\big\{\mbox{diagonal}\big(
                      \bM_{\qDens((\bSigmaLone)^{-1})}\big)\big\} + \{\nuSigmaLone\,\diag(
                      \sSigmaOneLone^2,\ldots,\sSigmaqLone^2)\}^{-1}$
            \item[] $ \LambdaqASigmaLtwo \thickarrow \diag\big\{\mbox{diagonal}\big(
                      \bM_{\qDens((\bSigmaLtwo)^{-1})}\big)\big\} + \{\nuSigmaLtwo\,\diag(
                        \sSigmaOneLtwo^2,\ldots,\sSigmaqLtwo^2)\}^{-1}$
            \item[] $\bM_{\qDens(\bASigmaLone^{-1})}\thickarrow \xi_{\qDens(\bASigmaLone)}\bLambda_{\qDens(\bASigmaLone)}^{-1}$
                    \ \ \ ;\ \ \
                    $\bM_{\qDens(\bASigmaLtwo^{-1})}\thickarrow \xi_{\qDens(\bASigmaLtwo)}\bLambda_{\qDens(\bASigmaLtwo)}^{-1}$.
          \end{itemize}
          \item[] until the increase in $\log\{\pDensUnder(\by;\qDens)\}$ is negligible.
          \item[] Outputs: $\bmu_{\qDens(\bbeta)}$,\ $\bSigma_{\qDens(\bbeta)}$,\
                  $\big\{\big(\bmu_{\qDens(\buLonei)},\bSigma_{\qDens(\buLonei)},
                  E_{\qDens}\{(\bbeta-\bmu_{\qDens(\bbeta)})(\buLonei-\bmuq{\buLonei})^T\}\big)
                  :1\le i\le m\}$,
          \item[] $ \qquad \qquad\big\{\big(\bmu_{\qDens(\buLtwoij)},\bSigma_{\qDens(\buLtwoij)},
                    E_{\qDens}\{(\bbeta-\bmu_{\qDens(\bbeta)})(\buLtwoij-\bmuq{\buLtwoij})^T\},$
          \item[] $ \qquad \qquad\quad E_{\qDens}\{(\buLonei-\bmuq{\buLonei})(\buLtwoij-\bmuq{\buLtwoij})^T\}\big)
                    :1 \le i \le m, 1 \le j \le n_{i} \big\},$
          \item[]$\qquad\qquad$ $\xi_{\qDens(\mysigeps^2)},\lambda_{\qDens(\mysigeps^2)},
                  \xi_{\qDens(\bSigmaLone)},\bLambda_{\qDens(\bSigmaLone)},
                  \xi_{\qDens(\bSigmaLtwo)},\bLambda_{\qDens(\bSigmaLtwo)}$
        \end{itemize}
      \end{small}
    \end{minipage}
  \end{center}
  \caption{\textbf{continued.} \textit{This is a continuation of the description of this algorithm that commences
on a preceding page.}}
\end{algorithm}

\subsection{Variational Message Passing}

\begin{figure}[!ht]
\centering
{\includegraphics[width=0.85\textwidth]{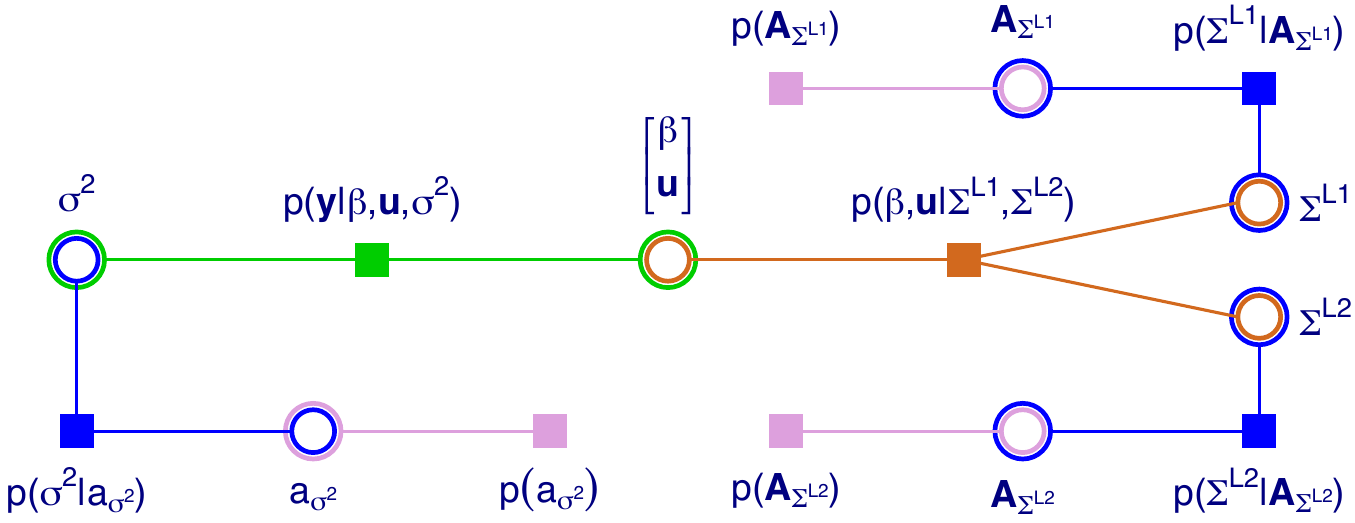}}
\caption{\it Factor graph representation of the Bayesian three-level 
Gaussian response linear mixed model (\ref{eq:threeLevelGaussRespBaye}).}
\label{fig:threeLevFacGraph} 
\end{figure}

For studying the variational message passing alternative we first
note that the joint density function of all of the 
random variables and random vectors in the Bayesian three-level Gaussian 
response linear mixed model (\ref{eq:threeLevelGaussRespBaye})
can be factorized as follows:
$$
\begin{array}{l}
\pDens(\by,\bbeta,\bu,\mysigeps^2,\bSigmaLone,\bSigmaLtwo,\asigsq,\bASigmaLone,\bASigmaLtwo)
=\pDens(\by|\bbeta,\bu,\mysigeps^2)\pDens(\mysigeps^2|\asigsq)\pDens(\asigsq)\\[0.5ex]
\null\qquad\qquad\qquad
\times\pDens(\bbeta,\bu|\bSigmaLone,\bSigmaLtwo)\pDens(\bSigmaLone|\bASigmaLone)\pDens(\bASigmaLone)
\pDens(\bSigmaLtwo|\bASigmaLtwo)\pDens(\bASigmaLtwo).
\end{array}
$$
Figure \ref{fig:threeLevFacGraph} provides the relevant factor graph 
with color-coding of fragment types.

As with the two-level case, each of these fragments in Figure \ref{fig:threeLevFacGraph} 
appear in Section 4.1 of Wand (2017). To achieve streamlined variational message passing 
for three-level random effects models we require tailored versions of the Gaussian likelihood 
fragment updates and Gaussian penalization fragment updates. These are provided in
the next two subsections as Algorithms \ref{alg:threeLevelVMPlik} and \ref{alg:threeLevelVMPpen}.
However, they each rely on the \ThreeLevelNaturalToCommonParameters\ algorithm,
which is listed as Algorithm \ref{alg:ThreeLevelNaturalToCommonParameters}.

\begin{algorithm}[!th]
  \begin{center}
    \begin{minipage}[t]{154mm}
      \begin{small}
        \begin{itemize}
          \setlength\itemsep{0pt}
          \item[] Inputs: $p, q_{1}, q_{2}, m, \{n_i: 1\le i\le m\}, \etaSUBqbetau$
          \item[] $\omegaAlgSixA\thickarrow\mbox{first $p$ entries of $\etaSUBqbetau$}$
          \item[] $\omegaAlgSixB\thickarrow\mbox{next $\smhalf\,p(p+1)$ entries of $\etaSUBqbetau$}$
                   \ ;\ $\OmegaAlgSixC\thickarrow -2\vecof^{-1}(\bD_p^{+T}\omegaAlgSixB)$
          \item[] $\iStt\thickarrow p + \smhalf\,p(p+1)+1$\ \ \ ;\ \ \ $\iEnd\thickarrow\iStt + q_{1}-1$
          \item[] For $i=1,\ldots,m$:
          \begin{itemize}
            \setlength\itemsep{0pt}
            \item[] $\omegaAlgSixD\thickarrow\mbox{sub-vector of}\ \etaSUBqbetau$\ \mbox{with entries $\iStt$ to $\iEnd$ inclusive}
            \item[] $\iStt\thickarrow\iEnd+1$\ \ \ ;\ \ \ $\iEnd\thickarrow\iStt + \smhalf\,q_{1}(q_{1}+1)-1$
            \item[] $\omegaAlgSixE\thickarrow\mbox{sub-vector of}\ \etaSUBqbetau$\ \mbox{with entries $\iStt$ to $\iEnd$ inclusive}
            \item[] $\iStt\thickarrow\iEnd+1$\ \ \ ;\ \ \ $\iEnd\thickarrow\iStt + pq_{1}-1$
            \item[] $\omegaAlgSixF\thickarrow\mbox{sub-vector of}\ \etaSUBqbetau$\ \mbox{with entries $\iStt$ to $\iEnd$ inclusive}
            \item[] $\iStt\thickarrow\iEnd+1$\ \ \ ;\ \ \ $\iEnd\thickarrow\iStt + q_{1}-1$
            \item[] $\OmegaAlgSixG\thickarrow\,-2\,\vecof^{-1}(\bD_{q_1}^{+T}\omegaAlgSixE)$\ \ \ ;\ \ \
                    $\OmegaAlgSixH\thickarrow\,-\vecof^{-1}_{p\times q_{1}}(\omegaAlgSixF)$
          \end{itemize}
          \item[] $\iEnd\thickarrow\iEnd - q_1 + q_{2}$
          \item[] \mbox{For $i=1, \hdots, m$:}
          \begin{itemize}
            \setlength\itemsep{0pt}
            \item[] \mbox{For $j=1, \hdots, n_{i}$:}
            \begin{itemize}
              \item[] $\omegaAlgSixI\thickarrow\mbox{sub-vector of}\ \etaSUBqbetau$\ \mbox{with entries $\iStt$ to $\iEnd$ inclusive}
              \item[] $\iStt\thickarrow\iEnd+1$\ \ \ ;\ \ \ $\iEnd\thickarrow\iStt + \smhalf\,q_{2}(q_{2}+1)-1$
              \item[] $\omegaAlgSixJ\thickarrow\mbox{sub-vector of}\ \etaSUBqbetau$\ \mbox{with entries $\iStt$ to $\iEnd$ inclusive}
              \item[] $\iStt\thickarrow\iEnd+1$\ \ \ ;\ \ \ $\iEnd\thickarrow\iStt + pq_{2}-1$
              \item[] $\omegaAlgSixK\thickarrow\mbox{sub-vector of}\ \etaSUBqbetau$\ \mbox{with entries $\iStt$ to $\iEnd$ inclusive}
              \item[] $\iStt\thickarrow\iEnd+1$\ \ \ ;\ \ \ $\iEnd\thickarrow\iStt + q_{1} q_{2}-1$
              \item[] $\omegaAlgSixL\thickarrow\mbox{sub-vector of}\ \etaSUBqbetau$\ \mbox{with entries $\iStt$ to $\iEnd$ inclusive}
              \item[] $\iStt\thickarrow\iEnd+1$\ \ \ ;\ \ \ $\iEnd\thickarrow\iStt + q_{2}-1$
              \item[] $\OmegaAlgSixM\thickarrow\,-2\,\vecof^{-1}(\bD_{q_{2}}^{+T}\omegaAlgSixJ)$\ \ \ ;\ \ \
                      $\OmegaAlgSixN\thickarrow\,-\vecof^{-1}_{p\times q_{2}}(\omegaAlgSixK)$
              \item[] $\OmegaAlgSixO\thickarrow\,-\vecof^{-1}_{q_{1}\times q_{2}}(\omegaAlgSixL)$
            \end{itemize}
          \end{itemize}
          \item[] $\SscAlgSix\thickarrow \SolveThreeLevelSparseMatrix\Big(\omegaAlgSixA,\OmegaAlgSixC,
                   \big\{(\omegaAlgSixD,\OmegaAlgSixG,\OmegaAlgSixH):1\le i\le m, $
          \item[] \qquad\qquad\qquad\qquad\qquad\qquad\qquad\qquad
                  $(\omegaAlgSixI,\OmegaAlgSixM,\OmegaAlgSixN,\OmegaAlgSixO):1\le i\le m,1\le j\le n_{i}\big\}\Big)$
          \item[] $\bmu_{\qDens(\bbeta)}\thickarrow\mbox{$\xveco$ component of $\SscAlgSix$}$\ \ ;\ \
                  $\bSigma_{\qDens(\bbeta)}\thickarrow\mbox{$\AUoo$ component of $\SscAlgSix$}$
          \item[] For $i=1,\ldots,m$:
          \begin{itemize}
            \setlength\itemsep{0pt}
            \item[] $\bmu_{\qDens(\buLone_i)}\thickarrow\mbox{$\xvectCi$ component of $\SscAlgSix$}$
                     \ \ ;\ \ $\bSigma_{\qDens(\buLone_i)}\thickarrow\mbox{$\AUttCi$ component of $\SscAlgSix$}$,
            \item[] $E_{\qDens}\{(\bbeta-\bmu_{\qDens(\bbeta)})\}(\buLone_i-\bmu_{\qDens(\buLone_i)})^T\}
                    \thickarrow\mbox{$\AUotCi$ component of $\SscAlgSix$}$
            \item[] \textsl{continued on a subsequent page}\ $\ldots$
          \end{itemize}
        \end{itemize}
      \end{small}
    \end{minipage}
  \end{center}
  \caption{
\textit{The} \ThreeLevelNaturalToCommonParameters\ \textit{algorithm. The algorithm description requires more than one page
and is continued on a subsequent page.}}
  \label{alg:ThreeLevelNaturalToCommonParameters}
\end{algorithm}

\setcounter{algorithm}{5}
\begin{algorithm}[!th]
  \begin{center}
    \begin{minipage}[t]{154mm}
      \begin{small}
        \begin{itemize}
          \item[]
            \begin{itemize}
              \item[] For $j=1,\ldots,n_{i}$:
              \begin{itemize}
                \setlength\itemsep{0pt}
                \item[] $\bmu_{\qDens(\bu_{ij}^{L2})}\thickarrow\mbox{$\xvectCij$ component of $\SscAlgSix$}$
                         \ \ ;\ \ $\bSigma_{\qDens(\bu_{ij}^{L2})}\thickarrow\mbox{$\AUttCij$ component of $\SscAlgSix$}$
                \item[] $E_{\qDens}\{(\bbeta-\bmu_{\qDens(\bbeta)})\}(\bu_{ij}^{L2}-\bmu_{\qDens(\bu_{ij}^{L2})})^T\}
                        \thickarrow\mbox{$\AUotCij$ component of $\SscAlgSix$}$
                \item[] $E_{\qDens}\{(\bu_i^{L1}-\bmu_{\qDens(\bu_i^{L1})})\}(\bu_{ij}^{L2}-\bmu_{\qDens(\bu_{ij}^{L2})})^T\}
                        \thickarrow\mbox{$\AUotCicommaj$ component of $\SscAlgSix$}$
              \end{itemize}
            \end{itemize}
          \item[] Outputs: $\bmu_{\qDens(\bbeta)},\bSigma_{\qDens(\bbeta)},
                           \big\{\big(\bmu_{\qDens(\bu_i^{L1})},\bSigma_{\qDens(\bu_i^{L1})},
                          E_{\qDens}\{(\bbeta-\bmu_{\qDens(\bbeta)}\}(\bu_i^{L1}-
                          \bmu_{\qDens(\bu_i^{L1})})^T\}):\ 1\le i\le m\big\}$,
          \item[] \qquad\qquad $\big\{\big(\bmu_{\qDens(\bu_{ij}^{L2})},\bSigma_{\qDens(\bu_{ij}^{L2})},
                  E_{\qDens}\{(\bbeta-\bmu_{\qDens(\bbeta)})\}(\bu_{ij}^{L2}-\bmu_{\qDens(\bu_{ij}^{L2})})^T\},$
          \item[] \qquad\qquad\quad $E_{\qDens}\{(\bu_i^{L1}-\bmu_{\qDens(\bu_i^{L1})})\}(\bu_{ij}^{L2}-
                  \bmu_{\qDens(\bu_{ij}^{L2})})^T\}\big):1\le i\le m,\ 1\le j\le n_i\big\}$
        \end{itemize}
      \end{small}
    \end{minipage}
  \end{center}
\caption{\textbf{continued.} \textit{This is a continuation of the description of this algorithm that commences
on a preceding page.}}
\end{algorithm}

\subsection{Streamlined Gaussian Likelihood Fragment Updates}

Streamlined updating for the Gaussian likelihood fragment
with three-level random effects structure is analogous
to the two-level case discussed in Section \ref{sec:streamGaussLikFrag}.
The relevant factor is shown in green in Figure \ref{fig:threeLevFacGraph}.
The message from the likelihood factor to the vector of fixed and random 
effects instead has the form
\null\vfill\eject
\null\vfill\eject

\begin{equation}
\begin{array}{l}
\mSUBpybetausigsqEpsTObetau=\\[1ex]
\null\qquad\qquad
\exp\left\{
\left[
{\setlength\arraycolsep{0pt}
\begin{array}{c}
\ \ \ \ \ \ \ \ \ \  \bbeta\\
\ \ \ \ \ \vech(\bbeta\bbeta^T)\\[1ex]
\displaystyle{\stack{1\le i\le m}}\left[
{\setlength\arraycolsep{0pt}
\begin{array}{c}
\buLone_i\\
\vech\big(\buLone_i(\buLone_i)^T\big)\\
\vecof\big(\bbeta(\buLone_i)^T\big)
\end{array}
}
\right]\\[5ex]
\displaystyle{\stack{1\le i\le m}}\left[
\displaystyle{\stack{1\le j\le n_i}}\left[
{\setlength\arraycolsep{0pt}
\begin{array}{c}
\buLtwo_{ij}\\[1ex]
\vech\big(\buLtwo_{ij}(\buLtwo_{ij})^T\big)\\[1ex]
\vecof\big(\bbeta(\buLtwo_{ij})^T\big)\\[1ex]
\vecof\big(\buLone_{ij}(\buLtwo_{ij})^T\big)
\end{array}
}
\right]
\right]
\end{array}
}
\right]^T\null\hskip-3mm\etaSUBpybetausigsqEpsTObetau
\right\}
\end{array}
\label{eq:threeLevLikToCoeffMsg}
\end{equation}
and we assume that $\mSUBbetauTOpybetausigsqEps$ is in the same 
exponential family. Result \ref{res:threeLevelVMPlik} points the
way to streamlining the fragment updates in the three-level case.
Its derivation is given in Section \ref{sec:resThreeLevelVMPlikDeriv}.

\jump
\noindent
\begin{result}
The variational message passing updates of 
the quantities $\bmu_{\qDens(\bbeta)}$, $\bmu_{\qDens(\buLone_i)}$, $1\le i\le m$,
$\bmu_{\qDens(\buLtwo_{ij})}$, $1\le i\le m$, $1\le j\le n_i$,
and the sub-blocks of $\bSigma_{\qDens(\bbeta,\bu)}$ 
corresponding to (\ref{eq:subBlocksThreeLev}) with $\qDens$-density 
expectations with respect to the normalization of
$$\mSUBpybetausigsqEpsTObetau\,\mSUBbetauTOpybetausigsqEps$$
are expressible as a three-level sparse matrix problem (see Appendix \ref{sec:threeLevSMA})
with
{\setlength\arraycolsep{1pt}
\begin{eqnarray*}
&&\AtLev=\\[1ex]
&&-2\,\left[
\begin{array}{cc}
\vecof^{-1}(\bD_p^{+T}\bdeta_{1,2}) 
& \Big(\smhalf\,\displaystyle{\stack{1\le i\le m}}\Big[\vecof_{p\times q_1}^{-1}(\bdeta_{2,3,i})^T
\displaystyle{\stack{1\le j\le n_i}}\{\vecof_{p\times q_2}^{-1}(\bdeta_{3,3,ij})^T\}\Big]\Big)^T   \\[1ex]
\begin{array}{l}
\smhalf\,\displaystyle{\stack{1\le i\le m}}\Big[\vecof_{p\times q_1}^{-1}(\bdeta_{2,3,i})^T \\
\displaystyle{\stack{1\le j\le n_i}}\{\vecof_{p\times q_2}^{-1}(\bdeta_{3,3,ij})^T\}\Big]
\end{array}
& {\displaystyle\blockdiag{1\le i\le m}}
\left[
       \begin{array}{cc}
       \vecof^{-1}(\bD_{q_1}^{+T}\bdeta_{2,2,i})  
& \Big[\smhalf\,\displaystyle{\stack{1\le j\le n_i}\{\vecof_{q_1\times q_2}^{-1}(\bdeta_{3,4,ij})^T\}}\Big]^T  \\[1ex]
 \smhalf\,\displaystyle{\stack{1\le j\le n_i}\{\vecof_{q_1\times q_2}^{-1}(\bdeta_{3,4,ij})^T\}} 
& {\displaystyle\blockdiag{1\le j\le n_i}}\{\vecof^{-1}(\bD_{q_2}^{+T}\bdeta_{3,2,ij})\}
       \end{array}
       \right]
\end{array}
\right]
\end{eqnarray*}
}
and
$$
\ba\equiv
\left[
\arraycolsep=2.2pt\def\arraystretch{1.6}
\begin{array}{c}
\setstretch{4.5}
\bdeta_{1,1} \\[1ex]
\displaystyle{\stack{1\le i\le m}}(\bdeta_{2,1,i})\\[1ex]
\displaystyle{\stack{1\le i\le m}}
\Big\{\displaystyle{\stack{1\le j\le n_i}}(\bdeta_{3,1,ij})\Big\}\\[1ex]
\end{array}
\right]
\quad\mbox{where}\quad
\left[
\begin{array}{l}
\bdeta_{1,1}\ (p\times 1)\\[1ex]
\bdeta_{1,2}\ (\smhalf p(p+1)\times 1)\\[1ex]
{\displaystyle\stack{1\le i\le m}}
\left[
\begin{array}{ll}
\bdeta_{2,1,i}&(q_1\times 1)\\
\bdeta_{2,2,i}&(\smhalf q_1(q_1+1)\times 1)\\
\bdeta_{2,3,i}&(pq_1\times 1)
\end{array}
\right]
\\[4ex]
{\displaystyle\stack{1\le i\le m}}
\left[
{\displaystyle\stack{1\le j\le n_i}}
\left[
\begin{array}{ll}
\bdeta_{3,1,ij}&(q_2\times 1)\\
\bdeta_{3,2,ij}&(\smhalf q_2(q_2+1)\times 1)\\
\bdeta_{3,3,ij}&(pq_2\times 1)\\
\bdeta_{3,4,ij}&(q_1q_2\times 1)
\end{array}
\right]
\right]
\end{array}
\right]
$$
is the partitioning of $\etaSUBpybetausigsqEpsCONNbetau$ that defines 
$\bdeta_{1,1}$, $\bdeta_{1,2}$, $\{(\bdeta_{2,1,i},\bdeta_{2,2,i},\bdeta_{2,3,i}):1\le i\le m\}$
and $\{(\bdeta_{3,1,ij},\bdeta_{3,2,ij},\bdeta_{3,3,ij},\bdeta_{3,4,ij}):1\le i\le m,1\le j\le n_i\}$.
The solutions are, according to notation illustrated by
(\ref{eq:3levSubBlockNotat1})--(\ref{eq:3levSubBlockNotat3}),
$\bmu_{\qDens(\bbeta)}=\xveco$, $\bSigma_{\qDens(\bbeta)}=\AUoo$ and
$$\bmu_{\qDens(\buLonei)}=\xvectCi,\ \bSigma_{\qDens(\buLonei)}=\AUttCi,\ 
E_{\qDens}\{(\bbeta-\bmu_{\qDens(\bbeta)})(\buLonei-\bmuq{\buLonei})^T\}=
\AUotCi\ \textit{for}\ 1\le i\le m$$
and
$$\bmu_{\qDens(\buLtwoij)}=\bx_{2, ij},\ 
\bSigma_{\qDens(\buLtwoij)}=\bA^{22, ij},\ 
E_{\qDens}\{(\bbeta-\bmu_{\qDens(\bbeta)})(\buLtwoij-\bmuq{\buLtwoij})^T\}
=\bA^{12, ij},$$
$$E_{\qDens}\{(\buLonei-\bmuq{\buLonei})(\buLtwoij-\bmuq{\buLtwoij})^T\}
=\bA^{12,\iCOMMAj}\ \textit{for}\ 1\le i\le m,\ 1\le j\le n_i.
$$
\label{res:threeLevelVMPlik}
\end{result}

The message from the likelihood factor to $\sigma^2$ has the form
as in the two-level case, as given by (\ref{eq:GaussLikToSigsqMsg}).
Streamlined Gaussian likelihood fragment updates for the messages from
$\pDens(\by|\bbeta,\bu,\sigma^2)$ to $(\bbeta,\bu)$ and $\sigma^2$ is 
encapsulated in Algorithm \ref{alg:threeLevelVMPlik}.
Note its use of the notation defined by (\ref{eq:doubArrowNotat}).
Its justification is described in Section \ref{sec:algThreeLevelVMPlikDeriv}.

\begin{algorithm}[!th]
  \begin{center}
    \begin{minipage}[t]{165mm}
      \textbf{Data Inputs:} $\by_{ij}(o_{ij}\times1),\ \bX_{ij}(o_{ij}\times p),\ \bZLone_{ij}(o_{ij}\times q_{1}),\
        \bZLtwo_{ij}(o_{ij}\times q_{2}), \ 1 \le i\le m$, \\
        \null\ \ \ \ \ \ \ \ \ \ \ \ \ \ \ \ \ \ \ \ \ \ \ \ \  \ \ $1\le j\le n_{i}$.
      \jump\noindent
      \textbf{Parameter Inputs:} $\etaSUBpybetausigsqEpsTObetau$,\ $\etaSUBbetauTOpybetausigsqEps$,\
       $\etaSUBpybetausigsqEpsTOsigsqEps$,\\
      \null\ \ \ \ \ \ \ \ \ \ \ \ \ \ \ \ \ \ \ \ \ \ \ \ \  \ \ $\etaSUBsigsqEpsTOpybetausigsqEps$\\
      \textbf{Updates:}
      \begin{itemize}
        \setlength\itemsep{0pt}
        \item[]$\mu_{\qDens(1/\mysigeps^2)}\thickarrow
                \Big(\big(\etaSUBpybetausigsqEpsCONNsigsqEps\big)_1+1
                \Big)\Big/\big(\etaSUBpybetausigsqEpsCONNsigsqEps\big)_2$
        \item[] $\SscAlgSeven\thickarrow
                 \ThreeLevelNaturalToCommonParameters\Big(p,q_{1},q_{2},m,\{n_i:1\le i\le m\},$
        \item[] \quad\quad\quad\quad $\etaSUBpybetausigsqEpsCONNbetau\Big)$
        \item[] $\bmu_{\qDens(\bbeta)}\thickarrow\mbox{$\bmu_{\qDens(\bbeta)}$ component of $\SscAlgSeven$}$\ \ ;\ \
                $\bSigma_{\qDens(\bbeta)}\thickarrow\mbox{$\bSigma_{\qDens(\bbeta)}$ component of $\SscAlgSeven$}$
        \item[]$\omegaAlgSevenA\thickarrow\bzero_p$\ \ ;\ \
               $\omegaAlgSevenB\thickarrow\bzero_{\frac{1}{2}\,p(p+1)}$\ \ ;\ \
               $\omegaAlgSevenC\thickarrow 0$
        \item[] For $i=1,\ldots,m$:
        \begin{itemize}
          \item[] $\bmu_{\qDens(\buLone_i)}\thickarrow\mbox{$\bmu_{\qDens(\buLone_i)}$ component of $\SscAlgSeven$}\ \ ; \ \
                   \bSigma_{\qDens(\buLone_i)}\thickarrow\mbox{$\bSigma_{\qDens(\buLone_i)}$ component of $\SscAlgSeven$}$
          \item[] $E_{\qDens}\{(\bbeta-\bmu_{\qDens(\bbeta)})(\buLone_i-\bmu_{\qDens(\buLone_i)})^T\}\thickarrow
                   E_{\qDens}\{(\bbeta-\bmu_{\qDens(\bbeta)})(\buLone_i-\bmu_{\qDens(\buLone_i)})^T\}$
          \item[] \quad\quad\quad\quad\quad\quad\quad\quad\quad\quad\quad\quad\quad\quad\quad\quad\quad\quad
                  $\mbox{component of $\SscAlgSeven$}$
          \item[] $\omegaAlgSevenD\thickarrow\bzero_{q_1}$\ \ ;\ \ $\omegaAlgSevenE\thickarrow\bzero_{\frac{1}{2}\,q_1(q_1+1)}$
                  \ \ ;\ \ $\omegaAlgSevenF\thickarrow\bzero_{p\,q_1}$
          \item[] For $j=1,\ldots,n_{i}$:
          \begin{itemize}
            \item[] $\omegaAlgSevenA\thickarrow\omegaAlgSevenA+\bX_{ij}^T\by_{ij}$\ \ \ ;\ \ \
                    $\omegaAlgSevenB\thickarrow\omegaAlgSevenB-\smhalf\bD_p^T\vecof(\bX_{ij}^T\bX_{ij})$
            \item[] $\omegaAlgSevenD\thickarrow\omegaAlgSevenD+(\bZLone_{ij})^T\by_{ij}$\ \ \ ;\ \ \
                   $\omegaAlgSevenE\thickarrow\omegaAlgSevenE-\smhalf\bD_{q_1}^T\vecof\big((\bZLone_{ij})^T\bZLone_{ij}\big)$
            \item[] $\omegaAlgSevenF\thickarrow\omegaAlgSevenF-\vecof\big(\bX_{ij}^T\bZLone_{ij}\big)$
            \item[] $\bmu_{\qDens(\buLtwo_{ij})}\thickarrow\mbox{$\bmu_{\qDens(\buLtwo_{ij})}$\ component of $\SscAlgSeven$}$
            \item[] $\bSigma_{\qDens(\buLtwo_{ij})}\thickarrow\mbox{$\bSigma_{\qDens(\buLtwo_{ij})}$\ component of $\SscAlgSeven$}$
            \item[] $E_{\qDens}\{(\bbeta-\bmu_{\qDens(\bbeta)})(\buLtwo_{ij}-\bmu_{\qDens(\buLtwo_{ij})})^T\}\thickarrow
                     E_{\qDens}\{(\bbeta-\bmu_{\qDens(\bbeta)})(\buLtwo_{ij}-\bmu_{\qDens(\buLtwo_{ij})})^T\}$
            \item[] \quad\quad\quad\quad\quad\quad\quad\quad\quad\quad\quad\quad\quad\quad\quad\quad\quad\quad
                    $\mbox{component of $\SscAlgSeven$}$
            \item[] $E_{\qDens}\{(\buLone_i-\bmu_{\qDens(\buLone_i)})(\buLtwo_{ij}-\bmu_{\qDens(\buLtwo_{ij})})^T\}\thickarrow$
            \item[] \quad\quad\quad\quad\quad\quad $E_{\qDens}\{(\buLone_i-\bmu_{\qDens(\buLone_i)})(\buLtwo_{ij}-\bmu_{\qDens(\buLtwo_{ij})})^T\}
                    \mbox{\ component of $\SscAlgSeven$}$
            \item[]$\omegaAlgSevenC\thickarrow\omegaAlgSevenC
                    -\smhalf\Vert\by_{ij}-\bX_{ij}\bmu_{\qDens(\bbeta)} - \bZLone_{ij}\bmu_{\qDens(\buLone_i)}
                    - \bZLtwo_{ij}\bmu_{\qDens(\buLtwo_{ij})}\Vert^2$
            \item[] $\omegaAlgSevenC\thickarrow\omegaAlgSevenC-\smhalf\tr(\bSigma_{\qDens(\bbeta)}\bX_{ij}^T\bX_{ij})
                     -\smhalf\tr(\bSigma_{\qDens(\buLone_i)}(\bZLone_{ij})^T\bZLone_{ij})$
            \item[] $\qquad\qquad\quad-\smhalf\tr(\bSigma_{\qDens(\buLtwo_{ij})}\bZ^{L2T}_{ij}\bZ^{L2}_{ij})$
            \item[] $\qquad\qquad\quad-\tr[\{(\bZLone_{ij})^T\bX_{ij}E_{\qDens}\{(\bbeta-\bmu_{\qDens(\bbeta)})(\buLone_i
                     -\bmu_{\qDens(\buLone_i)})^T\}]$
            \item[] $\qquad\qquad\quad-\tr[\{(\bZLtwo_{ij})^T\bX_{ij}E_{\qDens}\{(\bbeta-\bmu_{\qDens(\bbeta)})(\buLtwo_{ij}
                     -\bmu_{\qDens(\buLtwo_{ij})})^T\}]$
            \item[] $\qquad\qquad\quad-\tr[\{(\bZLtwo_{ij})^T\bZLone_{ij}E_{\qDens}\{(\buLone_i
                     -\bmu_{\qDens(\buLone_i)})(\buLtwo_{ij}
                      -\bmu_{\qDens(\buLtwo_{ij})})^T\}]$
            \item[] \textsl{continued on a subsequent page}\ $\ldots$
          \end{itemize}
        \end{itemize}
      \end{itemize}
    \end{minipage}
  \end{center}
  \caption{\it 
 The inputs, updates and outputs of the matrix algebraic streamlined
               Gaussian likelihood fragment for three-level models. The algorithm description 
requires more than one page and is continued on a subsequent page.}
  \label{alg:threeLevelVMPlik}
\end{algorithm}

\setcounter{algorithm}{6}
\begin{algorithm}[!th]
  \begin{center}
    \begin{minipage}[t]{165mm}
      \begin{itemize}
        \item[] $\etaSUBpybetausigsqEpsTObetau\thickarrow\mu_{\qDens(1/\mysigeps^2)}
                  \left[
                    \begin{array}{c}
                      \omegaAlgSevenA\\
                      \omegaAlgSevenB\\[2ex]
                      {\displaystyle\stack{1\le i\le m}}\left[
                        \begin{array}{c}
                          \omegaAlgSevenD\\
                          \omegaAlgSevenE\\
                          \omegaAlgSevenF\\
                        \end{array}
                      \right]\\[6ex]
                      {\displaystyle\stack{1\le i\le m}}\left[
                        {\displaystyle\stack{1\le j\le n_i}}\left[
                          \begin{array}{c}
                            (\bZLtwo_{ij})^T\by_{ij}\\
                            -\smhalf \bD_{q_{2}}^T\vecof((\bZLtwo_{ij})^T\bZLtwo_{ij})\\
                            -\vecof(\bX_{ij}^T\bZLtwo_{ij})\\
                            -\vecof\big((\bZLone_{ij})^T\bZLtwo_{ij}\big)\\
                          \end{array}\right]
                        \right]
                      \end{array}
                    \right]$
        \item[] $\etaSUBpybetausigsqEpsTOsigsqEps\thickarrow
                 \left[
                    \begin{array}{c}
                      -\smhalf{\displaystyle\sum_{i=1}^m\sum_{j=1}^{n_i}}\,o_{ij}\\[1ex]
                      \omegaAlgSevenC
                    \end{array}
                 \right]$
      \end{itemize}
      \textbf{Parameter Outputs:} $\etaSUBpybetausigsqEpsTObetau$, $\etaSUBpybetausigsqEpsTOsigsqEps$.
      \jump
    \end{minipage}
  \end{center}
\caption{\textbf{continued.} \textit{This is a continuation of the description of this algorithm that commences
on a preceding page.}}
\end{algorithm}

\subsection{Streamlined Gaussian Penalization Fragment Updates}

Here we treat the Gaussian penalization fragment for three-level random effects 
structure. This fragment is shown in brown in Figure \ref{fig:threeLevFacGraph}.
We assume that
$$\mSUBpbetauSigmaLoneLtwoTObetau\quad\mbox{and}\quad\mSUBpybetausigsqEpsTObetau$$ 
are in the same exponential family. In other words, $\mSUBpbetauSigmaLoneLtwoTObetau$ 
has the form given by the right-hand side of (\ref{eq:threeLevLikToCoeffMsg})
but with natural parameter vector
$$\etaSUBpybetausigsqEpsTObetau\quad\mbox{replaced by}\quad
\etaSUBpbetauSigmaLoneLtwoTObetau.
$$
The fragment's other factor to stochastic node messages are
$$\mSUBpbetauSigmaLoneLtwoTOSigmaLone=
\exp\left\{
\left[
\begin{array}{c}
\log|\bSigmaLone|\\[1ex]
\vech\big((\bSigmaLone)^{-1}\big)
\end{array}
\right]^T\etaSUBpbetauSigmaLoneLtwoTOSigmaLone
\right\}
$$
and
$$\mSUBpbetauSigmaLoneLtwoTOSigmaLtwo=
\exp\left\{
\left[
\begin{array}{c}
\log|\bSigmaLtwo|\\[1ex]
\vech\big((\bSigmaLtwo)^{-1}\big)
\end{array}
\right]^T\etaSUBpbetauSigmaLoneLtwoTOSigmaLtwo
\right\}.
$$
Streamlined updating of the three-level Gaussian penalization fragment is
aided by Result \ref{res:threeLevelVMPpen}:

\null\vfill\eject
\null\vfill\eject
\begin{result}
The variational message passing updates of 
the quantities $\bmu_{\qDens(\buLone_i)}, \bSigma_{\qDens(\buLone_i)}$, $1\le i\le m$,
and $\bmu_{\qDens(\buLtwo_{ij})},\bSigma_{\qDens(\buLtwo_{ij})}$, 
$1\le i\le m$, $1\le j\le n_i$, with $\qDens$-density 
expectations with respect to the normalization of
$$\mSUBpbetauSigmaLoneLtwoTObetau\,\mSUBbetauTOpbetauSigmaLoneLtwo$$
are expressible as a three-level sparse matrix problem (see Appendix \ref{sec:threeLevSMA})
with
{\setlength\arraycolsep{1pt}
\begin{eqnarray*}
&&\AtLev=\\[1ex]
&&-2\,\left[
\begin{array}{cc}
\vecof^{-1}(\bD_p^{+T}\bdeta_{1,2}) 
& \Big(\smhalf\,\displaystyle{\stack{1\le i\le m}}\Big[\vecof_{p\times q_1}^{-1}(\bdeta_{2,3,i})^T
\displaystyle{\stack{1\le j\le n_i}}\{\vecof_{p\times q_2}^{-1}(\bdeta_{3,3,ij})^T\}\Big]\Big)^T   \\[1ex]
\begin{array}{l}
\smhalf\,\displaystyle{\stack{1\le i\le m}}\Big[\vecof_{p\times q_1}^{-1}(\bdeta_{2,3,i})^T \\
\displaystyle{\stack{1\le j\le n_i}}\{\vecof_{p\times q_2}^{-1}(\bdeta_{3,3,ij})^T\}\Big]
\end{array}
& {\displaystyle\blockdiag{1\le i\le m}}
\left[
       \begin{array}{cc}
       \vecof^{-1}(\bD_{q_1}^{+T}\bdeta_{2,2,i})  
& \Big[\smhalf\,\displaystyle{\stack{1\le j\le n_i}\{\vecof_{q_1\times q_2}^{-1}(\bdeta_{3,4,ij})^T\}}\Big]^T  \\[1ex]
 \smhalf\,\displaystyle{\stack{1\le j\le n_i}\{\vecof_{q_1\times q_2}^{-1}(\bdeta_{3,4,ij})^T\}} 
& {\displaystyle\blockdiag{1\le j\le n_i}}\{\vecof^{-1}(\bD_{q_2}^{+T}\bdeta_{3,2,ij})\}
       \end{array}
       \right]
\end{array}
\right]
\end{eqnarray*}
}
and
$$
\ba\equiv
\left[
\arraycolsep=2.2pt\def\arraystretch{1.6}
\begin{array}{c}
\setstretch{4.5}
\bdeta_{1,1} \\[1ex]
\displaystyle{\stack{1\le i\le m}}(\bdeta_{2,1,i})\\[1ex]
\displaystyle{\stack{1\le i\le m}}
\Big\{\displaystyle{\stack{1\le j\le n_i}}(\bdeta_{3,1,ij})\Big\}\\[1ex]
\end{array}
\right]
\quad\mbox{where}\quad
\left[
\begin{array}{l}
\bdeta_{1,1}\ (p\times 1)\\[1ex]
\bdeta_{1,2}\ (\smhalf p(p+1)\times 1)\\[1ex]
{\displaystyle\stack{1\le i\le m}}
\left[
\begin{array}{ll}
\bdeta_{2,1,i}&(q_1\times 1)\\
\bdeta_{2,2,i}&(\smhalf q_1(q_1+1)\times 1)\\
\bdeta_{2,3,i}&(pq_1\times 1)
\end{array}
\right]
\\[4ex]
{\displaystyle\stack{1\le i\le m}}
\left[
{\displaystyle\stack{1\le j\le n_i}}
\left[
\begin{array}{ll}
\bdeta_{3,1,ij}&(q_2\times 1)\\
\bdeta_{3,2,ij}&(\smhalf q_2(q_2+1)\times 1)\\
\bdeta_{3,3,ij}&(pq_2\times 1)\\
\bdeta_{3,4,ij}&(q_1q_2\times 1)
\end{array}
\right]
\right]
\end{array}
\right]
$$
is the partitioning of $\etaSUBpbetauSigmaLoneLtwoCONNbetau$ that defines 
$\bdeta_{1,1}$, $\bdeta_{1,2}$, $\{(\bdeta_{2,1,i},\bdeta_{2,2,i},\bdeta_{2,3,i}):1\le i\le m\}$
and $\{(\bdeta_{3,1,ij},\bdeta_{3,2,ij},\bdeta_{3,3,ij},\bdeta_{3,4,ij}):1\le i\le m,1\le j\le n_i\}$.
The solutions are, according to notation illustrated by
(\ref{eq:3levSubBlockNotat1})--(\ref{eq:3levSubBlockNotat3}),
$$\bmu_{\qDens(\buLonei)}=\xvectCi,\ \bSigma_{\qDens(\buLonei)}=\AUttCi\quad\mbox{for}\quad 1\le i\le m$$
and
$$\bmu_{\qDens(\buLtwoij)}=\bx_{2, ij},\ 
\bSigma_{\qDens(\buLtwoij)}=\bA^{22, ij}
\quad\mbox{for}\quad 1\le i\le m,\ 1\le j\le n_i.
$$
\label{res:threeLevelVMPpen}
\end{result}

Algorithm \ref{alg:threeLevelVMPpen} provides the natural parameter vector 
updates for the three-level Gaussian penalization fragment based on
Result \ref{res:threeLevelVMPlik}.
Note that natural parameter vectors containing a $\leftrightarrow$ in their subscript, 
such as $\etaSUBpbetauSigmaLoneLtwoCONNSigmaLone$, are defined by (\ref{eq:doubArrowNotat}).

\begin{algorithm}[!th]
  \begin{center}
    \begin{minipage}[t]{165mm}
\textbf{Hyperparameter Inputs:} $\bmu_{\bbeta}(p\times 1)$, $\bSigma_{\bbeta}(p\times p)$,\\
\textbf{Parameter Inputs:} $\etaSUBpbetauSigmaLoneLtwoTObetau$,\ $\etaSUBbetauTOpbetauSigmaLoneLtwo$,
\ $\etaSUBpbetauSigmaLoneLtwoTOSigmaLone$,\\[1ex]
\null$\qquad\qquad\qquad\qquad\ \ $$\etaSUBSigmaLoneTOpbetauSigmaLoneLtwo$,\ $\etaSUBpbetauSigmaLoneLtwoTOSigmaLtwo$,
\ $\etaSUBSigmaLtwoTOpbetauSigmaLoneLtwo$\\
\textbf{Updates:}
      \begin{itemize}
        \setlength\itemsep{0pt}
      \item[] $\omegaAlgEightA\thickarrow\mbox{first entry of $\etaSUBpbetauSigmaLoneLtwoCONNSigmaLone$}$
     \item[] $\omegaAlgEightB\thickarrow\mbox{remaining entries of $\etaSUBpbetauSigmaLoneLtwoCONNSigmaLone$}$
     \item[] $\bM_{\qDens((\bSigmaLone)^{-1})}\thickarrow
     \big\{\omegaAlgEightA + \smhalf(q_1+1)\big\}\{\vecof^{-1}\big(\bD_{q_1}^{+T}\omegaAlgEightB\big)\}^{-1}$
      \item[] $\omegaAlgEightC\thickarrow\mbox{first entry of $\etaSUBpbetauSigmaLoneLtwoCONNSigmaLtwo$}$
      
     \item[] $\omegaAlgEightD\thickarrow\mbox{remaining entries of $\etaSUBpbetauSigmaLoneLtwoCONNSigmaLtwo$}$
     \item[] $\bM_{\qDens((\bSigmaLtwo)^{-1})}\thickarrow
     \big\{\omegaAlgEightC + \smhalf(q_2+1)\big\}\{\vecof^{-1}\big(\bD_{q_2}^{+T}\omegaAlgEightD\big)\}^{-1}$
        \item[] $\SscAlgEight\thickarrow
                 \ThreeLevelNaturalToCommonParameters$\\
        $\null\qquad\qquad\Big(p,q_{1},q_{2},m,\{n_i:1\le i\le m\},\etaSUBpybetausigsqEpsCONNbetau\Big)$
        \item[]
               $\omegaAlgEightE\thickarrow\bzero_{\frac{1}{2}\,q_1(q_1+1)}$\ \ ;\ \
               $\omegaAlgEightF\thickarrow \bzero_{\frac{1}{2}\,q_2(q_2+1)}$
        \item[] For $i=1,\ldots,m$:
        \begin{itemize}
        \setlength\itemsep{0pt}
          \item[] $\bmu_{\qDens(\buLone_i)}\thickarrow\mbox{$\bmu_{\qDens(\buLone_i)}$ component of $\SscAlgEight$}\ \ ; \ \
                   \bSigma_{\qDens(\buLone_i)}\thickarrow\mbox{$\bSigma_{\qDens(\buLone_i)}$ component of $\SscAlgEight$}$
          \item[] $\omegaAlgEightE\thickarrow\omegaAlgEightE -\smhalf\,
\bD_{q_1}^T\vecof\Big(\bmu_{\qDens(\buLone_i)}\bmu_{\qDens(\buLone_i)}^T +\bSigma_{\qDens(\buLone_i)}\Big)$
          \item[] For $j=1,\ldots,n_{i}$:
          \begin{itemize}
            \item[] $\bmu_{\qDens(\buLtwo_{ij})}\thickarrow\mbox{$\bmu_{\qDens(\buLtwo_{ij})}$\ component of $\SscAlgEight$}$
            \ ;\ $\bSigma_{\qDens(\buLtwo_{ij})}\thickarrow\mbox{$\bSigma_{\qDens(\buLtwo_{ij})}$\ component of $\SscAlgEight$}$
           \item[]$\omegaAlgEightF\thickarrow\omegaAlgEightF -\smhalf\,
\bD_{q_2}^T\vecof\Big(\bmu_{\qDens(\buLtwo_{ij})}\bmu_{\qDens(\buLtwo_{ij})}^T +\bSigma_{\qDens(\buLtwo_{ij})}\Big)$
          \end{itemize}
        \end{itemize}
\item[] $\etaSUBpbetauSigmaLoneLtwoTObetau\thickarrow
                  \left[
                    \begin{array}{c}
                    \bSigma_{\bbeta}^{-1}\bmu_{\bbeta}\\[2ex]
                     -\smhalf\bD_p^T\vecof(\bSigma_{\bbeta}^{-1})\\[2ex]
                      {\displaystyle\stack{1\le i\le m}}\left[
                        \begin{array}{c}
                          \bzero_{q_1}\\[1ex]
                          -\smhalf\bD_{q_{1}}^T\vecof\big(\bM_{\qDens((\bSigmaLone)^{-1})}\big)\\[2ex]
                          \bzero_{pq_1}\\
                        \end{array}
                      \right]\\[6ex]
                      {\displaystyle\stack{1\le i\le m}}\left[
                        {\displaystyle\stack{1\le j\le n_i}}\left[
                          \begin{array}{c}
                            \bzero_{q_2}\\
                            -\smhalf \bD_{q_{2}}^T\vecof\big(\bM_{\qDens((\bSigmaLtwo)^{-1})}\big)\\
                            \bzero_{pq_2}\\
                            \bzero_{q_1q_2}\\
                          \end{array}\right]
                        \right]
                      \end{array}
                    \right]$
       \item[] $\etaSUBpbetauSigmaLoneLtwoTOSigmaLone\thickarrow
                 \left[
                    \begin{array}{c}
                      -\smhalf m\\[2ex]
                      \omegaAlgEightE
                    \end{array}
                 \right]\ \ ;\ \ 
                 \etaSUBpbetauSigmaLoneLtwoTOSigmaLtwo\thickarrow
                 \left[
                    \begin{array}{c}
                      -\smhalf{\displaystyle\sum_{i=1}^m} n_i\\[2ex]
                      \omegaAlgEightF
                    \end{array}
                 \right]
              $
      \end{itemize}
      \textbf{Parameter Outputs:} $\etaSUBpbetauSigmaLoneLtwoTObetau$, $\etaSUBpbetauSigmaLoneLtwoTOSigmaLone$,
                                   $\etaSUBpbetauSigmaLoneLtwoTOSigmaLtwo$

    \end{minipage}
  \end{center}
  \caption{\it The inputs, updates and outputs of the matrix algebraic streamlined
               Gaussian penalization fragment for three-level models.}
\label{alg:threeLevelVMPpen}
\end{algorithm}

\subsection{$\qDens$-Density Determination After Variational Message Passing Convergence}

The advice given in Section \ref{sec:qDensTwoLev} for the two-level case extends
straightforwardly to the three-level case. The main change is that the steps that
we need to first carry out are:
{\setlength\arraycolsep{1pt}
\begin{eqnarray*}
\etaSUBqbetau&\thickarrow& \etaSUBpybetausigsqEpsTObetau+\etaSUBpbetauSigmaLoneLtwoTObetau\\[1ex]
\SscSecondInText&\thickarrow& \ThreeLevelNaturalToCommonParameters\Big(p,q_1,q_2,m,\{n_i:1\le i\le m\},
\etaSUBqbetau\Big).
\end{eqnarray*}
}

\section{Computational Complexity and Timing Results}\label{sec:timing}

Table \ref{tab:computComplex} summarizes and compares the large sample computational complexities
of streamlined mean field variational Bayes Algorithms \ref{alg:twoLevelMFVB} 
and \ref{alg:threeLevelMFVB} and the na\"{\i}ve implementation alternative.
To aid digestibility, in Table \ref{tab:computComplex}  
we are imposing the following balanced designs restrictions:  $n_i=n$ and $o_{ij}=o$ for all values
of the indices. The values of $m$, $n$ and $o$ are assumed to be diverging whilst $p$, $q$, $q_1$, $q_2$ 
and the numbers of mean field variational Bayes iterations are held fixed.
The entries of Table \ref{tab:computComplex} are justified by
results concerning the number of floating point operations for matrix multiplications
and QR decompositions given in, for example, Sections 1.2.4 and 5.5.9 of 
Golub \myand van Loan (1989).
We see from Table \ref{tab:computComplex} that the
floating point operation counts of Algorithms \ref{alg:twoLevelMFVB} 
and \ref{alg:threeLevelMFVB} are linear in the number of observations
and these streamlined algorithms offer quadratic improvements over na\"{\i}ve 
implementation.

\begin{table}[!ht]
\begin{center}
\begin{tabular}{cccc}
\hline\\[-2.5ex]
level             & na\"{\i}ve & streamlined  &  na\"{\i}ve/streamlined\\
\hline\\[-1.5ex]
two-level        &       $O(m^3n)$      &   $O(mn)$      &   $O(m^2)$    \\[0.5ex]
three-level      &       $O(m^3n^3o)$   &   $O(mno)$     &   $O(m^2n^2)$ \\
\hline
\end{tabular}
\end{center}
\caption{\textit{The order of magnitudes of the number of floating point operations 
for Algorithms \ref{alg:twoLevelMFVB} and \ref{alg:threeLevelMFVB} and na\"{\i}ve
implementation. The ratio of na\"{\i}ve to streamlined computation is also given.
The designs are assumed to be balanced and $m,n,o\to\infty$ whilst  
$p$, $q$, $q_1$, $q_2$ and the numbers of mean field variational Bayes iterations 
are fixed.}}
\label{tab:computComplex}
\end{table}

To assess finite sample performance, we obtained timing results for simulated data 
according to a version of model (\ref{eq:twoLevelGaussRespBaye})
for which both the fixed effects and random effects had dimension $2$, corresponding
to random intercepts and slopes for a single continuous predictor which was generated
from the Uniform distribution on the unit interval. The true parameter values were set to 
$$\bbeta_{\mbox{\tiny true}}=\left[
\begin{array}{c}
0.58\\
1.98
\end{array}
\right],
\qquad
\sigma^2_{\mbox{\tiny true}}=0.1
\qquad\mbox{and}\qquad
\bSigma_{\mbox{\tiny true}}=
\left[
\begin{array}{cc}
2.58 & 0.22\\
0.22 & 1.73
\end{array}
\right]
$$
and, throughout the study, the $n_i$ values were generated uniformly 
on the set $\{30,\ldots,60\}$. The study was run on a \texttt{MacBook Air} 
laptop with a 2.2 gigahertz processor and 8 gigabytes of random access memory.
The number of mean field iterations was fixed at $50$.

\begin{table}[!ht]
\begin{center}
\begin{tabular}{cccc}
\hline\\[-2.5ex]
$m$                & na\"{\i}ve & streamlined  &  na\"{\i}ve/streamlined\\
\hline\\[-1.5ex]
200      &       2.75  (0.0482)      &0.035 (0.00000)    &   78.5\\
400      &      22.30  (0.2490)      &0.070 (0.00148)    &  319.0\\
600      &      84.40  (0.4940)      &0.108 (0.00445)    &  782.0\\
800      &     213.00  (0.9160)      &0.143 (0.00445)    & 1490.0\\
1,600    &     427.00  (3.1000)      &0.183 (0.00741)    & 2340.0\\
\hline
\end{tabular}
\end{center}
\caption{\textit{Median (median absolute deviation) of elapsed computing times in seconds for 
fitting model (\ref{eq:twoLevelGaussRespBaye}) na\"{\i}vely versus with streamlining
via Algorithm \ref{alg:twoLevelMFVB}. The fourth column lists the ratios of the median
computing times.}}
\label{tab:StreamVsNaive}
\end{table}

The first phase of the study involved comparing the computational times of
the streamlined Algorithm \ref{alg:twoLevelMFVB} with its na\"{\i}ve
counterpart for which (\ref{eq:muSigmaMFVBupd}) was implemented directly.
To allow for maximal speed,  both approaches were implemented in the 
low-level language \texttt{Fortran 77}. The number of groups varied
over $m\in\{200,400,600,800,1000\}$ and $100$ replications were simulated
for each value of $m$. For the most demanding $m=1,000$ case the 
streamlined implementation had a median computing time of 0.183 seconds
and a maximum of 0.354 seconds. By comparison, the 
na\"{\i}ve approach had a median computing time of 7 minutes and, for 
a few replications, took several hours. Because of such outliers in
the na\"{\i}ve computational times our summary of this first phase,
given in Table \ref{tab:StreamVsNaive}, uses medians and median absolute
deviations. As the number of groups increases into the several hundreds
we see that streamlined variational inference becomes thousands of times faster
in terms of median performance.

The second phase of our timing study involved ramping up the number
of groups into the tens of thousands and recording computational
times for Algorithm \ref{alg:twoLevelMFVB}. We used the 
geometric progression $m\in\{400,1200,3600,10800,32400\}$
and another 100 replications. Table \ref{tab:StreamAbsolute} shows
that the average computing times increase approximately linearly
with $m$ and only around 7 seconds are required for handling
$m=32,400$ groups.

\begin{table}[!ht]
\begin{center}
\begin{tabular}{ccccc}
\hline\\[-1.5ex]
\hfill$m=400$\hfill&\hfill$m=1,200$\hfill&\hfill$m=3,600$\hfill&\hfill$m=10,800$\hfill&\hfill $m=32,400$\hfill\\
\hline\\[-1.5ex]
0.0781    &    0.2400   &    0.7140    &    2.30   &  6.980   \\
(0.0122)  &   (0.0343)  &    (0.0806)  &   (0.270) &  (0.857) \\
\hline
\end{tabular}
\end{center}
\caption{\textit{Average (standard deviation) of elapsed computing times in seconds for 
fitting model (\ref{eq:twoLevelGaussRespBaye}) with streamlining via Algorithm \ref{alg:twoLevelMFVB}.}}
\label{tab:StreamAbsolute}
\end{table}

In summary, the streamlined approach is vastly superior to na\"{\i}ve implementation
in terms of speed and scales well to large data multilevel data situations.

As a by-product of our timing studies we also recorded the empirical coverage percentages
for credible intervals with an advertized coverage of 95\%. The results are 
given in Table \ref{tab:coverRes} and based on $1,000$ replications. Apart from $\sigma$,
the parameters in Table \ref{tab:coverRes} are sub-components of $\bbeta$ and 
$\bSigma$ according to
$$\bbeta=\left[
\begin{array}{c}
\beta_0\\
\beta_1
\end{array}
\right]
\qquad\mbox{and}\qquad
\bSigma=
\left[
\begin{array}{cc}
\sigma_1^2           & \rho\sigma_1\sigma_2\\
\rho\sigma_1\sigma_2 & \sigma_2^2
\end{array}
\right].
$$

\begin{table}[!ht]
\begin{center}
{\setlength\tabcolsep{6pt}
\begin{tabular}{ccccccccccccccc}
\hline\\[-1.5ex]
parameter  & $m=100$&$m=200$&$m=400$&$m=800$&$m=1,600$\\
\hline\\[-1.5ex]
$\beta_0$  & 96.2& 95.0& 95.6& 94.7& 95.3 \\  
$\beta_1$  & 94.8& 95.2& 94.5& 95.4& 93.5 \\ 
$\sigma$   & 95.1& 94.0& 95.3& 95.1& 94.6 \\
$\sigma_1$ & 93.8& 93.6& 95.1& 95.2& 95.3 \\
$\sigma_2$ & 94.3& 94.3& 93.9& 95.5& 95.3 \\
$\rho$     & 93.9& 95.9& 95.1& 95.0& 93.8 \\
\hline
\end{tabular}
}
\end{center}
\caption{\textit{Empirical coverage percentages for advertized 95\% credible intervals
produced by Algorithm \ref{alg:twoLevelMFVB} for the simulation study described in the text.
The empirical coverage percentages are based on $1,000$ replications.}}
\label{tab:coverRes}
\end{table}

Taking into account the margins of error in percentage estimates based on $1,000$
replications, the empirical coverages are seen to be in keeping with the 95\% 
advertized level.

\begin{figure}[!ht]
\centering
{\includegraphics[width=\textwidth]{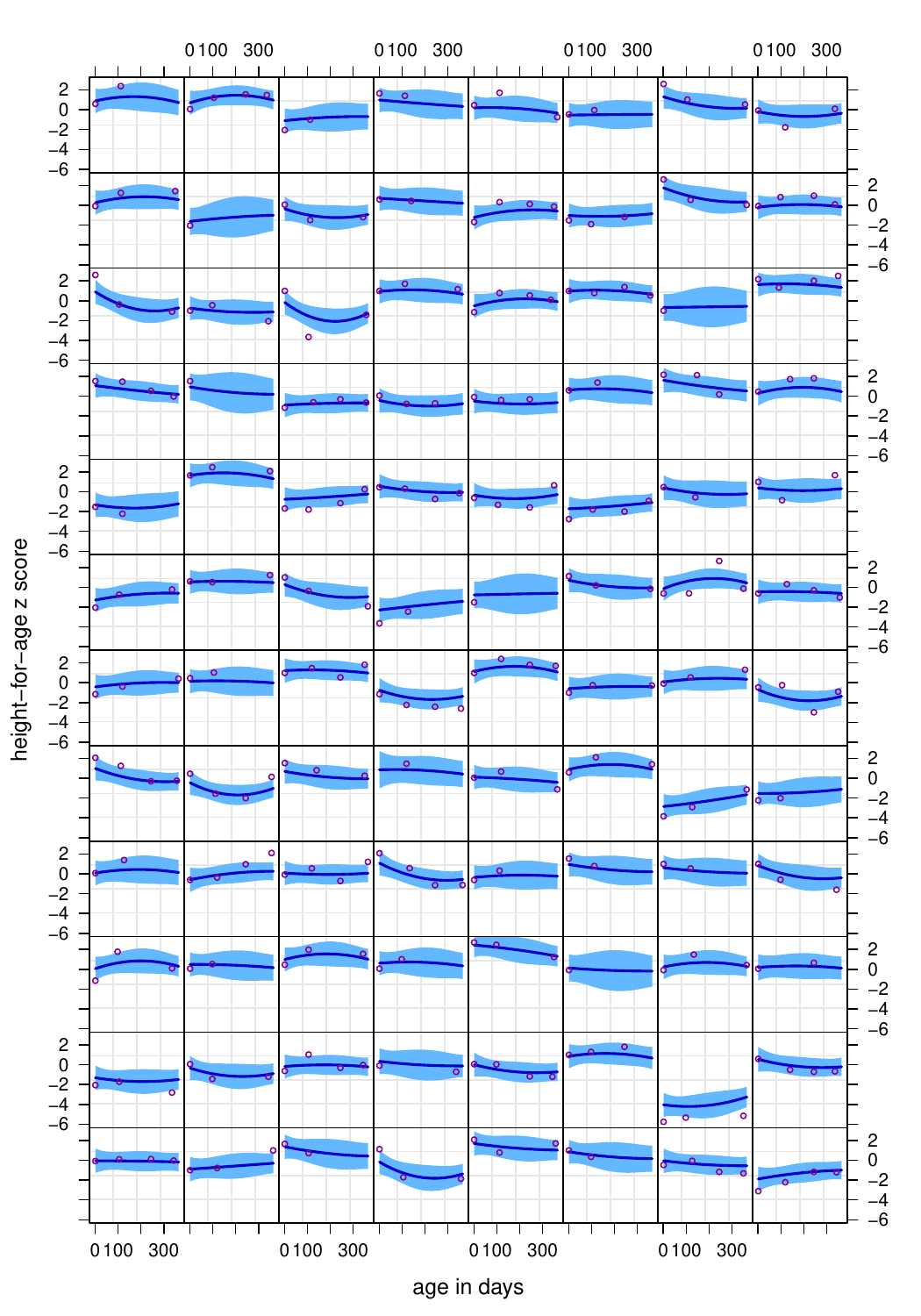}}
\caption{\it Fitted random quadratics for 96 randomly chosen infants from 
the streamlined mean field variational Bayes analysis of data from the 
Collaborative Perinatal Project for infants in the first year of life. The curves 
correspond to slices of the fitted surface according to the model defined by 
(\ref{eq:CPPmodel}) and (\ref{eq:CPPpriors}) with each of the other predictors 
set to its average value. The light blue shading corresponds to pointwise 95\% 
credible intervals.}
\label{fig:CPPanaAgeFits} 
\end{figure}

\section{Illustration for Data From a Large Longitudinal Perinatal Study}\label{sec:applic}

We now provide illustration for data from the Collaborative Perinatal Project,
a large longitudinal perinatal health study that was run in the United States 
of America during 1959--1974 (e.g. Klebanoff, 2009). The data are publicly available 
from the U.S. National Archives with identifier 606622. For our illustration in this 
section, which focuses on the first year of life, the number of infants followed
longitudinally is 44,708 and the number of fields is 125,564. We do not perform
a full-blown analysis of these data and eschew matters such as 
careful variable creation, model selection and interpretation. Instead we consider an illustrative 
Bayesian mixed model, with two-level random effects, and compare streamlined
mean field variational Bayes and Markov chain Monte Carlo fits.
Specifically, we consider the model 
\begin{equation}
\begin{array}{l}
y_{ij}|\beta_0,\ldots,\beta_7,\mysigeps^2\simind
N\Big(\beta_0+u_{i0}+(\beta_1+u_{i1})x_{1ij}+(\beta_1+u_{i2})x_{2ij}^2\\[0ex]
\qquad\qquad\qquad\qquad\qquad\qquad+\beta_3\,x_{3ij}+\ldots+\beta_7\,x_{7ij},\mysigeps^2\Big),\\[1ex]
\left[
\begin{array}{c}
u_{i0}\\
u_{i1}\\
u_{i2}
\end{array}
\right]\Bigg|\bSigma\simind N(\bzero,\bSigma),\ \mbox{for $1\le i\le 44,708$ and $1\le j\le n_i$}
\end{array}
\label{eq:CPPmodel}
\end{equation}
with priors
\begin{equation}
\begin{array}{c}
\beta_0,\ldots,\beta_7\simind N(0,10^{10}),\quad
\mysigeps^2|\asigsq\sim\mbox{Inverse-$\chi^2$}(1,1/\asigsq),\\[1ex]
\asigsq\sim\mbox{Inverse-$\chi^2$}(1,10^{-10}),\quad
\bSigma|\ASigma\sim\mbox{Inverse-G-Wishart}\big(\Gfull,6,\ASigma^{-1}\big),\\[1ex]
\ASigma\sim\mbox{Inverse-G-Wishart}(\Gdiag,1,2\times 10^{-10}\bI_3)
\end{array}
\label{eq:CPPpriors}
\end{equation}
where $y_{ij}$ denotes the $j$th response recording for the $i$th infant 
and a similar notation applies to the predictors $x_{1ij},\ldots,x_{7ij}$.
The response and predictor variables are:
{\setlength\arraycolsep{1pt}
\begin{eqnarray*}
y&\equiv&\mbox{height-for-age z-score (see below for details),}\\[0ex]
x_1&\equiv&\mbox{age of infant in days,}\\[0ex]
x_2&\equiv&\mbox{indicator that infant is male,}\\[0ex]
x_3&\equiv&\mbox{indicator that mother is Asian,}\\[0ex]
x_4&\equiv&\mbox{indicator that mother is Black,}\\[0ex]
x_5&\equiv&\mbox{indicator that mother is married,}\\[0ex]
x_6&\equiv&\mbox{indicator that mother smoked $10$ or more cigarettes per day}\\[0ex]
\mbox{and}\ x_7&\equiv&\mbox{indicator that mother attended $10$ or more 
ante-natal visits during pregnancy.}
\end{eqnarray*}
}
The height-for-age z-score is a World Health Organization standardized measure
for the height of children after accounting for age.
In the Bayesian analysis involving fitting (\ref{eq:CPPmodel}) with priors 
(\ref{eq:CPPpriors}) we divided the $y$ and $x_1$ data by the
respective sample standard deviations for each variable. We then 
convert to the original units for the reporting of results.

Model (\ref{eq:CPPmodel}) is an extension of the common random intercepts and slopes
model to quad\-ratic fitting, and allows each infant to have his or her own parabola
for the effect of age on height-for-age z-score. Figure \ref{fig:CPPanaAgeFits} 
shows the fits for 96 randomly chosen infants. It is apparent from 
Figure \ref{fig:CPPanaAgeFits} that the curvature in the age effects warrants
the extension to random quadratics.

\begin{figure}[!ht]
\centering
{\includegraphics[width=\textwidth]{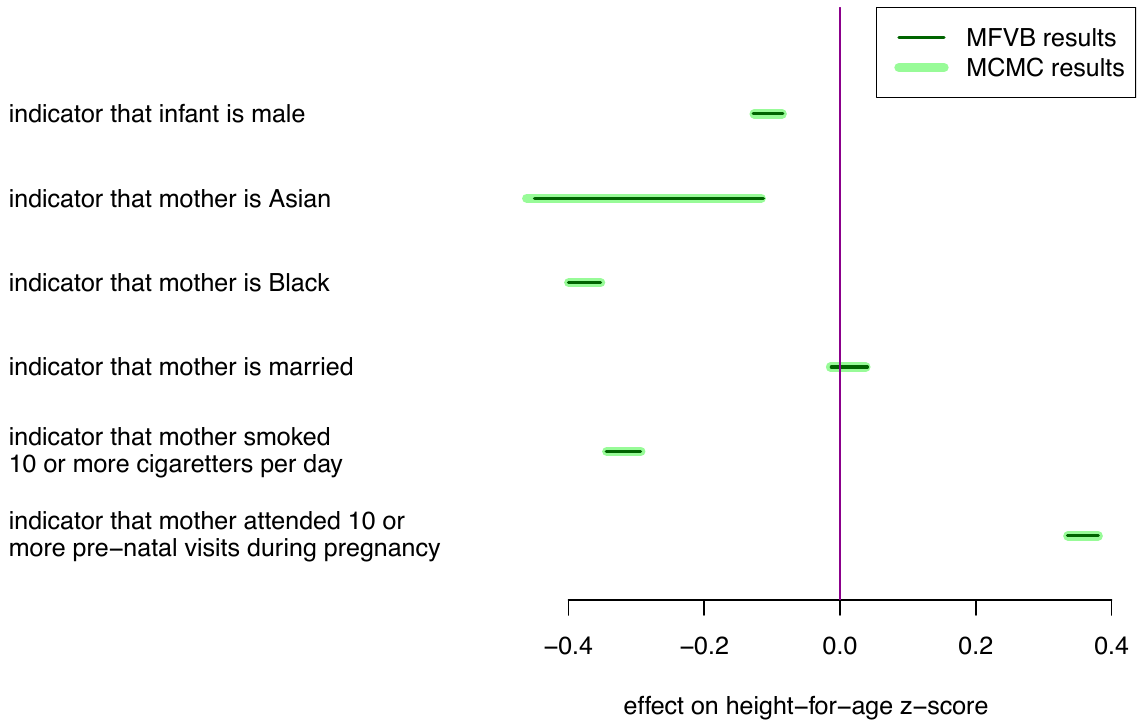}}
\caption{\it Approximate 95\% credible intervals for $\beta_3,\ldots,\beta_7$ 
for two approximate Bayesian inference fits to the model defined by (\ref{eq:CPPmodel}) 
and (\ref{eq:CPPpriors}) for the data from the Collaborative Perinatal
Project for infants in the first year of life. The thin dark green line segments display
credible intervals based on streamlined mean field variational Bayes. The thick
light green line segments display credible intervals based on
a version of Markov chain Monte Carlo.}
\label{fig:CPPeffects} 
\end{figure}

In Figure \ref{fig:CPPeffects} we summarize the approximate Bayesian inference
for $\beta_3,\ldots,\beta_7$ via 95\% credible intervals. The results for 
Markov chain Monte Carlo-based analysis using \textsf{rstan} ,
the \textsf{R} (R Core Team, 2020) interface to the \textsf{Stan} language
(Stan Development Team, 2019), are also shown. The number of 
mean field variational Bayes iterations is 100 and the Markov 
chain Monte Carlo results are based on a warmup sample of
size $1,000$ and a retained sample of size $1,000$.

It is apparent from Figure \ref{fig:CPPeffects} that streamlined mean field variational
Bayes and Markov chain Monte Carlo deliver very similar inference for the effects
of the binary predictors. As explained in Section 3.1 of Menictas \myand Wand (2013),
mean field variational Bayes tends to be very accurate for Gaussian response
models of the type being used in this example and the mild product restriction
(\ref{eq:producRestrict}). However, such high accuracy is not manifest in general.
Ignorance of important posterior dependencies via mean field restrictions often
lead to credible intervals being too small (e.g. Wang \myand Titterington, 2005).
In Figure \ref{fig:CPPeffects} there are pronounced negative effects due to ethnicity
and maternal smoking and a pronounced positive effect due to pre-natal care.

Even though streamlined mean field variational Bayes and Markov chain Monte Carlo 
deliver similar inference for this example, the former is 
significantly faster. However, it is difficult to quantify the speed
gains scientifically due to factors such as stopping criteria,
implementation language and quality of the chains. For the Figure \ref{fig:CPPeffects} 
fits, using the \texttt{MacBook Air} laptop described in 
Section \ref{sec:timing} the Markov chain Monte Carlo
fits required about 36 hours whilst the streamlined variational results 
took just 24 seconds. However, this comparison is based on a \emph{convenient} version 
of Markov chain Monte Carlo in which all the user has to do is specify the
model and let the \textsf{Stan} Bayesian inference engine do the work.
This convenience comes at the cost that general purpose Bayesian inference
engines tend to be slower than Markov chain Monte implementations for 
specific models. For the model and priors given by (\ref{eq:CPPmodel})
and (\ref{eq:CPPpriors}) Gibbs sampling involves standard
distributions and can be streamlined by sampling from the fixed
effects vector and then looping through the random effect vectors for 
each infant. After carrying out the requisite algebra, 
and programming streamlined Gibbs sampling in \textsf{R}, we found that 
Markov chain Monte Carlo fitting with the same chain sizes and laptop
required about $3.5$ hours. This is about 10 times faster than \textsf{Stan}, but took 
a lot longer to code. Lastly, we implemented streamlined Gibbs
sampling using the low-level \textsf{C++} language with the aid of
the \textsf{R} packages \textsf{Rcpp} (Eddelbuettal \textit{et al.}, 2019),
\textsf{RcppArmadillo} (Eddelbuettal \textit{et al.}, 2019) and 
\textsf{RcppDist} (Duck-Mayr, 2018). The coding time required by the
authors for this \textsf{C++} implementation was much longer than 
using \textsf{Stan}, but it resulted in a fitting time of just 4.9 minutes. 
Compared with \textsf{Stan}, the quality of the chains produced by 
these streamlined Gibbs sampling implementations is not as high and
larger warmup and kept sample sizes may be warranted in practice.

Table \ref{tab:CPPtimings} summarizes all of the timings
for this example. It shows that, depending on how 
Markov chain Monte Carlo is implemented, Bayesian 
linear mixed model analysis of the Collaborative Perinatal 
Project data is between several thousand times and a dozen times 
slower than streamlined variational inference. 

\begin{table}
\begin{center}
\begin{tabular}{llc}
\hline\\[-2.5ex]
approach                      &  computing time         & MCMC/(streamlined MFVB)\\
\hline\\[-1.5ex]
MCMC via \textsf{rstan}       &    36  hours     &   5,400    \\[0.5ex]
MCMC via \textsf{R} code      &     3.5 hours    &    514     \\[0.5ex]
MCMC via \textsf{C++} code    &    4.9 minutes   &   12.3     \\[0.5ex]
streamlined MFVB              &   24 seconds     &    ---     \\           
\hline
\end{tabular}
\end{center}
\caption{\textit{Computing times for four different approaches to approximate
Bayesian fitting of (\ref{eq:CPPmodel}) to the Collaborative Perinatal Project data.
The first three approaches are Markov chain Monte Carlo (MCMC) with a warmup
of length 1,000 and then 1,000 retained samples. The last approach is 
streamlined mean field variational Bayes (MFVB) with 100 iterations.
The ratios of the MCMC computing times to that of streamlined MFVB are
also shown.
}}
\label{tab:CPPtimings}
\end{table}

\section{Closing Remarks}\label{sec:closing}

We have provided comprehensive coverage of streamlined mean field variational Bayes
and variational message passing for two-level and three-level Gaussian response linear mixed
models. There are numerous extensions which cannot fit into a single article. 
One is the addition of penalized spline terms as treated in Lee \myand Wand (2016).
Another is non-Gaussian likelihood fragments. Group specific curve models
(e.g. Durban \textit{et al.}, 2005) also lend themselves to streamlining via
the \SolveTwoLevelSparseLeastSquares\ and \SolveThreeLevelSparseLeastSquares\ 
algorithms and Menictas \textit{et al.} (2019) provide full details.
Lastly, there are Gaussian response linear mixed models with more than
two levels of nesting. The present article provides a blueprint for which
these various extensions can be resolved systematically.

\section*{Acknowledgments}

We are grateful to  Hon Hwang, Robert Kohn, Cathy Lee,  Luca Maestrini, Chris Oates, Louise Ryan, 
the editor and two reviewers for their valuable contributions.
This research was supported by Australian Research Council Discovery Project 
DP180100597 and aided by the Knowledge integration project within the Bill and Melinda Gates 
Foundation.

\section*{References}

\bib
Atay-Kayis, A. \myand Massam, H. (2005).
A Monte Carlo method for computing marginal likelihood
in nondecomposable Gaussian graphical models.
\textit{Biometrika}, {\bf 92}, 317--335.

\bib
Baltagi, B.H. (2013). 
\textit{Econometric Analysis of Panel Data, Fifth Edition}.
Chichester, U.K.: John Wiley \& Sons.

\bib
Bishop, C.M. (2006). {\it Pattern Recognition and Machine Learning}.
New York: Springer.

\bib
Duck-Mayr, J.B. (2018). \textsf{RcppDist}: \textsf{Rcpp} integration of 
additional probability distributions. \textsf{R} package version 0.1.1.
\texttt{https://CRAN.R-project.org}

\bib
Durban, M., Harezlak, J., Wand, M.P. \myand Carroll, R.J. (2005).
Simple fitting of subject-specific curves for longitudinal data.
{\it Statistics in Medicine}, {\bf 24}, 1153--1167.

\bib
Edelbeuttel, D., Francois, R., Allaire, J.J., Ushey, K.,
Kou, Q., Russell, N., Bates, D. and Chambers, J. (2019). 
\textsf{Rcpp}: Seamless \textsf{R} and \textsf{C++}
integration. \textsf{R} package version 1.0.3.
\texttt{http://www.rcpp.org}

\bib
Edelbeuttel, D., Francois, R., Bates, D. and Ni, Binxiang. (2019). 
\textsf{RcppArmadillo}: \textsf{Rcpp} integration for the \textsf{Armadillo}
templated linear algebra library. \textsf{R} package version 0.9.800.1.0.
\texttt{https://CRAN.R-project.org}

\bib
Fitzmaurice, G., Davidian, M., Verbeke, G. \myand Molenberghs, G. (Editors) (2008).
{\it Longitudinal Data Analysis}. Boca Raton, Florida: Chapman \& Hall/CRC.

\bib
Gentle, J.E. (2007). \textit{Matrix Algebra}.
New York: Springer.

\bib
Goldstein, H. (2010). \textit{Multilevel Statistical Models, Fourth 
Edition}. Chichester, U.K.: John Wiley \& Sons. 

\bib
Golub, G.H. \myand Van Loan, C.F. (1989). \textit{Matrix Computations}, 
Baltimore: The Johns Hopkins University Press.

\bib
Harville, D.A. (2008). \textit{Matrix Algebra from a Statistician's
Perspective}. New York: Springer.

\bib
Klebanoff, M.A. (2009). The Collaborative Perinatal Project:
a 50-year retrospective. \textit{Paediatric and Perinatal Epidemiology},
{\bf 23(1)}, 2--8.

\bib
Lee, C.Y.Y. \myand Wand, M.P. (2016).
Streamlined mean field variational Bayes for longitudinal
and multilevel data analysis. \textit{Biometrical Journal},
{\bf 58}, 868--895.

\bib
Maestrini, L. \myand Wand, M.P. (2020).
The Inverse G-Wishart distribution and variational message passing.
Unpublished manuscript available at\\ 
\texttt{https://arxiv.org/abs/2005.09876}

\bib
McCulloch, C.E., Searle, S.R. \myand Neuhaus, J.M. (2008).
\textit{Generalized, Linear, and Mixed Models}.
Hoboken, New Jersey: John Wiley \& Sons.

\bib
Menictas, M., Nolan, T.H., Simpson, D.G. \myand Wand, M.P. (2019).
Streamlined variational inference for higher level group-specific curve models.
\textit{Statistical Modelling}, in press.

\bib
Menictas, M. \myand Wand, M.P. (2013).
Variational inference for marginal longitudinal semiparametric
regression. \textit{Stat}, {\bf 2}, 61--71.

\bib
Minka, T. (2005). 
Divergence measures and message passing.
\textit{Microsoft Research Technical Report Series}, 
{\bf MSR-TR-2005-173}, 1--17.

\bib
Nolan, T.H. \myand Wand, M.P. (2020). Streamlined solutions to 
multilevel sparse matrix problems. \textit{ANZIAM Journal}, in press.

\bib
Nolan, T.H. (2020). \textit{Variational Bayesian Inference: Message Passing
Schemes and Streamlined Multilevel Data Analysis.}
Doctor of Philosophy thesis, University of Technology Sydney.

\bib
Pinheiro, J.C. \myand Bates, D.M. (2000).
\textit{Mixed-Effects Models in} \textsf{S} 
\textit{and} \textsf{S-PLUS}. New York: Springer-Verlag.

\bib
\textsf{R} Core Team (2020). \textsf{R}: A language and
environment for statistical computing. \textsf{R} Foundation
for Statistical Computing, Vienna, Austria.
\texttt{https://www.R-project.org/}.

\bib
\textsf{Stan} Development Team (2019). \textsf{RStan}: the \textsf{R}
interface to \textsf{Stan}. \textsf{R} package version 2.19.2.
\texttt{https://mc-stan.org/}.

\bib
Rao, J.N.K. \myand Molina, I. (2015).
\textit{Small Area Estimation, Second Edition}.
Hoboken, New Jersey: John Wiley \& Sons.

\bib
Robinson, G.K. (1991). That BLUP is a good thing: the estimation
of random effects. {\it Statistical Science}, {\bf 6}, 15--51.

\bib
Rue, H. \myand Held, L. (2005). \textit{Gaussian Markov Random Fields}.
Boca Raton, Florida: Chapman \& Hall/CRC.

\bib
Wand, M.P. (2017).
Fast approximate inference for arbitrarily large semiparametric
regression models via message passing (with discussion).
\textit{Journal of the American Statistical Association}, 
{\bf 112}, 137--168.

\bib
Wand, M.P. \myand Ormerod, J.T. (2011).
Penalized wavelets: embedding wavelets
into semiparametric regression.
{\it Electronic Journal of Statistics},
{\bf 5}, 1654--1717.

\bib
Wang, B. \myand Titterington, D.M. (2005).
Inadequacy of interval estimates corresponding to variational 
Bayesian approximations. 
In {\it Proceedings of the 10th International Workshop
of Artificial Intelligence and Statistics}
(eds. R.G. Cowell \myand Z. Ghahramani), 373--380.

\null\vfill\eject

%
%
\appendix
\renewcommand{\thealgorithm}{A.\arabic{algorithm}}
\setcounter{algorithm}{0}


\centerline{\Large\textbf{Appendix}}

\section{Multilevel Sparse Matrix Problem Algorithms}\label{sec:matAlgs}

Algorithms \ref{alg:twoLevelMFVB}--\ref{alg:threeLevelVMPpen} rely on
four fundamental matrix algebraic algorithms that solve the two-level
and three-level versions of \emph{multilevel sparse matrix problems}.
This class of problems are defined in Nolan \myand Wand (2020).
These four algorithms:
\begin{center}
\begin{tabular}{lll}
\SolveTwoLevelSparseMatrix         &\ \ \ \ \ &Algorithm \ref{alg:SolveTwoLevelSparseMatrix} \\ 
\SolveTwoLevelSparseLeastSquares   &\ \ \ \ \ &Algorithm \ref{alg:SolveTwoLevelSparseLeastSquares} \\ 
\SolveThreeLevelSparseMatrix       &\ \ \ \ \ &Algorithm \ref{alg:SolveThreeLevelSparseMatrix} \\  
\SolveThreeLevelSparseLeastSquares &\ \ \ \ \ &Algorithm \ref{alg:SolveThreeLevelSparseLeastSquares} 
\end{tabular}
\end{center}
and their underpinnings are presented in this appendix.

\subsection{Two-Level Sparse Matrix Algorithms}\label{sec:twoLevSMA}

Two-level sparse matrix problems are described in Section 2 of 
Nolan \myand Wand (2020). The notation used there is also used
in this section. Here we present two algorithms, named 

\vskip5pt
\centerline{\SolveTwoLevelSparseMatrix\quad and\quad \SolveTwoLevelSparseLeastSquares}
\vskip5pt

\noindent
which are at the heart of streamlining variational inference for two-level models.

The \SolveTwoLevelSparseMatrix\ algorithm is concerned with solving general 
two-level sparse linear system problem $\bA\bx=\ba$, where
\begin{equation}
\AtLev\equiv
\left[
\arraycolsep=2.2pt\def\arraystretch{1.6}
\begin{array}{c|c|c|c|c}
\setstretch{4.5}
\ALoo     & \ALotCo & \ALotCt  &\ \ \cdots\ \ &\ALotCm \\
\hline
\ALotCo^T & \ALttCo & \bO      & \cdots   & \bO    \\
\hline
\ALotCt^T & \bO     & \ALttCt  & \cdots   & \bO    \\ 
\hline
\vdots    & \vdots  & \vdots   & \ddots   & \vdots   \\
\hline
\ALotCm^T & \bO     & \bO      & \cdots   &\ALttCm \\ 
\end{array}
\right],
\quad
\ba\equiv
\left[
\arraycolsep=2.2pt\def\arraystretch{1.6}
\begin{array}{c}
\setstretch{4.5}
\aveco     \\
\hline
\avectCo \\
\hline
\avectCt \\ 
\hline
\vdots    \\
\hline
\avectCm \\ 
\end{array}
\right]
\qquad\mbox{and}\qquad
\bx\equiv
\left[
\arraycolsep=2.2pt\def\arraystretch{1.6}
\begin{array}{c}
\setstretch{4.5}
\xveco     \\
\hline
\xvectCo \\
\hline
\xvectCt \\ 
\hline
\vdots    \\
\hline
\xvectCm \\ 
\end{array}
\right]
\label{eq:AtLevxa}
\end{equation}
and obtaining the sub-matrices corresponding to the non-zero blocks of $\AtLev$:
\begin{equation}
\AtLev^{-1}\equiv
\left[
\arraycolsep=2.2pt\def\arraystretch{1.6}
\begin{array}{c|c|c|c|c}
\setstretch{4.5}
\AUoo     & \AUotCo & \AUotCt  &\ \ \cdots\ \ &\AUotCm \\
\hline
\AUotCoT & \AUttCo & \bigX      & \cdots   & \bigX    \\
\hline
\AUotCtT & \bigX     & \AUttCt  & \cdots   & \bigX    \\ 
\hline
\vdots    & \vdots  & \vdots   & \ddots   & \vdots   \\
\hline
\AUotCmT & \bigX     & \bigX      & \cdots   &\AUttCm \\ 
\end{array}
\right].
\label{eq:AtLevInv}
\end{equation}
As will be elaborated upon later, the blocks represented by the $\bigX$ symbol 
are not of interest. \SolveTwoLevelSparseMatrix\ is listed as 
Algorithm \ref{alg:SolveTwoLevelSparseMatrix} and is justified by 
Theorem 2.2 of Nolan \myand Wand (2020).

\begin{algorithm}[!th]
\begin{center}
\begin{minipage}[t]{154mm}
\begin{small}
\begin{itemize}
\setlength\itemsep{4pt}
\item[] Inputs: $\Big(\aveco(p\times1),\ALoo(p\times p),\,
\big\{\big(\avectCi(q\times 1),\ALttCi(q\times q),\ALotCi(p\times q)\big):\ 1\le i\le m\big\}$
\item[] $\bomegaAlgAoneA\thickarrow\ba_1$\ \ \ ;\ \ \ $\bOmegaAlgAoneB\thickarrow\bA_{11}$
\item[] For $i=1,\ldots,m$:
\begin{itemize}
\setlength\itemsep{4pt}
\item[]$\bomegaAlgAoneA\thickarrow\bomegaAlgAoneA-\ALotCi\ALttCi^{-1}\avectCi$
\ \ \ ;\ \ \ 
$\bOmegaAlgAoneB\thickarrow\bOmegaAlgAoneB-\ALotCi\ALttCi^{-1}\ALotCi^T$
\end{itemize}
\item[] $\AUoo\thickarrow\bOmegaAlgAoneB^{-1}$
\ \ \ ;\ \ \ $\xveco\thickarrow\AUoo\bomegaAlgAoneA$
\item[] For $i=1,\ldots,m$:
\begin{itemize}
\setlength\itemsep{4pt}
\item[] $\xvectCi\thickarrow\ALttCi^{-1}(\avectCi-\ALotCi^T\xveco)$\ \ \ ;\ \ \ 
$\AUotCi\thickarrow\,-(\ALttCi^{-1}\ALotCi^T\AUoo)^T$
\item[] $\AUttCi\thickarrow\,\ALttCi^{-1}\big(\bI-\ALotCi^T\bA^{12,i}\big)$ 
\end{itemize}
\item[] Output: $\Big(\xveco,\AUoo,\big\{\big(\xvectCi,\AUttCi,\AUotCi):\ 1\le i\le m\big\}\Big)$
\end{itemize}
\end{small}
\end{minipage}
\end{center}
\caption{\textit{The} \SolveTwoLevelSparseMatrix\ \textit{algorithm for solving the two-level sparse
matrix problem $\bx=\AtLev^{-1}\ba$ and sub-blocks of $\AtLev^{-1}$ corresponding to the non-zero
sub-blocks of $\AtLev$. The sub-block notation is given by (\ref{eq:AtLevxa}) and (\ref{eq:AtLevInv}).}}
\label{alg:SolveTwoLevelSparseMatrix} 
\end{algorithm}

The \SolveTwoLevelSparseLeastSquares\ algorithm arises in the special case where 
$\bx$ is the minimizer of the least squares problem 
$\Vert\bb-\bB\bx\Vert^2\equiv(\bb-\bB\bx)^T(\bb-\bB\bx)$ where the matrix
$\bB$ and vector $\bb$ have the generic forms
\begin{equation}
\bB\equiv
\left[
\arraycolsep=2.2pt\def\arraystretch{1.6}
\begin{array}{c|c|c|c|c}
\setstretch{4.5}
\Bmato            &\Bmatdoto           &\bO   &\cdots&\bO\\ 
\hline
\Bmatt            &\bO              &\Bmatdott&\cdots&\bO\\ 
\hline
\vdots            &\vdots           &\vdots           &\ddots&\vdots\\
\hline
\Bmatm &\bO       &\bO              &\cdots           &\Bmatdotm
\end{array}
\right]
\quad\mbox{and}\quad
\bb\equiv\left[
\arraycolsep=2.2pt\def\arraystretch{1.6}
\begin{array}{c}
\setstretch{4.5}
\bveco  \\ 
\hline
\bvect \\ 
\hline
\vdots \\
\hline
\bvecm \\
\end{array}
\right].
\label{eq:BandbForms}
\end{equation}
In this case $\AtLev=\bB^T\bB$, $\ba=\bB^T\bb$ so that the
sub-blocks of $\AtLev$ and $\ba$ take the forms
$$\ALoo=\sumim\Bmati^T\Bmati,\ \ \ALotCi=\Bmati^T\Bmatdoti,\ \ \ALttCi=\Bmatdoti^T\Bmatdoti,\ \
\aveco=\sumim\Bmati^T\bveci\ \ \mbox{and}\ \ \avectCi=\Bmatdoti^T\bveci.
$$
As demonstrated in Section \ref{sec:twoLevMods}, these forms arise in two-level random effects models.
Theorem 2.3 of Nolan \myand Wand (2020) shows that this special form lends itself
to a QR decomposition (e.g. Harville, 2008; Section 6.4.d) approach which has speed and stability 
advantages in regression settings (e.g. Gentle, 2007; Section 6.7.2).

\SolveTwoLevelSparseLeastSquares\ is listed as Algorithm \ref{alg:SolveTwoLevelSparseLeastSquares}.
Note that we use $\nadj_i$, rather than $n_i$, to denote the number of 
rows in each of $\bveci$, $\Bmati$ and $\Bmatdoti$ to avoid a notational
clash with common grouped data dimension notation as used in Section \ref{sec:twoLevMods}.
In the first loop over the $m$ groups of data the upper triangular matrices 
$\bR_i$, $1\le i\le m$, are obtained via QR-decomposition; a standard procedure
within most computing environments. Following that, all matrix equations involve
$\bR_i^{-1}$, which can be achieved rapidly via back-solving.

\begin{algorithm}[!th]
\begin{center}
\begin{minipage}[t]{154mm}
\begin{small}
\begin{itemize}
\setlength\itemsep{4pt}
\item[] Input: $\big\{\big(\bveci(\nadj_i\times1),
\ \Bmati(\nadj_i\times p),\ \Bmatdoti(\nadj_i\times q)\big):\ 1\le i\le m\big\}$
\item[] $\bomegaAlgAtwoA\thickarrow\mbox{NULL}$\ \ \ ;\ \ \ $\bOmegaAlgAtwoB\thickarrow\mbox{NULL}$
\item[] For $i=1,\ldots,m$:
\begin{itemize}
\setlength\itemsep{4pt}
\item[] Decompose $\Bmatdoti=\bQ_i\left[\begin{array}{c}   
\bR_i\\
\bO
\end{array}
\right]$ such that $\bQ_i^{-1}=\bQ_i^T$ and $\bR_i$ is upper-triangular.
\item[] $\cveczi\thickarrow\bQ_i^T\bveci\ \ \ ;\ \ \ \Cmatzi\thickarrow\bQ_i^T\Bmati$
\item[] $\cvecoi\thickarrow\mbox{first $q$ rows of}\ \cveczi$\ \ \ ;\ \ \ 
$\cvecti\thickarrow\mbox{remaining rows of}\ \cveczi$\ \ \ ;\ \ \ 
$\bomegaAlgAtwoA\thickarrow
\left[
\begin{array}{c}
\bomegaAlgAtwoA\\
\cvecti
\end{array}
\right]$
\item[]$\Cmatoi\thickarrow\mbox{first $q$ rows of}\ \Cmatzi$\ \ \ ;\ \ \ 
$\Cmatti\thickarrow\mbox{remaining rows of}\ \Cmatzi$\ \ \ ;\ \ \  
$\bOmegaAlgAtwoB\thickarrow
\left[
\begin{array}{c}
\bOmegaAlgAtwoB\\
\Cmatti
\end{array}
\right]$
\end{itemize}
\item[] Decompose $\bOmegaAlgAtwoB=\bQ\left[\begin{array}{c}   
\bR\\
\bO
\end{array}
\right]$ such that $\bQ^{-1}=\bQ^T$ and $\bR$ is upper-triangular.
\item[] $\bc\thickarrow\mbox{first $p$ rows of $\bQ^T\bomegaAlgAtwoA$}$
\ \ \ ;\ \ \ $\xveco\thickarrow\bR^{-1}\bc$\ \ \ ;\ \ \ 
$\AUoo\thickarrow\bR^{-1}\bR^{-T}$
\item[] For $i=1,\ldots,m$:
\begin{itemize}
\setlength\itemsep{4pt}
\item[] $\xvectCi\thickarrow\bR_i^{-1}(\bc_{1i}-\Cmatoi\xveco)$\ \ \ ;\ \ \ 
$\AUotCi\thickarrow\,-\AUoo(\bR_i^{-1}\Cmatoi)^T$
\item[] $\AUttCi\thickarrow\bR_i^{-1}(\bR_i^{-T} - \Cmatoi\AUotCi)$
\end{itemize}
\item[] Output: $\Big(\xveco,\AUoo,\big\{\big(\xvectCi,\AUttCi,\AUotCi):\ 1\le i\le m\big\}\Big)$
\end{itemize}
\end{small}
\end{minipage}
\end{center}
\caption{\SolveTwoLevelSparseLeastSquares\ \textit{for solving the two-level sparse matrix
least squares problem: minimise $\Vert\bb-\bB\,\bx\Vert^2$ in $\bx$ and sub-blocks of $\bA^{-1}$
corresponding to the non-zero sub-blocks of $\bA=\bB^T\bB$. The sub-block notation is
given by (\ref{eq:AtLevxa}), (\ref{eq:AtLevInv}) and (\ref{eq:BandbForms}).}}
\label{alg:SolveTwoLevelSparseLeastSquares} 
\end{algorithm}

Note that in Algorithm \ref{alg:SolveTwoLevelSparseLeastSquares} 
calculations such as $\bQ_i^T\bB_i$ do not require storage of $\bQ_i$ and use
of ordinary multiplication. Standard matrix algebraic programming languages store information
concerning $\bQ_i$ in a compact form from which matrices such as $\bQ_i^T\bB_i$ 
can be efficiently obtained.

\subsection{Three-Level Sparse Matrix Algorithms}\label{sec:threeLevSMA}

Extension to the three-level situation is described in Section 3 of 
Nolan \myand Wand (2020). Theorems 3.2 and 3.3 given there lead to 
the algorithms

\vskip5pt
\centerline{\SolveThreeLevelSparseMatrix\quad and\quad \SolveThreeLevelSparseLeastSquares}
\vskip5pt

\noindent
which facilitate streamlining variational inference for three-level models.

An illustrative three-level sparse matrix is:
\begin{equation}
\bA =
\left[ \arraycolsep=2.2pt\def\arraystretch{1.6} 
   \begin{array}{c | c | c | c | c | c | c | c}
   \setstretch{4.5}
   \A{11} & \A{12,1} & \A{12,11} & \A{12,12} & \A{12,2} & \A{12,21} & \A{12,22} & \A{12,23} \\
   \hline
   \A{12,1}^T & \A{22,1} & \A{12,1,1} & \A{12,1,2} & \bO & \bO & \bO & \bO \\
   \hline
   \A{12,11}^T & \A{12,1,1}^T & \A{22,11} & \bO & \bO & \bO & \bO & \bO \\
   \hline
   \A{12,12}^T & \A{12,1,2}^T & \bO & \A{22,12} & \bO & \bO & \bO & \bO \\
   \hline
   \A{12,2}^T & \bO & \bO & \bO & \A{22,2} & \A{12,2,1} & \A{12,2,2} & \A{12,2,3} \\
   \hline
   \A{12,21}^T & \bO & \bO & \bO & \A{12,2,1}^T & \A{22,21} & \bO & \bO \\
   \hline
   \A{12,22}^T & \bO & \bO & \bO & \A{12,2,2}^T & \bO & \A{22,22} & \bO \\
   \hline
   \A{12,23}^T & \bO & \bO & \bO & \A{12,2,3}^T & \bO & \bO & \A{22,23}
\end{array} \right]
\label{eq:3levSubBlockNotat1}
\end{equation}
and corresponds to level 2 group sizes of $n_1=2$ and $n_2=3$,
and a level 3 group size of $m=2$. A general three-level sparse matrix $\bA$ 
consists of the following components:
\begin{itemize}
\item A $p \times p$ matrix $\A{11}$, which is designated the $(1,1)$-block position.
\item A set of partitioned matrices 
$\left\{\left[\begin{array}{c|c|c|c} \A{12,i} & \A{12,ij} 
& \dots & \A{12,in_i} \end{array} \right]: 1 \le i \le m \right\}$, 
which is designated the $(1, 2)$-block position. For each $1 \le i \le m$, $\A{12,i}$ 
is $p \times q_1$, and for each $1 \le j \le n_i$, $\A{12,ij}$ is $p \times q_2$.
\item A $(2,1)$-block, which is simply the transpose of the $(1,2)$-block.
\item A block diagonal structure along the $(2,2)$-block position, where each sub-block 
is a two-level sparse matrix, 
as defined in (\ref{eq:AtLevxa}). For each $1 \le i \le m$, $\A{22,i}$ is $q_1 \times q_1$, and for each 
$1 \le j \le n_i$, $\A{12,\iCOMMAj}$ is $q_1 \times q_2$ and $\A{22,ij}$ is $q_2 \times q_2$.
\end{itemize}
The three-level sparse linear system problem takes the form $\bA\,\bx=\ba$ 
where we partition the vectors $\ba$ and $\bx$ as follows:
\begin{equation}
   \ba \equiv
      \left[ \arraycolsep=2.2pt\def\arraystretch{1.6}
         \begin{array}{c}
         \setstretch{4.5}
         \ba_{1} \\
         \hline
         \ba_{2,1} \\
         \hline
         \ba_{2,11} \\
         \hline
         \ba_{2,12} \\
         \hline
         \ba_{2,2} \\
         \hline
         \ba_{2,21} \\
         \hline
         \ba_{2,22} \\
         \hline
         \ba_{2,23}
      \end{array} \right] \qquad
      \text{and} 
\qquad
   \bx \equiv
      \left[ 
         \arraycolsep=2.2pt\def\arraystretch{1.6}
         \begin{array}{c}
         \setstretch{4.5}
         \bx_{1} \\
         \hline
         \bx_{2,1} \\
         \hline
         \bx_{2,11} \\
         \hline
         \bx_{2,12} \\
         \hline
         \bx_{2,2} \\
         \hline
         \bx_{2,21} \\
         \hline
         \bx_{2,22} \\
         \hline
         \bx_{2,23}
      \end{array} \right].
\label{eq:3levSubBlockNotat2}
\end{equation}
Here $\ba_1$ and $\bx_1$ are $p \times 1$ vectors. 
Then, for each $1 \le i \le m$, $\ba_{2,i}$ and $\bx_{2,i}$ are $q_1 \times 1$ vectors.
Lastly, for each $1 \le i \le m$ and $1 \le j \le n_i$ the vectors  $\ba_{2,ij}$ and $\bx_{2,ij}$ 
have dimension $q_2 \times 1$.

The three-level sparse matrix inverse problem involves determination of the sub-blocks
of $\bA^{-1}$ corresponding to the non-zero sub-blocks of $\bA$. Our notation for
these sub-blocks is illustrated by 
\begin{equation}
\bA^{-1} =
\left[\arraycolsep=2.2pt\def\arraystretch{1.6}
   \begin{array}{c | c | c | c | c | c | c | c}
 \setstretch{4.5}
   \Ainv{11} & \Ainv{12,1} & \Ainv{12,11} & \Ainv{12,12} & \Ainv{12,2} & \Ainv{12,21} & \Ainv{12,22} & \Ainv{12,23} \\
   \hline
   \ATinv{12,1} & \Ainv{22,1} & \Ainv{12,1,1} & \Ainv{12,1,2} & \bigX & \bigX & \bigX & \bigX \\
   \hline
   \ATinv{12,11} & \ATinv{12,1,1} & \Ainv{22,11} & \bigX & \bigX & \bigX & \bigX & \bigX \\
   \hline
   \ATinv{12,12} & \ATinv{12,1,2} & \bigX & \Ainv{22,12} & \bigX & \bigX & \bigX & \bigX \\
   \hline
   \ATinv{12,2} & \bigX & \bigX & \bigX & \Ainv{22,2} & \Ainv{12,2,1} & \Ainv{12,2,2} & \Ainv{12,2,3} \\
   \hline
   \ATinv{12,21} & \bigX & \bigX & \bigX & \ATinv{12,2,1} & \Ainv{22,21} & \bigX & \bigX \\
   \hline
   \ATinv{12,22} & \bigX & \bigX & \bigX & \ATinv{12,2,2} & \bigX & \Ainv{22,22} & \bigX \\
   \hline
   \ATinv{12,23} & \bigX & \bigX & \bigX & \ATinv{12,2,3} & \bigX & \bigX & \Ainv{22,23}
\end{array} \right]
\label{eq:3levSubBlockNotat3}
\end{equation}
for the $m=2$, $n_1=2$ and $n_2=3$ case. 

\SolveThreeLevelSparseMatrix, which provides streamlined solutions for the general 
three-level sparse matrix problem, is listed as 
Algorithm \ref{alg:SolveThreeLevelSparseMatrix}.

\begin{algorithm}[!th]
\begin{center}
\begin{minipage}[t]{154mm}
\begin{small}
\begin{itemize}
\setlength\itemsep{4pt}
\item[] Input: $\Big(\ba_1(p\times1),\bA_{11}(p\times p),\,
\big\{\big(\ba_{2,i}(q_1\times1),\bA_{22,i}(q_1\times q_1),\bA_{12,i}(p\times q_1):\ 1\le i\le m\big\}$,\\
$\qquad\qquad\big\{\ba_{2,ij}(q_2\times1),\bA_{22,ij}(q_2\times q_2),\bA_{12,ij}(p\times q_2),
\bA_{12,\iCOMMAj}(q_1\times q_2)\big):\ 1\le i\le m,\ 1\le j\le n_i\big\}\Big)$.
\item[]$\bomegaAlgAthreeA\thickarrow\ba_1$\ \ \ ;\ \ \ $\bOmegaAlgAthreeB\thickarrow\bA_{11}$
\item[] For $i=1,\ldots,m$:
\begin{itemize}
\setlength\itemsep{4pt}
\item[] $\bh_{2,i}\thickarrow\ba_{2,i}$\ \ \ ;\ \ \ $\bH_{12,i}\thickarrow \bA_{12,i}$
\ \ \ ;\ \ \ $\bH_{22,i}\thickarrow \bA_{22,i}$
\item[] For $j=1,\ldots,n_i$:
\begin{itemize}
\setlength\itemsep{4pt}
\item[] $\bh_{2,i}\thickarrow\bh_{2,i} - \bA_{12,\iCOMMAj}\bA_{22,ij}^{-1}\ba_{2,ij}$\ \ \ ;\ \ \ 
$\bH_{12,i}\thickarrow\bH_{12,i} - \bA_{12,ij}\bA_{22,ij}^{-1}\bA_{12,\iCOMMAj}^T$\\[1ex]
$\bH_{22,i}\thickarrow\bH_{22,i} - \bA_{12,\iCOMMAj}\bA_{22,ij}^{-1}\bA_{12,\iCOMMAj}^T$
\item[]$\bomegaAlgAthreeA\thickarrow\bomegaAlgAthreeA-\bA_{12,ij}\bA_{22,ij}^{-1}\ba_{2,ij}$\ \ \ ;\ \ \  
$\bOmegaAlgAthreeB\thickarrow\bOmegaAlgAthreeB-\bA_{12,ij}\bA_{22,ij}^{-1}\bA_{12,ij}^T$
\end{itemize}
\item[] $\bomegaAlgAthreeA\thickarrow\bomegaAlgAthreeA-\bH_{12,i}\bH_{22,i}^{-1}\bh_{2,i}$\ \ \ ;\ \ \ 
$\bOmegaAlgAthreeB\thickarrow\bOmegaAlgAthreeB-\bH_{12,i}\bH_{22,i}^{-1}\bH_{12,i}^T$
\end{itemize}
\item[] $\bA^{11}\thickarrow \bOmegaAlgAthreeB^{-1}$\ \ \ ;\ \ \ $\bx_1\thickarrow\bA^{11}\bomegaAlgAthreeA$
\item[] For $i=1,\ldots,m$:
\begin{itemize}
\item[] $\bx_{2,i}\thickarrow\,\bH_{22,i}^{-1}(\bh_{2,i}-\bH_{12,i}^T\bx_1)$\ \ \ ;\ \ \ 
$\bA^{12,i}\thickarrow\,-(\bH_{22,i}^{-1}\bH_{12,i}^T\bA^{11})^T$
\item[] $\bA^{22,i}\thickarrow\,\bH_{22,i}^{-1}(\bI-\bH_{12,i}^T\bA^{12,i})$ 
\item[] For $j=1,\ldots,n_i$:
\begin{itemize}
\item[]$\bx_{2,ij}\thickarrow\bA_{22,ij}^{-1}\big(\ba_{2,ij}-\bA_{12,ij}^T\bx_1-\bA_{12,\iCOMMAj}^T\bx_{2,i}\big)$
\item[]$\bA^{12,ij}\thickarrow\, -\big\{\bA_{22,ij}^{-1}\big(\bA_{12,ij}^T\bA^{11}+\bA_{12,\iCOMMAj}^T\,\bA^{12,i\,T}\big)\big\}^T$ 
\item[]$\bA^{12,\iCOMMAj}\thickarrow\, -\big\{\bA_{22,ij}^{-1}\big(\bA_{12,ij}^T\,\bA^{12,i}+\bA_{12,\iCOMMAj}^T\,\bA^{22,i}\big)\big\}^T$ 
\item[]$\bA^{22,ij}\thickarrow \bA_{22,ij}^{-1}\big(\bI - \bA_{12,ij}^T\bA^{12,ij} - \bA_{12,\iCOMMAj}^T\bA^{12,\iCOMMAj}\big)$ 
\end{itemize}
\end{itemize}
\item[] Output: $\Big(\xveco,\AUoo,\big\{\big(\xvectCi,\AUttCi,\AUotCi):\ 1\le i\le m\big\},$\\
$\null\qquad\qquad\big\{\big(\bx_{2,ij},\bA^{22,ij},\bA^{12,ij},\bA^{12,\iCOMMAj},
\big):\ 1\le i\le m,\ 1\le j\le n_i\big\}\Big)$
\end{itemize}
\end{small}
\end{minipage}
\end{center}
\caption{\textit{The} \SolveThreeLevelSparseMatrix\ \textit{algorithm for solving the three-level sparse
matrix problem $\bx=\AtLev^{-1}\ba$ and sub-blocks of $\AtLev^{-1}$ corresponding to the non-zero
sub-blocks of $\AtLev$. The sub-block notation is given by (\ref{eq:3levSubBlockNotat1}),
(\ref{eq:3levSubBlockNotat2}) and (\ref{eq:3levSubBlockNotat3}).}}
\label{alg:SolveThreeLevelSparseMatrix} 
\end{algorithm}

\begin{algorithm}[!th]
\begin{center}
\begin{minipage}[t]{155mm}
\begin{small}
\begin{itemize}
\setlength\itemsep{4pt}
\item[] Input: $\big\{\big(\bb_{ij}(\oadj_{ij}\times1),
\ \bB_{ij}(\oadj_{ij}\times p),\ \bBdot_{ij}(\oadj_{ij}\times q_1),
\ \bBdotdot_{ij}(\oadj_{ij}\times q_2)\big):\ 1\le i\le m,\ 1\le j\le n_i\big\}$
\item[] $\bomegaAlgAfourA\thickarrow\mbox{NULL}$\ \ \ ;\ \ \ $\bOmegaAlgAfourB\thickarrow\mbox{NULL}$
\item[] For $i=1,\ldots,m$:
\begin{itemize}
\setlength\itemsep{4pt}
\item[]  $\bomegaAlgAfourC\thickarrow\mbox{NULL}$\ \ \ ;\ \ \ $\bOmegaAlgAfourD\thickarrow\mbox{NULL}$
\ \ \ ;\ \ \ $\bOmegaAlgAfourE\thickarrow\mbox{NULL}$
\item[] For $j=1,\ldots,n_i$:
\begin{itemize}
\setlength\itemsep{4pt}
\item[] Decompose $\bBdotdot_{ij}=\bQ_{ij}\left[\begin{array}{c}   
\bR_{ij}\\
\bO
\end{array}
\right]$ such that $\bQ_{ij}^{-1}=\bQ_{ij}^T$ and $\bR_{ij}$ is upper-triangular.
\item[] $\bd_{0ij}\thickarrow\bQ_{ij}^T\bb_{ij}\ \ \ ;\ \ \ \bD_{0ij}\thickarrow\bQ_{ij}^T\bB_{ij}
\ \ \ ;\ \ \ \bDdot_{0ij}\thickarrow\bQ_{ij}^T\bBdot_{ij}
$
\item[] $\bd_{1ij}\thickarrow\mbox{1st $q_2$ rows of}\ \bd_{0ij}$\ \ ;\ \ 
$\bd_{2ij}\thickarrow\mbox{remaining rows of}\ \bd_{0ij}$\ \ ;\ \ 
$\bomegaAlgAfourC\thickarrow
\left[
\begin{array}{c}
\bomegaAlgAfourC\\
\bd_{2ij}
\end{array}
\right]$
\item[]$\bD_{1ij}\thickarrow\mbox{1st $q_2$ rows of}\ \bD_{0ij}$\ ;\  
$\bD_{2ij}\thickarrow\mbox{remaining rows of}\ \bD_{0ij}$\ ;\   
$\bOmegaAlgAfourD\thickarrow
\left[
\begin{array}{c}
\bOmegaAlgAfourD\\
\bD_{2ij}
\end{array}
\right]$
\item[]$\bDdot_{1ij}\thickarrow\mbox{1st $q_2$ rows of}\ \bDdot_{0ij}$\ ;\ 
$\bDdot_{2ij}\thickarrow\mbox{remaining rows of}\ \bDdot_{0ij}$\ ;\  
$\bOmegaAlgAfourE\thickarrow
\left[
\begin{array}{c}
\bOmegaAlgAfourE\\
\bDdot_{2ij}
\end{array}
\right]$
\end{itemize}
\item[] Decompose $\bOmegaAlgAfourE=\bQ_i\left[\begin{array}{c}   
\bR_i\\
\bO
\end{array}
\right]$ such that $\bQ_i^{-1}=\bQ_i^T$ and $\bR_i$ is upper-triangular.
\item[] $\bc_{0i}\thickarrow\bQ_i^T\bomegaAlgAfourC\ \ \ ;\ \ \ \bC_{0i}\thickarrow\bQ_i^T\bOmegaAlgAfourD$
\item[] $\bc_{1i}\thickarrow\mbox{1st $q_1$ rows of}\ \bc_{0i}$\ \ ;\ \ 
$\bc_{2i}\thickarrow\mbox{remaining rows of}\ \bc_{0i}$\ \ ;\ \ 
$\bomegaAlgAfourA\thickarrow
\left[
\begin{array}{c}
\bomegaAlgAfourA\\
\bc_{2i}
\end{array}
\right]$
\item[] $\bC_{1i}\thickarrow\mbox{1st $q_1$ rows of}\ \bC_{0i}$\ \ ;\ \ 
$\bC_{2i}\thickarrow\mbox{remaining rows of}\ \bC_{0i}$\ \ ;\ \ 
$\bOmegaAlgAfourB\thickarrow
\left[
\begin{array}{c}
\bOmegaAlgAfourB\\
\bC_{2i}
\end{array}
\right]$
\end{itemize}
\item[] Decompose $\bOmegaAlgAfourB=\bQ\left[\begin{array}{c}   
\bR\\
\bO
\end{array}
\right]$ so that $\bQ^{-1}=\bQ^T$ and $\bR$ is upper-triangular.
\item[] $\bc\thickarrow\mbox{first $p$ rows of $\bQ^T\bomegaAlgAfourA$}$
\ \ \ ;\ \ \ $\xveco\thickarrow\bR^{-1}\bc$\ \ \ ;\ \ \ 
$\AUoo\thickarrow\bR^{-1}\bR^{-T}$
\item[] For $i=1,\ldots,m$:
\begin{itemize}
\setlength\itemsep{4pt}
\item[] $\xvectCi\thickarrow\bR_i^{-1}(\bc_{1i}-\bC_{1i}\xveco)$\ \ \ ;\ \ \ 
$\AUotCi\thickarrow\,-\AUoo(\bR_i^{-1}\Cmatoi)^T$
\item[] $\AUttCi\thickarrow\bR_i^{-1}(\bR_i^{-T} - \Cmatoi\AUotCi)$
\item[] For $j=1,\ldots,n_i$:
\begin{itemize}
\item[] $\bx_{2,ij}\leftarrow\bR_{ij}^{-1} (\bd_{1ij} - \bD_{1ij} \bx_1 - \bDdot_{1ij} \bx_{2,i})$
\item[] $\Ainv{12,ij}\leftarrow - \left\{ \bR_{ij}^{-1}(\bD_{1ij} \Ainv{11} + \bDdot_{1ij} \ATinv{12,i}) \right\}^T$
\item[] $\Ainv{12,\iCOMMAj}\leftarrow - \left\{\bR_{ij}^{-1}(\bD_{1ij} \Ainv{12,i} + \bDdot_{1ij} \Ainv{22,i}) \right\}^T$
\item[] $\Ainv{22,ij}\leftarrow\bR_{ij}^{-1}\big(\bR_{ij}^{-T}-\bD_{1ij}\Ainv{12,ij}-\bDdot_{1ij}\Ainv{12,\iCOMMAj}               \big)$

\end{itemize}
\end{itemize}
\item[] Output: $\Big(\xveco,\AUoo,\big\{\big(\xvectCi,\AUttCi,\AUotCi):\ 1\le i\le m\big\}\Big)$\\
$\null\qquad\qquad\big\{\big(\bx_{2,ij},\bA^{22,ij},\bA^{12,ij},\bA^{12,\iCOMMAj}
\big):\ 1\le i\le m,\ 1\le j\le n_i\big\}\Big)$
\end{itemize}
\end{small}
\end{minipage}
\end{center}
\caption{\SolveThreeLevelSparseLeastSquares\ \textit{for solving the three-level sparse matrix
least squares problem: minimise $\Vert\bb-\bB\,\bx\Vert^2$ in $\bx$ and sub-blocks of $\bA^{-1}$
corresponding to the non-zero sub-blocks of $\bA=\bB^T\bB$. The sub-block notation is
given by (\ref{eq:AtLevInv}).}}
\label{alg:SolveThreeLevelSparseLeastSquares} 
\end{algorithm}

Next, consider the special case where a three-level sparse matrix problem arises
as a least squares problem where $\bx$ is the minimizer of the least squares problem 
$\Vert\bb-\bB\bx\Vert^2\equiv(\bb-\bB\bx)^T(\bb-\bB\bx)$ where $\bB$ is such that
$\bA=\bB^T\bB$ has three-level sparse structure. For the special case of 
$m=2$, $n_1=2$ and $n_2=3$ the forms of the $\bB$ and $\bb$ matrices are
\begin{equation}
   \bB \equiv
      \left[\arraycolsep=2.2pt\def\arraystretch{1.6}
         \begin{array}{c | c | c | c | c | c | c | c}
         \setstretch{4.5}
         \B{11} & \dB{11} & \ddB{11} & \bO & \bO & \bO & \bO & \bO \\
         \hline
         \B{12} & \dB{12} & \bO & \ddB{12} & \bO & \bO & \bO & \bO \\
         \hline
         \B{21} & \bO & \bO & \bO & \dB{21} & \ddB{21} & \bO & \bO \\
         \hline
         \B{22} & \bO & \bO & \bO & \dB{22} & \bO & \ddB{22} & \bO \\
         \hline
         \B{23} & \bO & \bO & \bO & \dB{23} & \bO & \bO & \ddB{23}
      \end{array} \right]
   \quad \text{and} \quad
   \bb \equiv
      \left[\arraycolsep=2.2pt\def\arraystretch{1.6}
         \begin{array}{c}
         \setstretch{4.5}
         \bb_{11} \\
         \hline
         \bb_{12} \\
         \hline
         \bb_{21} \\
         \hline
         \bb_{22} \\
         \hline
         \bb_{23}
      \end{array} \right].
\label{eq:BmatsThreeLevIllus}
\end{equation}
For general $1\le i\le m$ and $1 \le j \le n_i$, the dimensions of the sub-blocks of $\bb$ and $\bB$ are:
\begin{equation}
\bb_{ij}\ \text{is}\ \oadj_{ij}\times 1,\quad
\B{ij}\ \text{is}\ \oadj_{ij}\times p,\quad\dB{ij}\ \text{is}\ \oadj_{ij}\times q_1,
\quad\mbox{and}\quad\ddB{ij}\ \text{is}\ \oadj_{ij}\ \times q_2.
\label{eq:BmatsThreeLev}
\end{equation}
Here we use $\oadj_{ij}$ rather than $o_{ij}$ to avoid a notational clash with 
common grouped data dimension notation as used in Section \ref{sec:threeLevMods}.
The general forms of $\bB$ and $\bb$ in the three-level case are
\begin{equation}
{\setlength\arraycolsep{0pt}
\begin{array}{ll}
&\bB\equiv\Big[{\displaystyle\stack{1\le i\le m}}\Big\{{\displaystyle\stack{1\le j\le n_i}}(\B{ij})\Big\}\ \Big\vert
{\displaystyle\blockdiag{1\le i\le m}}\Big\{\Big[{\displaystyle\stack{1\le j\le n_i}}(\dB{ij})\ \big\vert
{\displaystyle\blockdiag{1\le j\le n_i}}(\ddB{ij})\Big]\Big\} \Big]\\[2ex]
\mbox{and}\quad
&\bb\equiv{\displaystyle\stack{1\le i\le m}}\Big\{{\displaystyle\stack{1\le j\le n_i}}(\bb_{ij})\Big\}.
\end{array}
}
\label{eq:BandbGenThreeLev}
\end{equation}

Algorithm \ref{alg:SolveThreeLevelSparseLeastSquares} provides a QR decomposition-based
solution to the three-level sparse matrix least squares problems when the inputs
are the matrices listed in (\ref{eq:BmatsThreeLev}).

\section{Derivations}\label{sec:derivations}

\subsection{Derivation of Result \ref{res:twoLevelMFVB}}\label{sec:resTwoDeriv}

It is straightforward to verify that the $\bmu_{\qDens(\bbeta,\bu)}$ 
and $\bSigma_{\qDens(\bbeta,\bu)}$ updates, given at (\ref{eq:muSigmaMFVBupd}), may be written as
$$\bmu_{\qDens(\bbeta,\bu)}\thickarrow(\bB^T\bB)^{-1}\bB^T\bb=\bA^{-1}\ba
\quad\mbox{and}\quad
\bSigma_{\qDens(\bbeta,\bu)}\thickarrow(\bB^T\bB)^{-1}=\bA^{-1}
$$
where $\bB$ and $\bb$ have the forms (\ref{eq:BandbForms})
with
$$
\bveci\equiv
\left[
\begin{array}{c}
\mu_{\qDens(1/\mysigeps^2)}^{1/2}\by_i\\[1ex]
m^{-1/2}\bSigma_{\bbeta}^{-1/2}\bmu_{\bbeta}\\[1ex]
\bzero
\end{array}
\right],
\quad
\Bmati\equiv
\left[
\begin{array}{c}
\mu_{\qDens(1/\mysigeps^2)}^{1/2}\bX_i\\[1ex]
m^{-1/2}\bSigma_{\bbeta}^{-1/2}\\[1ex]
\bO
\end{array}
\right]
\quad\mbox{and}\quad
\Bmatdoti\equiv
\left[
\begin{array}{c}
\mu_{\qDens(1/\mysigeps^2)}^{1/2}\bZ_i\\[1ex]
\bO              \\[1ex]
\bM_{\qDens(\bSigma^{-1})}^{1/2}
\end{array}
\right].
$$

\subsection{Derivation of Algorithm \ref{alg:twoLevelMFVB}}\label{sec:DerivTwoLevMFVB}

We first provide expressions for the $\qDens$-densities for mean
field variational Bayesian inference for the parameters in (\ref{eq:twoLevelGaussRespBaye}),
with product density restriction (\ref{eq:producRestrict}). 
Arguments analogous to those given in, for example, Appendix C of 
Wand \myand Ormerod (2011) lead to: 
$$\qDens(\bbeta,\bu)\ \mbox{is a $N(\bmu_{\qDens(\bbeta,\bu)},\bSigma_{\qDens(\bbeta,\bu)})$ density function}$$
where
$$\bSigma_{\qDens(\bbeta,\bu)}=(\bC^T\RMFVB^{-1}\bC+\DMFVB)^{-1}
\quad
\mbox{and}
\quad
\bmu_{\qDens(\bbeta,\bu)}=\bSigma_{\qDens(\bbeta,\bu)}(\bC^T\RMFVB^{-1}\by + \oMFVB)
$$
with $\RMFVB$, $\DMFVB$ and $\oMFVB$ defined via (\ref{eq:MFVBmatDefns}),
$$\qDens(\mysigeps^2)\ \mbox{is an $\mbox{Inverse-$\chi^2$}
\left(\xi_{\qDens(\mysigeps^2)},\lambda_{\qDens(\mysigeps^2)}\right)$ density function}
$$
where $\xi_{\qDens(\mysigeps^2)}=\nu_{\sigma^2}+\sumim n_i$ and
\begin{eqnarray*}
\lambda_{\qDens(\mysigeps^2)}&=&\mu_{\qDens(1/\asigsq)}+\sumim\,E_{\qDens}\{\Vert\by_i-\bX_i\bbeta-\bZ_i\bu_i\Vert^2\}\\[1ex]
&=&\mu_{\qDens(1/\asigsq)}+\sumim\,\big[\Vert\,E_{\qDens}(\by_i-\bX_i\bbeta-\bZ_i\bu_i)\Vert^2
+\tr\{\Cov_\qDens(\bX_i\bbeta+\bZ_i\bu_i)\}\big]\\[1ex]
&=&\mu_{\qDens(1/\asigsq)}+\sumim\,\Big(\Vert\,E_{\qDens}(\by_i-\bX_i\bbeta-\bZ_i\bu_i)\Vert^2
+\tr(\bX_i^T\bX_i\bSigma_{\qDens(\bbeta)})+\tr(\bZ_i^T\bZ_i\bSigma_{\qDens(\bu_i)})\\
&&\qquad\qquad\qquad +2\,\mbox{tr}\big[\bZ_i^T\bX_iE_{\qDens}\{(\bbeta-\bmu_{\qDens(\bbeta)})
                 (\bu_i-\bmuq{\bu_i})^T\}\big]\Big)
\end{eqnarray*}
with reciprocal moment $\mu_{\qDens(1/\mysigeps^2)}=\xi_{\qDens(\mysigeps^2)}/\lambda_{\qDens(\mysigeps^2)},$
$$\qDens(\bSigma)\ \mbox{is an $\mbox{Inverse-G-Wishart}
\left(\Gfull,\xi_{\qDens(\bSigma)},\bLambda_{\qDens(\bSigma)}\right)$ density function}
$$
where $\xi_{\qDens(\bSigma)}=\nuSigma+2q-2+m$ and
$$\bLambda_{\qDens(\bSigma)}=\bM_{\qDens(\bA_{\bSigma}^{-1})}
+\sumim\left(\bmu_{\qDens(\bu_i)}\bmu_{\qDens(\bu_i)}\trans + \bSigma_{\qDens(\bu_i)}\right)$$
with inverse moment $\bM_{\qDens(\bSigma^{-1})}=(\xi_{\qDens(\bSigma)}-q+1)\bLambda_{\qDens(\bSigma)}^{-1}$,
$$\qDens(\asigsq)\ \mbox{is an $\mbox{Inverse-$\chi^2$}
(\xi_{\qDens(\asigsq)},\lambda_{\qDens(\asigsq)})$ density function}$$
where $\xi_{\qDens(\asigsq)}=\nusigsq+1$,
$$\lambda_{\qDens(\asigsq)}=\mu_{\qDens(1/\mysigeps^2)}+1/(\nusigsq\ssigsq^2)$$
with reciprocal moment $\mu_{\qDens(1/\asigsq)}=\xi_{\qDens(\asigsq)}/\lambda_{\qDens(\asigsq)}$ and 
$$\qDens(\ASigma)\ \mbox{is an $\mbox{Inverse-G-Wishart}
\left(\Gdiag,\xi_{\qDens(\ASigma)},\bLambda_{\qDens(\ASigma)}\right)$ density function}
$$
where $\xi_{\qDens(\ASigma)}=\nuSigma+q$,
$$\bLambda_{\qDens(\ASigma)}=\diag\big\{\mbox{diagonal}\big(\bM_{\qDens(\bSigma^{-1})}\big)\big\}
+\bLambda_{\ASigma}$$
with inverse moment $\bM_{\qDens(\ASigma^{-1})}=\xi_{\qDens(\ASigma)}\bLambda_{\qDens(\ASigma)}^{-1}$.

The $\qDens$-density parameters are interdependent and their Kullback-Leibler divergence
optimal values can be found via a coordinate ascent iterative algorithm,
which corresponds to Algorithm 2 of Lee \myand Wand (2016) for the special
case of $L=0$ in the notation used there. However, as explained there, 
na\"{\i}ve updating of $\bmu_{\qDens(\bbeta,\bu)}$ and $\bSigma_{\qDens(\bbeta,\bu)}$ 
has massive computational and storage costs when the number of groups is large.
Result \ref{res:twoLevelMFVB} asserts that we can instead use 
\SolveTwoLevelSparseLeastSquares\ (Algorithm \ref{alg:SolveTwoLevelSparseLeastSquares})
to obtain $\bmu_{\qDens(\bbeta,\bu)}$ and relevant sub-blocks of 
$\bSigma_{\qDens(\bbeta,\bu)}$.

\subsection{Derivation of Result \ref{res:twoLevelVMPlik}}

Note that
{\setlength\arraycolsep{1pt}
\begin{eqnarray*}
\qDens(\bbeta,\bu)&\propto&\mSUBpybetausigsqEpsTObetau\,\mSUBbetauTOpybetausigsqEps\\[1ex]
&=&\exp\left\{
\left[
{\setlength\arraycolsep{0pt}
\begin{array}{c}
\ \ \ \ \ \ \ \ \ \  \bbeta\\
\ \ \ \ \ \vech(\bbeta\bbeta^T)\\[1ex]
{\displaystyle\stack{1\le i\le m}}\left[
{\setlength\arraycolsep{0pt}
\begin{array}{c}
\bu_i\\
\vech(\bu_i\bu_i^T)\\
\vecof(\bbeta\bu_i^T)
\end{array}
}
\right]
\end{array}
}
\right]^T\null\hskip-3mm\etaSUBpybetausigsqEpsCONNbetau
\right\}\\[1ex]
&=&\exp\left\{
\left[
\begin{array}{l}
\bbeta\\[1ex]
\bu
\end{array}
\right]^T
\ba
-\smhalf
\left[
\begin{array}{l}
\bbeta\\[1ex]
\bu
\end{array}
\right]^T\bA
\left[
\begin{array}{l}
\bbeta\\[1ex]
\bu
\end{array}
\right]
\right\}\\[1ex]
\end{eqnarray*}
}
where $\ba$ and $\bA$ as given in Result \ref{res:twoLevelVMPlik}
and the last step uses facts such as $\vech(\bM)=\bD_d^+\vecof(\bM)$
for any symmetric $d\times d$ matrix $\bM$. Standard manipulations 
then lead to
$$\bmu_{\qDens(\bbeta,\bu)}=\bA^{-1}\ba
\quad\mbox{and}\quad
\bSigma_{\qDens(\bbeta,\bu)}=\bA^{-1}.
$$
Result \ref{res:twoLevelVMPlik} then follows from 
extraction of the sub-blocks of $\bx=\bA^{-1}\ba$
and the important sub-blocks of $\bA^{-1}$ according to
(\ref{eq:CovMFVB}).

\subsection{Derivation of Algorithm \ref{alg:TwoLevelNaturalToCommonParameters}}\label{sec:DerivTwoLevNatToComm}

The two-level reduced exponential family form is 
{\setlength\arraycolsep{3pt}
\begin{eqnarray*}
\qDens(\bbeta,\bu)&\propto&
\exp\left\{
\left[
{\setlength\arraycolsep{0pt}
\begin{array}{c}
\ \ \ \ \ \ \ \ \ \  \bbeta\\
\ \ \ \ \ \vech(\bbeta\bbeta^T)\\[1ex]
{\displaystyle\stack{1\le i\le m}}\left[
{\setlength\arraycolsep{0pt}
\begin{array}{c}
\bu_i\\
\vech(\bu_i\bu_i^T)\\
\vecof(\bbeta\bu_i^T)
\end{array}
}
\right]
\end{array}
}
\right]^T\null\hskip-3mm\bdeta_{\qDens(\bbeta,\bu)}
\right\}\\[1ex]
&=&
\exp\left\{
\left[
\begin{array}{l}
\bbeta\\[1ex]
\bu
\end{array}
\right]^T
\ba
-\smhalf
\left[
\begin{array}{l}
\bbeta\\[1ex]
\bu
\end{array}
\right]^T\bA
\left[
\begin{array}{l}
\bbeta\\[1ex]
\bu
\end{array}
\right]
\right\}
\end{eqnarray*}
}
where $\bA$ and $\ba$ are as defined in Result \ref{res:twoLevelVMPlik} with 
$\etaSUBpybetausigsqEpsCONNbetau$ replaced by $\bdeta_{\qDens(\bbeta,\bu)}$
with $\bA$ having two-level sparse structure.
As with the derivation of Result \ref{res:twoLevelVMPlik}, we have the
relationships
\begin{equation}
\bmu_{\qDens(\bbeta,\bu)}=\bA^{-1}\ba
\quad\mbox{and}\quad
\bSigma_{\qDens(\bbeta,\bu)}=\bA^{-1}.
\label{eq:muSigmaAgain}
\end{equation}
The first part of Algorithm \ref{alg:TwoLevelNaturalToCommonParameters} is such that
the entries of $\etaSUBqbetau$ are sequentially unpacked and stored in
the vectors $\bomegaAlgTwoA$ and $\bomegaAlgTwoD$, $1\le i\le m$, corresponding
to the $\ba$ vector according to the partitioning in (\ref{eq:AtLevxa}) and the matrices 
$\bOmegaAlgTwoC$ and $\bOmegaAlgTwoG,\bOmegaAlgTwoH$, $1\le i\le m$, 
corresponding to the non-zero sub-blocks of $\bA$ in (\ref{eq:AtLevxa}).

Next, $\SscAlgTwo$ stores the streamlined solution to (\ref{eq:muSigmaAgain})
according to the \SolveTwoLevelSparseMatrix\ algorithm
(Algorithm \ref{alg:SolveTwoLevelSparseMatrix}).
The remainder of Algorithm \ref{alg:TwoLevelNaturalToCommonParameters}
is plucking off the relevant common parameter sub-blocks of 
$\bmu_{\qDens(\bbeta,\bu)}$ and $\bSigma_{\qDens(\bbeta,\bu)}$
based (\ref{eq:muSigmaAgain}) and keeping in mind that
(\ref{eq:muSigmaAgain}) represents a two-level sparse matrix problem.

\subsection{Derivation of Algorithm \ref{alg:twoLevelVMPlik}}\label{sec:drvTwoLevLikFrag}

First note that the logarithm of the fragment factor is, as a function of $(\bbeta,\bu)$: 
{\setlength\arraycolsep{1pt}
\begin{eqnarray*}
\log\,\pDens(\by|\bbeta,\bu,\mysigeps^2)&=&\,-\frac{1}{2\mysigeps^2}\sumim\Vert\by_i-\bX_i\bbeta-\bZ_i\bu_i\Vert^2+\const\\[1ex]
&=&(1/\mysigeps^2)\left[
{\setlength\arraycolsep{0pt}
\begin{array}{c}
\bbeta\\[2ex]
\vech(\bbeta\bbeta^T)\\[2ex]
{\displaystyle\stack{1\le i\le m}}
\left[
{\setlength\arraycolsep{0pt}
\begin{array}{c}
\bu_i\\[2ex]
\vech(\bu_i\bu_i^T)\\[2ex]
\vecof(\bbeta\bu_i^T)
\end{array}
}
\right]
\end{array}
}
\right]^T
\left[
{\setlength\arraycolsep{0pt}
\begin{array}{c}
{\displaystyle\sumim}\bX_i^T\by_i\\[1ex]
-\smhalf{\displaystyle\sumim}\bD_p^T\vecof(\bX_i^T\bX_i)\\[3ex]
{\displaystyle\stack{1\le i\le m}}
\left[
{\setlength\arraycolsep{0pt}
\begin{array}{c}
\bZ_i^T\by_i\\[1ex]
-\smhalf\bD_q^T\vecof(\bZ_i^T\bZ_i)\\[1ex]
-\vecof(\bX_i^T\bZ_i)
\end{array}
}
\right]
\end{array}
}
\right]+\mbox{const}.
\end{eqnarray*}
}
Therefore, from equations (8) and (9) of Wand (2017),
$$\mSUBpybetausigsqEpsTObetau\thickarrow
\exp\left\{
\left[
{\setlength\arraycolsep{0pt}
\begin{array}{c}
\bbeta\\[2ex]
\vech(\bbeta\bbeta^T)\\[2ex]
{\displaystyle\stack{1\le i\le m}}
\left[
{\setlength\arraycolsep{0pt}
\begin{array}{c}
\bu_i\\[2ex]
\vech(\bu_i\bu_i^T)\\[2ex]
\vecof(\bbeta\bu_i^T)
\end{array}
}
\right]
\end{array}
}
\right]^T
\etaSUBpybetausigsqEpsTObetau
\right\}
$$
where
$$\etaSUBpybetausigsqEpsTObetau\equiv\mu_{\qDens(1/\mysigeps^2)}\,\left[
{\setlength\arraycolsep{0pt}
\begin{array}{l}
\ \ \ \ \ \ \ \ \ \ {\displaystyle\sumim}\bX_i^T\by_i\\[1ex]
\ \ \ \ \ -\smhalf{\displaystyle\sumim}\bD_p^T\vecof(\bX_i^T\bX_i)\\[3ex]
{\displaystyle\stack{1\le i\le m}}
\left[
{\setlength\arraycolsep{0pt}
\begin{array}{c}
\bZ_i^T\by_i\\[1ex]
-\smhalf\bD_q^T\vecof(\bZ_i^T\bZ_i)\\[1ex]
-\vecof(\bX_i^T\bZ_i)
\end{array}
}
\right]
\end{array}
}
\right]
$$
and $\mu_{\qDens(1/\mysigeps^2)}$ denotes expectation of $1/\mysigeps^2$ with respect to 
the normalization of 
$$\mSUBpybetausigsqEpsTOsigsqEps\,\mSUBsigsqEpsTOpybetausigsqEps$$
which is an Inverse $\chi^2$ density function with natural parameter vector
$\etaSUBpybetausigsqEpsCONNsigsqEps$ and, according to Table S.1 in 
the online supplement of Wand (2017), leads to 
$$\mu_{\qDens(1/\mysigeps^2)}\thickarrow\Big(\big(\etaSUBpybetausigsqEpsCONNsigsqEps\big)_1+1\Big)/
\big(\etaSUBpybetausigsqEpsCONNsigsqEps\big)_2.
$$

The other factor to stochastic node message update is
$$
\mSUBpybetausigsqEpsTOsigsqEps
\thickarrow\exp\left\{\left[
\begin{array}{c}
\log(\mysigeps^2)\\[2ex]
1/\mysigeps^2
\end{array}
\right]^T\etaSUBpybetausigsqEpsTOsigsqEps
\right\}
$$
where
$$
\etaSUBpybetausigsqEpsTOsigsqEps
\equiv
\left[
\begin{array}{c}
-\smhalf{\displaystyle\sumim}\, n_i\\[3ex]
-\smhalf{\displaystyle\sumim}\,E_{\qDens}\{\Vert\by_i-\bX_i\bbeta-\bZ_i\bu_i\Vert^2\}
\end{array}
\right]
$$
with $E_{\qDens}$ denoting expectation with respect to the normalization of 
$$\mSUBpybetausigsqEpsTObetau\,\mSUBbetauTOpybetausigsqEps.$$
Then note that 
{\setlength\arraycolsep{1pt}
\begin{eqnarray*}
E_{\qDens}\{\Vert\by_i-\bX_i\bbeta-\bZ_i\bu_i\Vert^2\}
&=&\Vert\by_i-\bX_i\,\bmu_{\qDens(\bbeta)}-\bZ_i\,\bmu_{\qDens(\bu_i)}\Vert^2
+\tr(\bX_i^T\bX_i\bSigma_{\qDens(\bbeta)})\\
&&\,+\tr(\bZ_i^T\bZ_i\bSigma_{\qDens(\bu_i)})
+2\,\mbox{tr}\big[\bZ_i^T\bX_iE_{\qDens}\{(\bbeta-\bmu_{\qDens(\bbeta)})
                 (\bu_i-\bmuq{\bu_i})^T\}\big]
\end{eqnarray*}
}
where, for example, $\bmu_{\qDens(\bbeta)}\equiv E_{\qDens}(\bbeta)$ and $\bSigma_{\qDens(\bu_i)}\equiv\Cov_\qDens(\bu_i)$.
Result \ref{res:twoLevelVMPlik} links sub-blocks of $\etaSUBpybetausigsqEpsCONNbetau$ with the required sub-vectors 
of $\bmu_{\qDens(\bbeta,\bu)}$ and sub-blocks of $\bSigma_{\qDens(\bbeta,\bu)}$. These matrices are
extracted from  $\etaSUBpybetausigsqEpsCONNbetau$ in the call to
\TwoLevelNaturalToCommonParameters\ algorithm (Algorithm \ref{alg:TwoLevelNaturalToCommonParameters}).

\subsection{Derivation of Result \ref{res:twoLevelVMPpen}}

The derivation of Result \ref{res:twoLevelVMPpen} is very 
similar to that for Result \ref{res:twoLevelVMPlik}.

\subsection{Derivation of Algorithm \ref{alg:twoLevelVMPpen}}\label{sec:drvTwoLevPenFrag}

The logarithm on the fragment factor is, as a function of $(\bbeta,\bu)$: 
{\setlength\arraycolsep{1pt}
\begin{eqnarray*}
\log\,\pDens(\bbeta,\bu|\,\bSigma)&=&\,-\smhalf(\bbeta-\bmu_{\bbeta})^T\bSigma_{\bbeta}^{-1}(\bbeta-\bmu_{\bbeta})
-\frac{1}{2}\sumim\bu_i^T\bSigma^{-1}\bu_i+\const\\[1ex]
&=&\left[
{\setlength\arraycolsep{0pt}
\begin{array}{l}
\ \ \ \ \ \ \ \ \ \  \bbeta\\[2ex]
\ \ \ \ \ \vech(\bbeta\bbeta^T)\\[2ex]
{\displaystyle\stack{1\le i\le m}}\left[
{\setlength\arraycolsep{0pt}
\begin{array}{c}
\bu_i\\[2ex]
\vech(\bu_i\bu_i^T)\\[2ex]
\vecof(\bbeta\bu_i^T)
\end{array}
}
\right]
\end{array}
}
\right]^T
\left[
{\setlength\arraycolsep{0pt}
\begin{array}{c}
\bSigma_{\bbeta}^{-1}\bmu_{\bbeta}\\[2ex]
-\smhalf\bD_p^T\vecof(\bSigma_{\bbeta}^{-1})\\[2ex]
{\displaystyle\stack{1\le i\le m}}\left[
{\setlength\arraycolsep{0pt}
\begin{array}{c}
\bzero_q\\[2ex]
-\smhalf\bD_q^T\vecof(\bSigma^{-1})\\[2ex]
\bzero_{pq}
\end{array}
}
\right]
\end{array}
}
\right]+\mbox{const}.
\end{eqnarray*}
}
Therefore, from equations (8) and (9) of Wand (2017),
$$\mSUBpbetauSigmaTObetau\thickarrow
\exp\left\{
\left[
{\setlength\arraycolsep{0pt}
\begin{array}{c}
\bbeta\\[2ex]
\vech(\bbeta\bbeta^T)\\[2ex]
{\displaystyle\stack{1\le i\le m}}
\left[
{\setlength\arraycolsep{0pt}
\begin{array}{c}
\bu_i\\[2ex]
\vech(\bu_i\bu_i^T)\\[2ex]
\vecof(\bbeta\bu_i^T)
\end{array}
}
\right]
\end{array}
}
\right]^T
\etaSUBpbetauSigmaTObetau
\right\}
$$
where
$$\etaSUBpbetauSigmaTObetau\equiv\,
\left[
{\setlength\arraycolsep{0pt}
\begin{array}{c}
\bSigma_{\bbeta}^{-1}\bmu_{\bbeta}\\[2ex]
-\smhalf\bD_p^T\vecof(\bSigma_{\bbeta}^{-1})\\[2ex]
{\displaystyle\stack{1\le i\le m}}\left[
{\setlength\arraycolsep{0pt}
\begin{array}{c}
\bzero_q\\[2ex]
-\smhalf\bD_q^T\vecof\big(\bM_{\qDens(\bSigma^{-1})}\big)\\[2ex]
\bzero_{pq}
\end{array}
}
\right]
\end{array}
}
\right]
$$
and $\bM_{\qDens(\bSigma^{-1})}$ denotes expectation of $\bSigma^{-1}$ with respect to 
the normalization of 
$$\mSUBpbetauSigmaTOSigma\,\mSUBSigmaTOpbetauSigma$$
which is an Inverse G-Wishart density function with natural parameter vector
$\etaSUBpbetauSigmaCONNSigma$ and, according to Table S.1 in 
the online supplement of Wand (2017), leads to 
$$\bM_{\qDens(\bSigma^{-1})}\thickarrow\{\omegaAlgFourA+\smhalf(q+1)\}\{\vecof^{-1}(\bomegaAlgFourB)\}^{-1}$$
where $\omegaAlgFourA$ is the first entry of $\etaSUBpbetauSigmaCONNSigma$
and $\bomegaAlgFourB$ is the vector containing the remaining entries of 
$\etaSUBpbetauSigmaCONNSigma$.

The other factor to stochastic node message update is
$$
\mSUBpbetauSigmaTOSigma
\thickarrow\exp\left\{\left[
\begin{array}{c}
\log|\bSigma|\\[2ex]
\vech(\bSigma^{-1})
\end{array}
\right]^T\etaSUBpbetauSigmaTOSigma
\right\}
$$
where
$$
\etaSUBpbetauSigmaTOSigma
\equiv
\left[
\begin{array}{c}
-\smhalf\,m\\[3ex]
-\smhalf{\displaystyle\sumim}\,\bD_q^T\vecof\{E_{\qDens}(\bu_i\bu_i^T)\}
\end{array}
\right]
$$
with $E_{\qDens}$ denoting expectation with respect to the normalization of 
$$\mSUBpbetauSigmaTObetau\,\mSUBbetauTOpbetauSigma.$$
Then note that 
$$E_{\qDens}(\bu_i\bu_i^T)=\bmu_{\qDens(\bu_i)}\bmu_{\qDens(\bu_i)}^T+\bSigma_{\qDens(\bu_i)}$$
where, as before, $\bmu_{\qDens(\bu_i)}\equiv E_{\qDens}(\bu_i)$ and $\bSigma_{\qDens(\bu_i)}\equiv\Cov_\qDens(\bu_i)$.
Result \ref{res:twoLevelVMPpen} links sub-blocks of $\etaSUBpbetauSigmaTObetau$ with the required sub-vectors 
of $\bmu_{\qDens(\bbeta,\bu)}$ and sub-blocks of $\bSigma_{\qDens(\bbeta,\bu)}$. 
We then call upon Algorithm \ref{alg:TwoLevelNaturalToCommonParameters} 
to obtain $\bmu_{\qDens(\bu_i)}$ and $\bSigma_{\qDens(\bu_i)}$, $1\le i\le m$.

\subsection{Derivation of Result \ref{res:threeLevelMFVB}}

Routine matrix algebraic steps can verify that the $\bmu_{\qDens(\bbeta,\bu)}$ 
and $\bSigma_{\qDens(\bbeta,\bu)}$ updates, 
$$\bmu_{\qDens(\bbeta,\bu)}\leftarrow(\bC^T\RMFVB^{-1}\bC+\DMFVB)^{-1}(\bC^T\RMFVB^{-1}\by + \oMFVB)
\quad \mbox{and}\quad 
\bSigma_{\qDens(\bbeta,\bu)}\leftarrow(\bC^T\RMFVB^{-1}\bC+\DMFVB)^{-1},
$$
with $\bC$, $\DMFVB$ and $\RMFVB$ as defined by 
$$
\bC\equiv[\bX\ \bZ],\ 
\DBLUP\equiv\left[
{\setlength\arraycolsep{4pt}
\begin{array}{cc}
\bSigma_{\bbeta}^{-1} & \bO               \\[1ex]
\bO & {\displaystyle\blockdiag{1\le i\le m}}\left[
{\setlength\arraycolsep{2pt}
\begin{array}{cc}
\bM_{\qDens((\bSigmaLone)^{-1})} &  \bO \\
\bO                & \bI_{n_i}\otimes\bM_{\qDens((\bSigmaLtwo)^{-1})} 
\end{array}}
                     \right]
\end{array}}
\right]\ \mbox{and}\ \RBLUP\equiv\sigma^2\bI
$$
may be written as
$$\bmu_{\qDens(\bbeta,\bu)}\thickarrow(\bB^T\bB)^{-1}\bB^T\bb=\bA^{-1}\ba
\quad\mbox{and}\quad
\bSigma_{\qDens(\bbeta,\bu)}\thickarrow(\bB^T\bB)^{-1}=\bA^{-1}
$$
where $\bB$ and $\bb$ have the sparse three-level forms given by (\ref{eq:BandbGenThreeLev})
with
$$\bb_{ij}\equiv
\left[\begin{array}{c}   
\mu_{\qDens(1/\mysigeps^2)}^{1/2}\by_{ij}\\[2ex]
\Big(\displaystyle{\sum_{i=1}^{m}}\,n_i\Big)^{-1/2}\bSigma_{\bbeta}^{-1/2}\bmu_{\bbeta}\\[2ex]
\bzero\\[1ex]
\bzero
\end{array}
\right],
\quad\bB_{ij}\equiv
\left[\begin{array}{c}   
\mu_{\qDens(1/\mysigeps^2)}^{1/2}\bX_{ij}\\[2ex]
\Big(\displaystyle{\sum_{i=1}^{m}}\,n_i\Big)^{-1/2}\bSigma_{\bbeta}^{-1/2}\\[2ex]
\bO\\[2ex]
\bO
\end{array}
\right],
$$
$$
\bBdot_{ij}\equiv
\left[\begin{array}{c}   
\mu_{\qDens(1/\mysigeps^2)}^{1/2}\bZLone_{ij}\\[2ex]
\bO\\[2ex]
n_i^{-1/2}\Big(\bM_{\qDens(\bSigmaLoneMinusOne)}\Big)^{1/2}\\[2ex]
\bO
\end{array}
\right]
\quad\mbox{and}\quad
\bBdotdot_{ij}\equiv
\left[\begin{array}{c}   
\mu_{\qDens(1/\mysigeps^2)}^{1/2}\bZLtwo_{ij}\\[2ex]
\bO\\[2ex]
\bO\\[2ex]
\Big(\bM_{\qDens(\bSigmaLtwoMinusOne)}\Big)^{1/2}
\end{array}
\right].
$$

\subsection{Derivation of Algorithm \ref{alg:threeLevelMFVB}}

Algorithm \ref{alg:threeLevelMFVB} is the three-level counterpart of
Algorithm \ref{alg:twoLevelMFVB} and its derivation is analogous
to that given for Algorithm \ref{alg:twoLevelMFVB} in Section \ref{sec:DerivTwoLevMFVB}.

The first difference is that the $\bmu_{\qDens(\bbeta,\bu)}$ and 
$\bSigma_{\qDens(\bbeta,\bu)}$ updates are expressible as
three-level sparse matrix least squares problems and so 
the \SolveThreeLevelSparseLeastSquares\ algorithm
(Algorithm \ref{alg:SolveThreeLevelSparseLeastSquares}) is
used for streamlined updating of their relevant sub-blocks.

We still have $\qDens(\sigma^2)$ optimally being an Inverse Chi-Squared density function
but with shape parameter
$$\xi_{\qDens(\sigma^2)}=\nu_{\sigma^2}+\smhalf\sumim\sum_{j=1}^{n_i}\,o_{ij}$$
and rate parameter
{\setlength\arraycolsep{0pt}
\begin{eqnarray*}
\lambda_{\qDens(\sigma^2)}&=&\mu_{\qDens(1/\asigsq)}+
\smhalf\sumim\sum_{j=1}^{n_i}E_{\qDens}\{\Vert\by_{ij}-\bX_{ij}\bbeta-\bZLone_{ij}\buLone_i-\bZLtwo_{ij}\buLtwo_{ij}\Vert^2\}\\[1ex]
&=&\mu_{\qDens(1/\asigsq)}+\smhalf\sumim\sum_{j=1}^{n_i}
\Big(\Vert\by_{ij}-\bX_{ij}\,\bmu_{\qDens(\bbeta)}-\bZLone_{ij}\,\bmu_{\qDens(\buLone_i)}
-\bZLtwo_{ij}\,\bmu_{\qDens(\buLtwo_{ij})}\Vert^2\\
&&\qquad\qquad+\tr(\bX_{ij}^T\bX_{ij}\bSigma_{\qDens(\bbeta)})+\tr\{(\bZLone_{ij})^T\bZLone_{ij}\bSigma_{\qDens(\buLone_i)}\}
+\tr\{(\bZLtwo_{ij})^T\bZLtwo_{ij}\bSigma_{\qDens(\buLtwo_{ij})}\}\\
&&\qquad\qquad+2\,\mbox{tr}\big[(\bZLone_{ij})^T\bX_{ij}E_{\qDens}\{(\bbeta-\bmu_{\qDens(\bbeta)})
                 (\buLone_i-\bmuq{\buLone_i})^T\}\big]\\
&&\qquad\qquad+2\,\mbox{tr}\big[(\bZLtwo_{ij})^T\bX_{ij}E_{\qDens}\{(\bbeta-\bmu_{\qDens(\bbeta)})
                 (\buLtwo_{ij}-\bmuq{\buLtwo_{ij}})^T\}\big]\\
&&\qquad\qquad+2\,\mbox{tr}\big[(\bZLtwo_{ij})^T\bZLone_{ij}E_{\qDens}\{(\buLone_i-\bmuq{\buLone_i})
                 (\buLtwo_{ij}-\bmuq{\buLtwo_{ij}})^T\}\big]\Big).
\end{eqnarray*}
}

The optimal $\qDens(\asigsq)$ density function is unaffected by the change from the two-level case
to the three-level situation.

The random effects covariance matrices are such that
$$\qDens(\bSigmaLone)\ \mbox{is an $\mbox{Inverse-G-Wishart}
\left(\Gfull,\xi_{\qDens(\bSigmaLone)},\bLambda_{\qDens(\bSigmaLone)}\right)$ density function}
$$
where $\xi_{\qDens(\bSigmaLone)}=\nuSigmaLone+2q_1-2+m$ and
$$\bLambda_{\qDens(\bSigmaLone)}=\bM_{\qDens(\bA_{\bSigmaLone}^{-1})}
+\sumim\left(\bmu_{\qDens(\buLone_i)}\bmu_{\qDens(\buLone_i)}\trans + \bSigma_{\qDens(\buLone_i)}\right),$$
whilst
$$\qDens(\bSigmaLtwo)\ \mbox{is an $\mbox{Inverse-G-Wishart}
\left(\Gfull,\xi_{\qDens(\bSigmaLtwo)},\bLambda_{\qDens(\bSigmaLtwo)}\right)$ density function}
$$
where $\xi_{\qDens(\bSigmaLtwo)}=\nuSigmaLtwo+2q_2-2+\sumim n_i$ and
$$\bLambda_{\qDens(\bSigmaLtwo)}=\bM_{\qDens(\bA_{\bSigmaLtwo}^{-1})}
+\sumim\sum_{j=1}^{n_i}\left(\bmu_{\qDens(\buLtwo_{ij})}\bmu_{\qDens(\buLtwo_{ij})}\trans 
+ \bSigma_{\qDens(\buLtwo_{ij})}\right).
$$

The optimal $\qDens(\bASigmaLone)$ and $\qDens(\bASigmaLtwo)$ density functions
have the same derivations and forms as $\qDens(\bASigma)$ in the two-level 
case.

Algorithm \ref{alg:threeLevelMFVB} is a streamlined iterative coordinate ascent for determination
of Kullback-Leibler optimal values of each of the $\qDens$-density parameters in
the Bayesian three-level mixed model (\ref{eq:threeLevelGaussRespBaye}).

\subsection{Derivation of Algorithm \ref{alg:ThreeLevelNaturalToCommonParameters}}

Algorithm \ref{alg:ThreeLevelNaturalToCommonParameters} is the three-level
counterpart of Algorithm \ref{alg:TwoLevelNaturalToCommonParameters} 
and they each use the same logic. Therefore, the 
Algorithm \ref{alg:ThreeLevelNaturalToCommonParameters} follows from
arguments similar to those given in Section \ref{sec:DerivTwoLevNatToComm}.

\subsection{Derivation of Result \ref{res:threeLevelVMPlik}}\label{sec:resThreeLevelVMPlikDeriv}

Note that
{\setlength\arraycolsep{1pt}
\begin{eqnarray*}
\qDens(\bbeta,\bu)&\propto&\mSUBpybetausigsqEpsTObetau\,\mSUBbetauTOpybetausigsqEps\\[1ex]
&=&\exp\left\{
\left[
{\setlength\arraycolsep{0pt}
\begin{array}{c}
\ \ \ \ \ \ \ \ \ \  \bbeta\\
\ \ \ \ \ \vech(\bbeta\bbeta^T)\\[1ex]
\displaystyle{\stack{1\le i\le m}}\left[
{\setlength\arraycolsep{0pt}
\begin{array}{c}
\buLone_i\\
\vech\big(\buLone_i(\buLone_i)^T\big)\\
\vecof\big(\bbeta(\buLone_i)^T\big)
\end{array}
}
\right]\\[5ex]
\displaystyle{\stack{1\le i\le m}}\left[
\displaystyle{\stack{1\le j\le n_i}}\left[
{\setlength\arraycolsep{0pt}
\begin{array}{c}
\buLtwo_{ij}\\[1ex]
\vech\big(\buLtwo_{ij}(\buLtwo_{ij})^T\big)\\[1ex]
\vecof\big(\bbeta(\buLtwo_{ij})^T\big)\\[1ex]
\vecof\big(\buLone_{i}(\buLtwo_{ij})^T\big)
\end{array}
}
\right]
\right]
\end{array}
}
\right]^T
\null\hskip-3mm\etaSUBpybetausigsqEpsTObetau
\right\}\\
&=&\exp\left\{
\left[
\begin{array}{l}
\bbeta\\[1ex]
\bu
\end{array}
\right]^T
\ba
-\smhalf
\left[
\begin{array}{l}
\bbeta\\[1ex]
\bu
\end{array}
\right]^T\bA
\left[
\begin{array}{l}
\bbeta\\[1ex]
\bu
\end{array}
\right]
\right\}\\[1ex]
\end{eqnarray*}
}
where $\ba$ and $\bA$ are as given in Result \ref{res:threeLevelVMPlik}.
The last step uses facts such as $\vech(\bM)=\bD_d^+\vecof(\bM)$
for any symmetric $d\times d$ matrix $\bM$. Standard manipulations 
then lead to
$$\bmu_{\qDens(\bbeta,\bu)}=\bA^{-1}\ba
\quad\mbox{and}\quad
\bSigma_{\qDens(\bbeta,\bu)}=\bA^{-1}
$$
and Result \ref{res:threeLevelVMPlik} then follows from 
extraction of the sub-blocks of $\bx=\bA^{-1}\ba$
and the sub-blocks of $\bA^{-1}$ corresponding
to the non-zero positions of $\bA$.

\subsection{Derivation of Algorithm \ref{alg:threeLevelVMPlik}}\label{sec:algThreeLevelVMPlikDeriv}

As a function of $(\bbeta,\bu)$, the logarithm of the fragment factor is: 
{\setlength\arraycolsep{1pt}
\begin{eqnarray*}
&&\log\,\pDens(\by|\bbeta,\bu,\mysigeps^2)=\,-\frac{1}{2\mysigeps^2}\sumim\sum_{j=1}^{n_i}
\Vert\by_{ij}-\bX_{ij}\bbeta-\bZLone_{ij}\buLone_i-\bZLtwo_{ij}\buLtwo_{ij}\Vert^2+\const\\[1ex]
&&\quad=(1/\mysigeps^2)
\left[
{\setlength\arraycolsep{0pt}
\begin{array}{c}
\ \ \ \ \ \ \ \ \ \  \bbeta\\
\ \ \ \ \ \vech(\bbeta\bbeta^T)\\[1ex]
\displaystyle{\stack{1\le i\le m}}\left[
{\setlength\arraycolsep{0pt}
\begin{array}{c}
\buLone_i\\
\vech\big(\buLone_i(\buLone_i)^T\big)\\
\vecof\big(\bbeta(\buLone_i)^T\big)
\end{array}
}
\right]\\[5ex]
\displaystyle{\stack{1\le i\le m}}\left[
\displaystyle{\stack{1\le j\le n_i}}\left[
{\setlength\arraycolsep{0pt}
\begin{array}{c}
\buLtwo_{ij}\\[1ex]
\vech\big(\buLtwo_{ij}(\buLtwo_{ij})^T\big)\\[1ex]
\vecof\big(\bbeta(\buLone_{ij})^T\big)\\[1ex]
\vecof\big(\buLone_{i}(\buLtwo_{ij})^T\big)
\end{array}
}
\right]
\right]
\end{array}
}
\right]^T
\left[
 \begin{array}{c}
   {\displaystyle\sumim\sum_{j=1}^{n_i}}\bX_{ij}^T\by_{ij}\\
   -\smhalf{\displaystyle\sumim\sum_{j=1}^{n_i}}\bD_p^T\vecof(\bX_{ij}^T\bX_{ij})\\[2ex]
   {\displaystyle\stack{1\le i\le m}}\left[
     \begin{array}{c}
       {\displaystyle\sum_{j=1}^{n_i}}(\bZLone_{ij})^T\by_{ij}\\
       -\smhalf{\displaystyle\sum_{j=1}^{n_i}}\bD_{q_1}^T\vecof\big((\bZLone_{ij})^T\bZLone_{ij}\big)\\
       -{\displaystyle\sum_{j=1}^{n_i}}\vecof\big(\bX_{ij}^T\bZLone_{ij}\big)\\
     \end{array}
   \right]\\[6ex]
   {\displaystyle\stack{1\le i\le m}}\left[
     {\displaystyle\stack{1\le j\le n_i}}\left[
       \begin{array}{c}
         (\bZLtwo_{ij})^T\by_{ij}\\
         -\smhalf \bD_{q_{2}}^T\vecof((\bZLtwo_{ij})^T\bZLtwo_{ij})\\
         -\vecof(\bX_{ij}^T\bZLtwo_{ij})\\
         -\vecof\big((\bZLone_{ij})^T\bZLtwo_{ij}\big)\\
       \end{array}\right]
     \right]
   \end{array}
\right]
+\mbox{const}.
\end{eqnarray*}
}
Therefore, from equations (8) and (9) of Wand (2017),
$$\mSUBpybetausigsqEpsTObetau\thickarrow
\exp\left\{
\left[
{\setlength\arraycolsep{0pt}
\begin{array}{c}
\ \ \ \ \ \ \ \ \ \  \bbeta\\
\ \ \ \ \ \vech(\bbeta\bbeta^T)\\[1ex]
\displaystyle{\stack{1\le i\le m}}\left[
{\setlength\arraycolsep{0pt}
\begin{array}{c}
\buLone_i\\
\vech\big(\buLone_i(\buLone_i)^T\big)\\
\vecof\big(\bbeta(\buLone_i)^T\big)
\end{array}
}
\right]\\[5ex]
\displaystyle{\stack{1\le i\le m}}\left[
\displaystyle{\stack{1\le j\le n_i}}\left[
{\setlength\arraycolsep{0pt}
\begin{array}{c}
\buLtwo_{ij}\\[1ex]
\vech\big(\buLtwo_{ij}(\buLtwo_{ij})^T\big)\\[1ex]
\vecof\big(\bbeta(\buLone_{ij})^T\big)\\[1ex]
\vecof\big(\buLone_{i}(\buLtwo_{ij})^T\big)
\end{array}
}
\right]
\right]
\end{array}
}
\right]^T
\etaSUBpybetausigsqEpsTObetau
\right\}
$$
where
$$\etaSUBpybetausigsqEpsTObetau\equiv\mu_{\qDens(1/\mysigeps^2)}\,
\left[
 \begin{array}{c}
   {\displaystyle\sumim\sum_{j=1}^{n_i}}\bX_{ij}^T\by_{ij}\\
   -\smhalf{\displaystyle\sumim\sum_{j=1}^{n_i}}\bD_p^T\vecof(\bX_{ij}^T\bX_{ij})\\[2ex]
   {\displaystyle\stack{1\le i\le m}}\left[
     \begin{array}{c}
       {\displaystyle\sum_{j=1}^{n_i}}(\bZLone_{ij})^T\by_{ij}\\
       -\smhalf{\displaystyle\sum_{j=1}^{n_i}}\bD_{q_1}^T\vecof\big((\bZLone_{ij})^T\bZLone_{ij}\big)\\
       -{\displaystyle\sum_{j=1}^{n_i}}\vecof\big(\bX_{ij}^T\bZLone_{ij}\big)\\
     \end{array}
   \right]\\[6ex]
   {\displaystyle\stack{1\le i\le m}}\left[
     {\displaystyle\stack{1\le j\le n_i}}\left[
       \begin{array}{c}
         (\bZLtwo_{ij})^T\by_{ij}\\
         -\smhalf \bD_{q_{2}}^T\vecof((\bZLtwo_{ij})^T\bZLtwo_{ij})\\
         -\vecof(\bX_{ij}^T\bZLtwo_{ij})\\
         -\vecof\big((\bZLone_{ij})^T\bZLtwo_{ij}\big)\\
       \end{array}\right]
     \right]
   \end{array}
\right]
$$
and $\mu_{\qDens(1/\mysigeps^2)}$ denotes expectation of $1/\mysigeps^2$ with respect to 
the normalization of
$$\mSUBpybetausigsqEpsTOsigsqEps\,\mSUBsigsqEpsTOpybetausigsqEps.$$
This is an Inverse $\chi^2$ density function with natural parameter vector
$\etaSUBpybetausigsqEpsCONNsigsqEps$ and, from Table S.1 in 
the online supplement of Wand (2017), we have
$$\mu_{\qDens(1/\mysigeps^2)}\thickarrow\Big(\big(\etaSUBpybetausigsqEpsCONNsigsqEps\big)_1+1\Big)/
\big(\etaSUBpybetausigsqEpsCONNsigsqEps\big)_2.
$$

The other factor to stochastic node message update is
$$
\mSUBpybetausigsqEpsTOsigsqEps
\thickarrow\exp\left\{\left[
\begin{array}{c}
\log(\mysigeps^2)\\[2ex]
1/\mysigeps^2
\end{array}
\right]^T\etaSUBpybetausigsqEpsTOsigsqEps
\right\}
$$
where
$$
\etaSUBpybetausigsqEpsTOsigsqEps
\equiv
\left[
\begin{array}{c}
-\smhalf{\displaystyle\sumim\sum_{j=1}^{n_i}}\, o_{ij}\\[3ex]
-\smhalf{\displaystyle\sumim}\,E_{\qDens}\{\Vert\by_{ij}-\bX_{ij}\bbeta-\bZLone_{ij}\buLone_i-\bZLtwo_{ij}\buLtwo_{ij}\Vert^2\}
\end{array}
\right]
$$
with $E_{\qDens}$ denoting expectation with respect to the normalization of 
$$\mSUBpybetausigsqEpsTObetau\,\mSUBbetauTOpybetausigsqEps.$$
Observing that
{\setlength\arraycolsep{1pt}
\begin{eqnarray*}
&&E_{\qDens}\{\Vert\by_{ij}-\bX_{ij}\bbeta-\bZLone_{ij}\buLone_i-\bZLtwo_{ij}\buLtwo_{ij}\Vert^2\}\\[1ex]
&&\qquad=\Vert\by_{ij}-\bX_{ij}\,\bmu_{\qDens(\bbeta)}-\bZLone_{ij}\,\bmu_{\qDens(\buLone_i)}
-\bZLtwo_{ij}\,\bmu_{\qDens(\buLtwo_{ij})}\Vert^2
+\tr(\bX_{ij}^T\bX_{ij}\bSigma_{\qDens(\bbeta)})\\
&&\qquad\qquad+\tr\{(\bZLone_{ij})^T\bZLone_{ij}\bSigma_{\qDens(\buLone_i)}\}
+\tr\{(\bZLtwo_{ij})^T\bZLtwo_{ij}\bSigma_{\qDens(\buLtwo_{ij})}\}\\
&&\qquad\qquad+2\,\mbox{tr}\big[(\bZLone_{ij})^T\bX_{ij}E_{\qDens}\{(\bbeta-\bmu_{\qDens(\bbeta)})
                 (\buLone_i-\bmuq{\buLone_i})^T\}\big]\\
&&\qquad\qquad+2\,\mbox{tr}\big[(\bZLtwo_{ij})^T\bX_{ij}E_{\qDens}\{(\bbeta-\bmu_{\qDens(\bbeta)})
                 (\buLtwo_{ij}-\bmuq{\buLtwo_{ij}})^T\}\big]\\
&&\qquad\qquad+2\,\mbox{tr}\big[(\bZLtwo_{ij})^T\bZLone_{ij}E_{\qDens}\{(\buLone_i-\bmuq{\buLone_i})
                 (\buLtwo_{ij}-\bmuq{\buLtwo_{ij}})^T\}\big]
\end{eqnarray*}
}
Result \ref{res:threeLevelVMPlik} shows how the sub-blocks of $\etaSUBpybetausigsqEpsCONNbetau$ 
are related to the required sub-vectors 
of $\bmu_{\qDens(\bbeta,\bu)}$ and sub-blocks of $\bSigma_{\qDens(\bbeta,\bu)}$.
These matrices are obtained from  $\etaSUBpybetausigsqEpsCONNbetau$ in the call 
to \ThreeLevelNaturalToCommonParameters\ algorithm 
(Algorithm \ref{alg:ThreeLevelNaturalToCommonParameters}).

\subsection{Derivation of Result \ref{res:threeLevelVMPpen}}

The derivation of Result \ref{res:threeLevelVMPpen} is very 
similar to that for Result \ref{res:threeLevelVMPlik}.

\subsection{Derivation of Algorithm \ref{alg:threeLevelVMPpen}}

The logarithm on the fragment factor is, as a function of $(\bbeta,\bu)$: 
{\setlength\arraycolsep{1pt}
\begin{eqnarray*}
&&\log\,\pDens(\bbeta,\bu|\,\bSigmaLone,\bSigmaLtwo)
=\,\smhalf(\bbeta-\bmu_{\bbeta})^T\bSigma_{\bbeta}^{-1}(\bbeta-\bmu_{\bbeta})
-\smhalf\sumim(\buLone_i)^T(\bSigmaLone)^{-1}\buLone_i\\
&&\qquad\qquad\qquad\qquad\qquad\qquad
-\smhalf\sumim\sum_{j=1}^{n_i}(\buLtwo_{ij})^T(\bSigmaLtwo)^{-1}\buLtwo_{ij}
+\const\\[1ex]
&&\qquad\qquad=\left[
{\setlength\arraycolsep{0pt}
\begin{array}{c}
\ \ \ \ \ \ \ \ \ \  \bbeta\\
\ \ \ \ \ \vech(\bbeta\bbeta^T)\\[1ex]
\displaystyle{\stack{1\le i\le m}}\left[
{\setlength\arraycolsep{0pt}
\begin{array}{c}
\buLone_i\\
\vech\big(\buLone_i(\buLone_i)^T\big)\\
\vecof\big(\bbeta(\buLone_i)^T\big)
\end{array}
}
\right]\\[5ex]
\displaystyle{\stack{1\le i\le m}}\left[
\displaystyle{\stack{1\le j\le n_i}}\left[
{\setlength\arraycolsep{0pt}
\begin{array}{c}
\buLtwo_{ij}\\[1ex]
\vech\big(\buLtwo_{ij}(\buLtwo_{ij})^T\big)\\[1ex]
\vecof\big(\bbeta(\buLtwo_{ij})^T\big)\\[1ex]
\vecof\big(\buLone_{i}(\buLtwo_{ij})^T\big)
\end{array}
}
\right]
\right]
\end{array}
}
\right]^T
\left[
 \begin{array}{c}
 \bSigma_{\bbeta}^{-1}\bmu_{\bbeta}\\[2ex]
  -\smhalf\bD_p^T\vecof(\bSigma_{\bbeta}^{-1})\\[2ex]
   {\displaystyle\stack{1\le i\le m}}\left[
     \begin{array}{c}
       \bzero_{q_1}\\[1ex]
       -\smhalf\bD_{q_{1}}^T\vecof\big((\bSigmaLone)^{-1}\big)\\[2ex]
       \bzero_{pq_1}\\
     \end{array}
   \right]\\[6ex]
   {\displaystyle\stack{1\le i\le m}}\left[
     {\displaystyle\stack{1\le j\le n_i}}\left[
       \begin{array}{c}
         \bzero_{q_2}\\
         -\smhalf \bD_{q_{2}}^T\vecof\big((\bSigmaLtwo)^{-1}\big)\\
         \bzero_{pq_2}\\
         \bzero_{q_1q_2}\\
       \end{array}\right]
     \right]
   \end{array}
 \right].\\
\end{eqnarray*}
}
Therefore, from equations (8) and (9) of Wand (2017),
$$\mSUBpbetauSigmaLoneLtwoTObetau\thickarrow
\exp\left\{
\left[
{\setlength\arraycolsep{0pt}
\begin{array}{c}
\ \ \ \ \ \ \ \ \ \  \bbeta\\
\ \ \ \ \ \vech(\bbeta\bbeta^T)\\[1ex]
\displaystyle{\stack{1\le i\le m}}\left[
{\setlength\arraycolsep{0pt}
\begin{array}{c}
\buLone_i\\
\vech\big(\buLone_i(\buLone_i)^T\big)\\
\vecof\big(\bbeta(\buLone_i)^T\big)
\end{array}
}
\right]\\[5ex]
\displaystyle{\stack{1\le i\le m}}\left[
\displaystyle{\stack{1\le j\le n_i}}\left[
{\setlength\arraycolsep{0pt}
\begin{array}{c}
\buLtwo_{ij}\\[1ex]
\vech\big(\buLtwo_{ij}(\buLtwo_{ij})^T\big)\\[1ex]
\vecof\big(\bbeta(\buLtwo_{ij})^T\big)\\[1ex]
\vecof\big(\buLone_{i}(\buLtwo_{ij})^T\big)
\end{array}
}
\right]
\right]
\end{array}
}
\right]^T
\etaSUBpbetauSigmaLoneLtwoTObetau
\right\}
$$
where
$$\etaSUBpbetauSigmaLoneLtwoTObetau\equiv\,
\left[
 \begin{array}{c}
 \bSigma_{\bbeta}^{-1}\bmu_{\bbeta}\\[2ex]
  -\smhalf\bD_p^T\vecof(\bSigma_{\bbeta}^{-1})\\[2ex]
   {\displaystyle\stack{1\le i\le m}}\left[
     \begin{array}{c}
       \bzero_{q_1}\\[1ex]
       -\smhalf\bD_{q_{1}}^T\vecof\big(\bM_{\qDens((\bSigmaLone)^{-1})}\big)\\[2ex]
       \bzero_{pq_1}\\
     \end{array}
   \right]\\[6ex]
   {\displaystyle\stack{1\le i\le m}}\left[
     {\displaystyle\stack{1\le j\le n_i}}\left[
       \begin{array}{c}
         \bzero_{q_2}\\
         -\smhalf \bD_{q_{2}}^T\vecof\big(\bM_{\qDens((\bSigmaLtwo)^{-1})}\big)\\
         \bzero_{pq_2}\\
         \bzero_{q_1q_2}\\
       \end{array}\right]
     \right]
   \end{array}
 \right].
$$
Here $\bM_{\qDens((\bSigmaLone)^{-1})}$ denotes expectation of $(\bSigmaLone)^{-1}$ with respect to 
the normalization of 
$$\mSUBpbetauSigmaLoneLtwoTOSigmaLone\,\mSUBSigmaLoneTOpbetauSigmaLoneLtwo$$
which is an Inverse G-Wishart density function with natural parameter vector
$\etaSUBpbetauSigmaLoneLtwoCONNSigmaLone$ and, according to Table S.1 in 
the online supplement of Wand (2017), leads to
$$\bM_{\qDens((\bSigmaLone)^{-1})}\thickarrow\{\omegaAlgEightA+\smhalf(q_1+1)\}\{\vecof^{-1}(\omegaAlgEightB)\}^{-1}$$
where $\omegaAlgEightA$ is the first entry of $\etaSUBpbetauSigmaLoneLtwoCONNSigmaLone$
and $\omegaAlgEightB$ is the vector containing the remaining entries of 
$\etaSUBpbetauSigmaLoneLtwoCONNSigmaLone$. The treatment of $\bM_{\qDens((\bSigmaLtwo)^{-1})}$
is analogous.

The message from $\pDens(\bbeta,\bu|\bSigmaLone,\bSigmaLtwo)$ to $\bSigmaLone$ is 
$$
\mSUBpbetauSigmaLoneLtwoTOSigmaLone
\thickarrow\exp\left\{\left[
\begin{array}{c}
\log|\bSigmaLone|\\[2ex]
\vech\big((\bSigmaLone)^{-1}\big)
\end{array}
\right]^T\etaSUBpbetauSigmaLoneLtwoTOSigmaLone
\right\}
$$
where
$$
\etaSUBpbetauSigmaLoneLtwoTOSigmaLone
\equiv
\left[
\begin{array}{c}
-\smhalf\,m\\[3ex]
-\smhalf{\displaystyle\sumim}\,\bD_{q_1}^T\vecof[E_{\qDens}\{\buLone_i(\buLone_i)^T\}]
\end{array}
\right]
$$
with $E_{\qDens}$ denoting expectation with respect to the normalization of 
$$\mSUBpbetauSigmaLoneLtwoTObetau\,\mSUBbetauTOpbetauSigmaLoneLtwo.$$
Similarly, the message from $\pDens(\bbeta,\bu|\bSigmaLone,\bSigmaLtwo)$ to $\bSigmaLtwo$ is 
$$
\mSUBpbetauSigmaLoneLtwoTOSigmaLtwo
\thickarrow\exp\left\{\left[
\begin{array}{c}
\log|\bSigmaLtwo|\\[2ex]
\vech\big((\bSigmaLtwo)^{-1}\big)
\end{array}
\right]^T\etaSUBpbetauSigmaLoneLtwoTOSigmaLtwo
\right\}
$$
where
$$
\etaSUBpbetauSigmaLoneLtwoTOSigmaLtwo
\equiv
\left[
\begin{array}{c}
-\smhalf\,{\displaystyle\sum_{i=1}^{m}}\,n_i\\[3ex]
-\smhalf{\displaystyle\sumim\sum_{j=1}^{n_i}}\,\bD_{q_2}^T\vecof[E_{\qDens}\{\buLtwo_{ij}(\buLtwo_{ij})^T\}]
\end{array}
\right].
$$

Now note that 
$$E_{\qDens}\{\buLone_i(\buLone_i)^T\}=\bmu_{\qDens(\buLone_i)}\bmu_{\qDens(\buLone_i)}^T+\bSigma_{\qDens(\buLone_i)}
\quad\mbox{and}\quad
E_{\qDens}\{\buLtwo_{ij}(\buLtwo_{ij})^T\}=\bmu_{\qDens(\buLtwo_{ij})}\bmu_{\qDens(\buLtwo_{ij})}^T+\bSigma_{\qDens(\buLtwo_{ij})}
$$
where, similar to before, 
$\bmu_{\qDens(\buLone_i)}\equiv E_{\qDens}(\buLone_i)$, $\bSigma_{\qDens(\buLone_i)}\equiv\Cov_\qDens(\buLone_i)$
and $\bmu_{\qDens(\buLtwo_{ij})}$ and $\bSigma_{\qDens(\buLtwo_{ij})}$ is defined similarly.
Result \ref{res:threeLevelVMPpen} links sub-blocks of $\etaSUBpbetauSigmaLoneLtwoCONNbetau$ with the required sub-vectors 
of $\bmu_{\qDens(\bbeta,\bu)}$ and sub-blocks of $\bSigma_{\qDens(\bbeta,\bu)}$. 
We then call upon Algorithm \ref{alg:ThreeLevelNaturalToCommonParameters} 
to obtain $\bmu_{\qDens(\buLone_i)}$ and $\bSigma_{\qDens(\buLone_i)}$, $1\le i\le m$,
as well as $\bmu_{\qDens(\buLtwo_{ij})}$ and $\bSigma_{\qDens(\buLtwo_{ij})}$, $1\le i\le m$, $1\le j\le n_i$.

Algorithm \ref{alg:threeLevelVMPpen} is a proceduralization of each of these results.

\end{document}